\documentclass[aps,twocolumn,prd,showpacs,showkeys,preprintnumbers,superscriptaddress,bibnotes,floatfix,longbibliography,notitlepage,nofootinbib]{revtex4-2}
\usepackage[colorlinks=true,breaklinks=true]{hyperref}
\usepackage[utf8]{inputenc}
\usepackage{url}
\usepackage{amssymb}
\usepackage{gensymb}
\usepackage{mathtools}
\usepackage[dvipsnames]{xcolor}
\definecolor{darkred}{rgb}{0.5,0,0}
\definecolor{darkblue}{rgb}{0,0,0.5}
\definecolor{firebrick}{rgb}{0.75,0.125,0.125}
\definecolor{darkgreen}{rgb}{0,0.5,0}
\hypersetup{urlcolor=darkblue,
	    citecolor=darkgreen,
	    linkcolor=firebrick}

\usepackage{orcidlink}
\usepackage{siunitx}
\usepackage{xspace}
\usepackage{fontawesome5}
\usepackage{booktabs}
\usepackage{multirow}
\usepackage{makecell}
\usepackage[textsize=small]{todonotes}
\setuptodonotes{inline}
\usepackage{enumitem}
\usepackage{placeins}
\usepackage{balance}
\usepackage[pagewise]{lineno}

\long\def\exclude#1{}

\newcommand{\ie}{{i.e.}}

\newcommand{\eg}{{e.g.}}

\newcommand{\eq}{Eq.}

\newcommand{\fig}{Fig.}

\newcommand{\Refe}{Ref.}
\newcommand{\Refes}{Refs.}
\newcommand{\equ}[1]{\eq~(\ref{equ:#1})}
\newcommand{\figu}[1]{\fig~\ref{fig:#1}}
\newcommand{\txs}{TXS~0506+056\xspace}
\newcommand{\ngc}{NGC~1068\xspace}
\newcommand{\plenum}{PLE$\nu$M\xspace}

\newcommand{\orcid}[1]{\href{https://orcid.org/#1}{\includegraphics[width=10pt]{orcid.pdf}}}
\bibpunct{[}{]}{,}{n}{}{,}
\newcommand{\drv}[1]{\ensuremath{{\rm d}#1}}
\newcommand{\aeff}{\ensuremath{A_{\rm eff}}}

\newcommand{\ecut}{\ensuremath{E_{\rm cut}}}

\newcommand{\ec}{\ensuremath{{\rm PLC}}}
\newcommand{\plw}{\ensuremath{{\rm PL}}}

\newcommand{\phiB}{\ensuremath{{\rm \Phi_{atm}}}}

\newcommand{\phizero}{\ensuremath{{\rm \Phi_{0}}}}

\DeclareSIUnit{\EeV}{EeV}
\DeclareSIUnit{\PeV}{PeV}
\DeclareSIUnit{\TeV}{TeV}
\DeclareSIUnit{\Mpc}{Mpc}
\DeclareSIUnit{\Gpc}{Gpc}
\DeclareSIUnit{\erg}{erg}
\DeclareSIUnit{\year}{yr}

\begin{document}

\title{Beyond first light: Global monitoring for high-energy neutrino astronomy~\href{https://github.com/PLEnuM-group/Plenum}{\faGithubSquare} 
}

\author{Lisa Johanna Schumacher
\orcidlink{0000-0001-8945-6722}} 
\email{lisa.j.schumacher@fau.de}
\affiliation{Erlangen Centre for Astroparticle Physics (ECAP), Friedrich-Alexander-Universität Erlangen-Nürnberg, 91058 Erlangen, Germany}

\author{Mauricio Bustamante
\orcidlink{0000-0001-6923-0865}} 
\email{mbustamante@nbi.ku.dk}
\affiliation{Niels Bohr International Academy, Niels Bohr Institute, University of Copenhagen, 2100 Copenhagen, Denmark}

\author{Matteo Agostini
\orcidlink{0000-0003-1151-5301}} 
\email{matteo.agostini@ucl.ac.uk}
\affiliation{Department of Physics and Astronomy, University College London, Gower Street, London WC1E 6BT, UK}

\author{Foteini Oikonomou
\orcidlink{0000-0002-0525-3758}} 
\email{foteini.oikonomou@ntnu.no}
\affiliation{Institutt for fysikk, Norwegian University of Science and Technology, Høgskoleringen 5, NO-7491 Trondheim, Norway}

\author{Elisa Resconi
\orcidlink{0000-0003-0705-2770}} 
\email{elisa.resconi@tum.de}
\affiliation{Technische Universit\"at München, TUM School of Natural Sciences, Physics Department, James-Frank-Stra\ss e 1, D-85748 Garching
bei M\"unchen, Germany}

\date{\today}

\begin{abstract}

Decades of progress have culminated in \textit{first light} for high-energy neutrino astronomy: the identification of the first astrophysical sources of TeV--PeV neutrinos by the IceCube neutrino telescope, the active galactic nuclei \ngc and \txs.
Today, the prospect of going \textit{beyond first light} to build high-energy neutrino astronomy in earnest by discovering many more neutrino sources is hampered by the relatively low rate of neutrino detection and the limited view of the sky afforded by IceCube, the single cubic-kilometer-scale neutrino telescope in operation.  Yet, this will not stand for much longer.
Already today, and over the next 10--20 years, the combined observations of new neutrino telescopes, larger and distributed around the world, will have the potential for transformative progress.
Together, they will increase the global rate of neutrino detection by up to 30 times and continuously monitor the entire sky. 
Within a new joint analysis network---the Planetary Neutrino Monitoring network (\plenum)~\href{https://github.com/PLEnuM-group/Plenum}{\faGithubSquare}---we make detailed forecasts for the discovery of steady-state astrophysical sources of high-energy neutrinos.
We show that a combined analysis of global data will expedite source discovery---in some cases, by decades---and enable the detection of fainter sources anywhere in the sky, discovering up to tens of new neutrino sources.

\end{abstract}

\maketitle


\section{Introduction}

High-energy astrophysical neutrinos, with TeV--PeV energies, hold the potential to answer long-standing open questions in astrophysics~\cite{Ackermann:2019ows, AlvesBatista:2019tlv, AlvesBatista:2021eeu, Ackermann:2022rqc, MammenAbraham:2022xoc} and particle physics~\cite{Ahlers:2018mkf, Ackermann:2019cxh, Arguelles:2019rbn, AlvesBatista:2021eeu, Adhikari:2022sve, Ackermann:2022rqc}: notably, what are the sources of ultra-high-energy cosmic rays and how does fundamental physics behave at the highest energies.  Answers to these questions would represent transformative progress.  Yet, a decade after the discovery of high-energy astrophysical neutrinos by the IceCube neutrino telescope~\cite{IceCube:2013low}, progress, while steady, is bounded by the experimental limitations that are natural in a nascent field.  

IceCube---still the largest neutrino telescope in operation---while enormously successful, has a relatively low detection rate of high-energy astrophysical neutrinos and a limited view of the sky with them, both of which stall progress. KM3NeT-ARCA~\cite{KM3Net:2016zxf}, which has recently reported the first observation of an astrophysical neutrino with more than 100 PeV~\cite{KM3Net:2016zxf, KM3NeT:2025npi}, and Baikal-GVD~\cite{Baikal-GVD:2019kwy} which has recently observed the diffuse astrophysical neutrino flux at over 3$\sigma$~\cite{Baikal-GVD:2022fis}, are still under construction.

We are now at a pivotal time for the field: an upcoming new generation of high-energy neutrino telescopes, currently under construction and planning, will address the above  limitations~\cite{Ackermann:2022rqc, MammenAbraham:2022xoc, Guepin:2022qpl}.  Because some of the planned detectors will be larger than IceCube, they will provide higher detection rates.  Because they will be located elsewhere, they will observe neutrinos coming from different regions of the sky.  However, there is still a risk that any one of these detectors, individually, may be insufficient to give definitive answers to the above questions.  

Combined, however, the detectors will all but eliminate this risk.  In this paper, we show via detailed projections based on estimated detector capabilities how much analyses that use their combined detection will outperform analyses that use any single one of them.  We focus on a high-priority science case: the discovery of new sources of high-energy astrophysical neutrinos---of which, today, we know less than a handful---and the characterization of new and known sources.  Finding many and possibly diverse sources is an essential step to building high-energy neutrino astronomy in earnest.

\begin{figure*}[t!]
 \centering
 \includegraphics[width=\textwidth, trim={1.2cm 0.5cm 0 0}, clip]{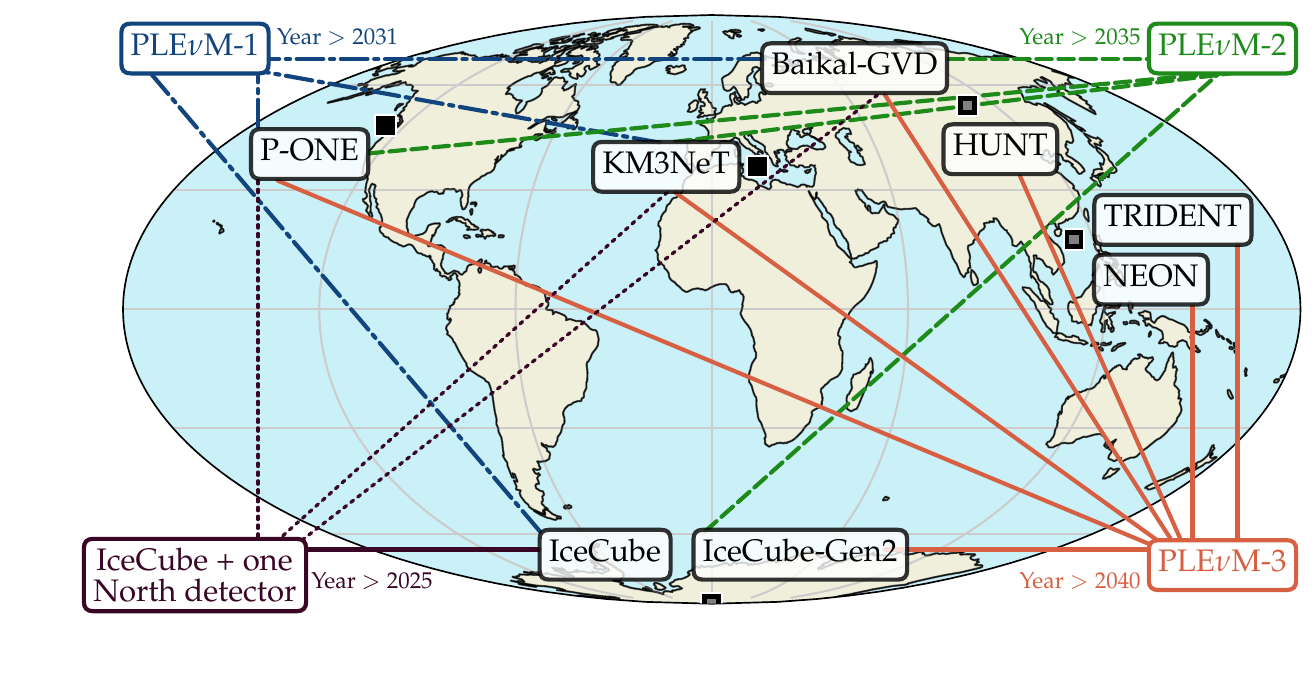}
 \caption{\textbf{\textit{Locations of the neutrino telescopes that make up \plenum.}} PLE$\nu$M-1 consists of IceCube, plus IceCube-sized telescopes placed at the locations of KM3NeT, P-ONE, and Baikal-GVD. PLE$\nu$M-2 is the same but with IceCube replaced by a detector 7.5 times larger, akin to IceCube-Gen2. PLE$\nu$M-3 adds the three planned Chinese detectors to \plenum-2.}
 \label{fig:skymap_detector_locations}
\end{figure*}

Figure~\ref{fig:skymap_detector_locations} shows the present and future high-energy neutrino telescopes we consider.  This comprises several in-ice and in-water neutrino telescopes based on the same detection strategy as IceCube, but of different sizes and built at different geographical locations: Baikal-GVD~\cite{Baikal-GVD:2019kwy} in Lake Baikal, KM3NeT~\cite{KM3Net:2016zxf} in the Mediterranean Sea, both under construction; P-ONE~\cite{P-ONE:2020ljt} in Cascadia Basin and IceCube-Gen2~\cite{IceCube-Gen2:2020qha} at the South Pole, planned for the 2030s; and NEON~\cite{Zhang:2024slv} and TRIDENT~\cite{Ye:2022vbk} in the South China Sea, and HUNT~\cite{Huang:2023mzt}, possibly in Lake Baikal, planned for the 2040s.  

To assess their combined power, we introduce the Planetary Neutrino Monitoring network (\plenum), a joint analysis framework to combine the observations of present and future high-energy neutrino telescopes, and to extract physical insight from them. We consider different detector combinations that represent the different stages in the development of upcoming telescopes (\figu{skymap_detector_locations}): the early 2030s (\plenum-1), the mid-2030s (\plenum-2), and the 2040s (\plenum-3).  We accompany our calculations with the publicly available \plenum software tool~\href{https://github.com/PLEnuM-group/Plenum}{\faGithubSquare}~\cite{github-plenum} that implements our methods and that has built-in flexibility to extend them. 

With \plenum, our goal is to motivate the community of high-energy neutrino physics and astrophysics to consider the future potential of the field globally, not limited by the capabilities of individual detectors.  Building on a history of collaboration between IceCube and ANTARES~\cite{ANTARES:2015moa, ANTARES:2017bia, ANTARES:2018nyb, ANTARES:2019itg, ANTARES:2020srt, ANTARES:2020leh} and the forthcoming collaboration between IceCube and KM3NeT, we wish to motivate and prepare for future cross-experiment analyses on an even larger scale, following the examples of classical observational astronomy and gravitational-wave detection.

The rest of this paper is organized as follows.  Section~\ref{sec:synopsis} gives a synopsis of our work. Section~\ref{sec:method} presents our working assumptions and methods.  Section~\ref{sec:flux_models} introduces the models of astrophysical neutrino flux that we use as benchmarks.  Section~\ref{sec:statistics} introduces the statistical methods we use to compute our projections.  Section~\ref{sec:results} shows our results on neutrino source discovery and characterization of their energy spectra.  Section~\ref{sec:summary} summarizes and concludes.


\section{Synopsis}
\label{sec:synopsis}

We illustrate the power of \plenum by making forecasts of the discovery potential of point-like high-energy neutrino sources, one of the most prominent science goals of the field~\cite{Ackermann:2019ows, Ackermann:2022rqc, Guepin:2022qpl}.  

Most high-energy astrophysical neutrinos detected by IceCube originate in so-far unresolved extragalactic sources.  Together, these neutrinos make up the diffuse flux that IceCube detects in the TeV--PeV energy range~\cite{IceCube:2020wum, Abbasi:2021qfz}.  Viable candidate source classes include starburst galaxies~\cite{Murase:2013rfa, Liu:2013wia, Katz:2013ooa, Tamborra:2014xia, Anchordoqui:2014yva, Chang:2014hua, Senno:2015tra}, galaxy clusters~\cite{Murase:2013rfa, Murase:2008yt, Kotera:2009ms, Fang:2017zjf}, and multiple types of active galactic nuclei~\cite{Kimura:2014jba, Alvarez-Muniz:2004xlu, Stecker:2013fxa, Kalashev:2014vya, Wang:2016vbf, Lamastra:2017iyo, Liu:2017bjr}, among others. Further, neutrino sources could be transient in their emission---like flaring blazars or gamma-ray bursts---or steady-state, at least on the time scales over which we observe them---like some active galaxies.

So far, despite numerous searches~\cite{IceCube:2019cia, IceCube:2019yml, ANTARES:2020srt, ANTARES:2020vzs, IceCube:2020wum, ANTARES:2020dpd, IceCube:2020nig, IceCube:2020mzw, ANTARES:2020zng, AMONTeam:2020otr, IceCube:2021xar, IceCube:2021waz, IceCube:2021slf, IceCube:2021pgw, IceCube:2022jpz, IceCube:2022ham, IceCube:2022heu, IceCube:2022mjy, IceCube:2023esf, IceCube:2023htm, IceCube:2023oqe, IceCube:2023myz}, only a handful of individual candidate high-energy neutrino sources have been identified: the flaring blazar TXS 0506+056~\cite{IceCube:2018cha, IceCube:2018dnn}---a transient source---the Seyfert galaxy NGC 1068---a steady-state source~\cite{IceCube:2022der}---and, possibly, tidal disruption events AT2019dsg~\cite{Stein:2020xhk}, AT2019fdr~\cite{Reusch:2021ztx}, and AT2019aalc~\cite{vanVelzen:2021zsm}.  The absence of many prominent sources has led us to conclude that neutrino sources are likely abundant, but that most are individually  weak~\cite{Silvestri:2009xb, Murase:2016gly, IceCube:2022ham}, making their detection in present telescopes challenging.  Further, the main strategy adopted by neutrino telescopes to search for sources uses \textit{through-going muon tracks} (Sec.~\ref{sec:event_rate-how_detected}) that reach them \textit{through} the Earth, leaving roughly half of the sky comparatively less closely inspected---in the case of IceCube, the Southern Hemisphere.

To demonstrate how \plenum will overcome both of the above limitations, we forecast the capability to discover steady-state sources from the present to the year 2050. We adopt a tentative timeline for when future detectors may come online, though the message of our work does not hinge on its being followed precisely.  We study sources like \ngc, steady-state analogs of \txs, and others with a different neutrino brightness and emission spectrum located elsewhere in the sky.  (There is also preliminary work on the use of \plenum in looking for 
transient sources~\cite{Schumacher:2023mxz} and measuring the diffuse flux~\cite{Schumacher:2021hhm}.) 
If not explicitly labeled as using experimental data, all calculations are based on simulated data.

\textbf{\textit{Already in the early 2030s, with \plenum we will be able to discover high-energy neutrino sources that are half as bright as \ngc, as bright as \txs, or significantly dimmer than both, anywhere in the sky, characterize their energy spectrum, and put models of neutrino production to the test.}}  With IceCube alone, achieving the same would require taking data past the year 2050.


\section{High-energy neutrinos in \plenum}
\label{sec:method}


\subsection{Exploring future possible scenarios}
\label{sec:method_what_is_plenum}

We compare the performance of IceCube alone \textit{vs.}~IceCube combined with the other future, similar in-ice and in-water Cherenkov detectors listed earlier: Baikal-GVD, KM3NeT, P-ONE, IceCube-Gen2, HUNT, NEON, and TRIDENT.  
We place HUNT in Lake Baikal, but its location is still being decided, and it might be placed instead in the South China Sea, too.
To produce the results in this paper, these are taken to be mock detectors modeled after IceCube, except for their location and size.  We elaborate on this simplification in Sec.~\ref{sec:method_similar_detectors}.  We do not comment on the technological or logistical feasibility of building these detectors.

Figure~\ref{fig:skymap_detector_locations} shows the locations of the detectors. We compare their performance in five possible scenarios of the future of high-energy neutrino telescopes:
\begin{description}
 \item[\textbf{IceCube-only}]
  The only future neutrino telescope in operation is IceCube. This scenario is counterfactual since Baikal-GVD and KM3NeT already operate in partial configurations today, and it is intended solely as a baseline against which to compare the other scenarios.
 \item[\textbf{IceCube + one Northern detector}]
  In addition to IceCube, we place one IceCube-sized detector at the location of KM3NeT in the Northern Hemisphere. Our conclusions would be the same when adding P-ONE or Baikal-GVD instead since they are located at similar latitudes.
 \item[\textbf{\plenum-1 (early 2030s)}]
  This consists of IceCube, plus three detectors in the Northern Hemisphere, each IceCube-sized, placed at the locations of Baikal-GVD, KM3NeT, and P-ONE.
 \item[\textbf{\plenum-2 (mid-2030s)}]
  This consists of a detector 7.5~times larger than IceCube at the South Pole, akin to IceCube-Gen2, plus three detectors in the Northern Hemisphere, each IceCube-sized, placed at the locations of Baikal-GVD, KM3NeT, and P-ONE. 
  \item[\textbf{\plenum-3 (2040s)}]
  This is \plenum-2 plus three large detectors: TRIDENT~\cite{Ye:2023dch_TRIDENT} (7.5~times IceCube), NEON~\cite{Zhang:2024slv} (10~times IceCube), and HUNT~\cite{Huang:2023mzt_HUNT} (30~times IceCube).
\end{description}
These are the same definitions of the \plenum configurations as in \Refe~\cite{Schumacher:2021hhm}, with the addition of  TRIDENT, NEON, and HUNT in \plenum-3.

In Figs.~\ref{fig:signal_model}, \ref{fig:source_discovery_vs_declination}, \ref{fig:ngc_contour}, and \ref{fig:discrimination_soft_spectrum_vs_declination}, we assume a live time of 3576.1~days, about 10~years, for each detector in the above scenarios other than IceCube, for which we assume the current accumulated live time of 14 years.
Our choice of using 10 years in these figures is motivated by the 2008--2018 IceCube data sample~\cite{IceCube:2019cia} on which we base our event-rate computations (Sec.~\ref{sec:method_effective_area}). 

Thus, in these figures, we estimate the performance of \plenum by adding the future data collected by 10 years of \plenum-1, \plenum-2, or \plenum-3 to the 14 years of IceCube data. This means that the exposure of \plenum-1 in these figures is around 3.9 times that of 14 years of IceCube alone; the exposure of \plenum-2, around 8.5 times; and the exposure of \plenum-3, around 42 times.  Evidently, this is a simplified scenario where all the detectors in each \plenum configuration start taking data simultaneously.

In Figs.~\ref{fig:time_evolution_discovery_soft_spectrum}, \ref{fig:time_evolution_discrimination_pl_vs_plc}, and \ref{fig:time_evolution_discovery_hard_spectrum}, we show instead a more realistic scenario where each detector starts operations at different times, following a tentative timeline (Table~\ref{tab:detectors}).


\subsection{Modeling the detectors}
\label{sec:method_similar_detectors}

The detection capabilities of different neutrino telescopes depend on their specific features, such as the detector geometry, interaction medium, and spacing between detector strings.  
Presently, however, detailed information on this---as represented by the effective area of the detector and by its energy and angular resolution---is publicly unavailable or available only partially for most of the upcoming detectors that we consider.
There is ongoing progress on this, especially from KM3NeT~\cite{km3net_irf} 
(see also \Refe~\cite{Twagirayezu:2023cpv} for P-ONE). 

To ensure a straightforward comparison between detectors, we assume that all of them have identical detection performance as IceCube, \ie, identical effective area, energy, and angular resolution (Sec.~\ref{sec:event_rate}), but different sizes and locations. 
In the absence of detailed detector simulations for all detectors, our assumption of identical detectors is sufficient to provide illustrative, baseline predictions of their combined reach.
For the sake of simplicity, and given the unknowns in the final detector proportions, we scale the effective areas with the expected volumes relative to the volume of IceCube. Appendix~\ref{app:aeff_scaling} contains a detailed note justifying our choice or scaling of the effective area, compared to alternatives.

While a future analysis based on real data recorded by different detectors must incorporate the features that are specific to each detector, the conclusions that we garner below from our forecasts would be broadly unaffected by incorporating them.

Table~\ref{tab:detectors} summarizes the information on the location, size, and start date of the detectors that we consider.  We detail our assumptions below, in Sec.~\ref{sec:event_rate}.


\subsection{Computing the rate of detected neutrinos}
\label{sec:event_rate}


\subsubsection{How are high-energy neutrinos detected?}
\label{sec:event_rate-how_detected}

\textbf{\textit{Neutrino telescopes.---}}High-energy neutrino telescopes, like IceCube, consist of cubic-kilometer-scale arrays of vertical strings of photomultipliers deployed kilometers deep below the surface within a transparent medium, \ie, ice or water~\cite{Markov:1961tyz}. 

At neutrino energies above the TeV scale, a neutrino interacting with matter most often undergoes deep inelastic neutrino-nucleon ($\nu N$) scattering (DIS).  In it, the neutrino interacts with a constituent parton of the nucleon---a quark or a gluon---and, in so doing, breaks up the nucleon. The products of the interaction include final-state hadrons---created in the hadronization of the destroyed nucleon---and a lepton---a neutrino when the interaction is neutral-current (\ie, mediated by a $Z$ boson) and a charged lepton when it is charged-current (\ie, mediated by a $W$ boson).  

The charged final-state products radiate Cherenkov light that propagates through the medium and is collected by the photomultipliers.
The amount of detected Cherenkov light and its temporal and spatial profiles are used to infer the energy and direction of the secondary particles.
From that, the energy and direction of the parent neutrino are reconstructed.
(Above about 100~PeV, other detection techniques---involving detecting instead fluorescence light and radio from the showers---become more efficient; see, \eg, \Refes~\cite{MammenAbraham:2022xoc, Ackermann:2022rqc} for reviews.)

Because the flux of high-energy astrophysical neutrinos is small, contemporary neutrino telescopes need large detector volumes. IceCube, currently the largest neutrino telescope in operation, instruments about 1~km$^3$ of Antarctic ice at the geographic South Pole.  Other neutrino telescopes that we consider for \plenum (Table~\ref{tab:detectors}) that are presently under construction (Baikal-GVD, KM3NeT) and initial testing (P-ONE), plan to instrument similar volumes of natural water. Future detectors (IceCube-Gen2, TRIDENT, NEON, HUNT) envision instrumenting volumes 7.5--30 times larger than IceCube.
The size and shape of the instrumented volume determine the expected number of detected neutrinos; this is captured in the detector effective area (Sec.~\ref{sec:method_effective_area}).

\begin{table*}[t!]
 \begin{ruledtabular}
  \caption{\label{tab:detectors}\textbf{\textit{Neutrino telescopes considered and their combinations considered in this analysis.}}  We consider present and future in-ice and in-water TeV--PeV neutrino telescopes.  In our simplified analysis, we treat future detectors as scaled-up versions of IceCube, translated and rotated to the location of each detector.  See \figu{skymap_detector_locations} for a graphical representation of detector locations and their combinations, and Sec.~\ref{sec:method} for details.}
  \centering
  \renewcommand{\arraystretch}{1.3}
  \begin{tabular}{llccccc}
   \multirow{2}{*}{Neutrino telescope} &
   \multirow{2}{*}{\makecell{Location\footnote{Exact used locations can be found in the GitHub repository (file \texttt{settings.py}).}}}   &
   \multirow{2}{*}{\makecell{Size relative \\ to IceCube\footnote{Approximate size of the final detector configuration that is used in this paper.}}}   &
   \multirow{2}{*}{\makecell{Start date\footnote{Approximate dates when the final configuration of the detector is expected to be completed, used in this analysis but subject to change.}}}   &
   \multicolumn{3}{c}{\makecell{Included in \plenum}}   \\
   \cline{5-7}
   & 
   &
   &
   &
   \plenum-1 &
   \plenum-2 &
   \plenum-3 \\
   \hline
   \multicolumn{7}{l}{\textbf{Ongoing:}} \\
   IceCube &
   South Pole, (\ang{0} E, \ang{-90} N) &
   1 &
   2011 &
   \checkmark
   &
   \\
   \hline
   \multicolumn{7}{l}{\textbf{Under construction:}} \\
   KM3NeT &
   Mediterranean Sea, (\ang{16,6} E, \ang{36,27} N)&
   1 &
   2025 &
   \checkmark &
   \checkmark &
   \checkmark   
   \\
   Baikal-GVD &
   Lake Baikal, (\ang{108,17} E, \ang{53,56} N) &
   1 &
   2027 &
   \checkmark &
   \checkmark &
   \checkmark   
   \\
   \hline
   \multicolumn{7}{l}{\textbf{Under prototyping, design, planning:}} \\
   P-ONE &
   Cascadia Basin, (\ang{-127,73} E, \ang{47,74} N) &
   1 &
   2031 &
   \checkmark &
   \checkmark &
   \checkmark   
   \\
   IceCube-Gen2 &
   South Pole, (\ang{0} E, \ang{-90} N) &
   7.5 &
   2035 &
   &
   \checkmark &
   \checkmark   
   \\
   TRIDENT &
   South China Sea, (\ang{114,0} E, \ang{17.4} N) &
   7.5 &
   2040 &
   &
   &
   \checkmark   
   \\
   NEON &
   South China Sea, (\ang{114,0} E, \ang{17.4} N)  &
   10 &
   2040 &
   &
   &
   \checkmark   
   \\
   HUNT &
   Lake Baikal, (\ang{108,17} E, \ang{53,56} N) \footnote{Two possible locations have been proposed for HUNT: Lake Baikal and the South China Sea.  We use the former in our analysis since NEON and TRIDENT are already planned for the latter.} &
   30 &
   2040 &
   &
   &
   \checkmark   
  \end{tabular}
 \end{ruledtabular}
\end{table*}

\smallskip

\textbf{\textit{Neutrino signatures.---}}
A neutrino telescope detects neutrinos predominantly as two types of events, each with a different shape of the light profile: \textit{cascades} and \textit{tracks}.  Cascades are electromagnetic and hadronic particle showers made mainly by the charged-current DIS of $\nu_e$ and $\nu_\tau$ (\ie, $\nu_l + N \to l + X$, where $l = e, \tau$, and $X$ are final-state hadrons) and also by the neutral-current DIS of neutrinos of all flavors (\ie, $\nu_l + N \to \nu_l + X$, where now $l = e, \mu, \tau$).  Tracks are made by the charged-current DIS of $\nu_\mu$ (\ie, $\nu_\mu + N \to \mu + X$), where the final-state muon is sufficiently energetic to leave a kilometer-length track of Cherenkov light in its wake.  
In addition, PeV-scale $\nu_\tau$ may be detected via a \textit{double bang}, consisting of two spatially separated showers seen in tandem: one due to the charged-current DIS of the $\nu_\tau$ and a later one due to the decay of the final-state tau it produces.

In a DIS event, the final-state hadrons receive a fraction $y$ of the parent neutrino energy---the inelasticity---and the final-state lepton receives the remaining fraction, $1-y$.
At TeV energies, the average value of the inelasticity is 0.4 for anti-neutrinos and 0.5 for neutrinos~(see, \eg, \Refes~\cite{Cooper-Sarkar:2011jtt, IceCube:2018pgc, Weigel:2024gzh}).
However, in any given neutrino-nucleon scattering, the value of $y$ is random and sampled from a distribution that peaks at $y = 0$ but is wide; see, \eg, Fig.~3 in \Refe~\cite{Weigel:2024gzh}. 
At PeV energies, the average value of the inelasticity becomes 0.25---making tracks due to final-state muons more energetic---and the DIS cross section and inelasticity distribution are nearly the same for neutrinos and anti-neutrinos of all flavors.  

The neutral-current cross section is about one-third of the charged-current one, but showers made by one or the other are largely indistinguishable on an event-by-event basis (see, however, \Refe~\cite{Li:2016kra}).  Similarly, events due to neutrinos and anti-neutrinos are indistinguishable (except around 6.3~PeV, due to the Glashow resonance of $\bar{\nu}_e$~\cite{Glashow:1960zz, IceCube:2021rpz}).  Therefore, in our calculations, we always consider the sum of neutrino and anti-neutrino fluxes.

\smallskip

\textbf{\textit{Muon tracks to search for sources.---}}Because of their elongated light profiles, tracks offer the sub-degree angular resolution suitable to search for astrophysical neutrino sources.  In contrast, cascades have more spherical light profiles and a poorer angular resolution of typically tens of degrees, though there is ongoing progress in reducing this~\cite{Abbasi:2021ryj, IceCube:2023ame}.  Thus, in our results below, we exclusively use muon tracks to search for sources. 

Most of the muon tracks detected by neutrino telescopes are \textit{through-going}, \ie, they are made in neutrino interactions that occur outside the instrumented detector volume, and where only a segment of the track crosses and exits it. The energy of the muon making the track is reconstructed from the energy deposited as light by the track segment that crosses the detector, with a typical error of about 20\% in $\log_{10}(E_\mu^{\rm rec}/{\rm GeV})$~\cite{IceCube:2013dkx}, where $E_\mu^{\rm rec}$ is the reconstructed muon energy

From this, accurately inferring the energy of the parent neutrino, $E_\nu$, requires detailed simulations of $\nu_\mu$ interaction and muon propagation that account for the properties of the detector medium and the detector geometry. 
Due to the stochastic nature of the inelasticity in DIS, there is an intrinsic uncertainty when inferring the parent neutrino energy from the through-going muon.
Also, due to the kinematics of the interaction, the final-state muon will have a different direction than the parent neutrino, called the kinematic angle.
This angle is, like the inelasticity, stochastic.
At \SI{1}{TeV}, the mean angle is around $1^\circ$, but becomes negligible at \SI{100}{TeV} and beyond.
In our work, we account for these complications by using descriptions of the detector response (Secs.~\ref{sec:method_effective_area} and~\ref{sec:method_resolution}) produced in dedicated simulations by the IceCube Collaboration.

The above complications limit not only the precision with which the neutrino energy spectrum emitted by an astrophysical source can be reconstructed but also our ability to separate it from the background of atmospheric neutrinos, which have a different energy spectrum.  Nonetheless, both tasks \textit{are} possible already today and will be enhanced with the combination of detectors that make \plenum.  We show this explicitly below when computing the expected rate of neutrino-induced events at a neutrino telescope (Secs.~\ref{sec:event_rate-calculation} and \ref{sec:method_resolution}) and in our forecasts for the discovery and spectral characterization of an astrophysical neutrino source (Sec.~\ref{sec:statistics}). 


\subsubsection{Detector effective area}
\label{sec:method_effective_area}

\begin{figure*}[t!]
 \centering
 \includegraphics[width=\textwidth]{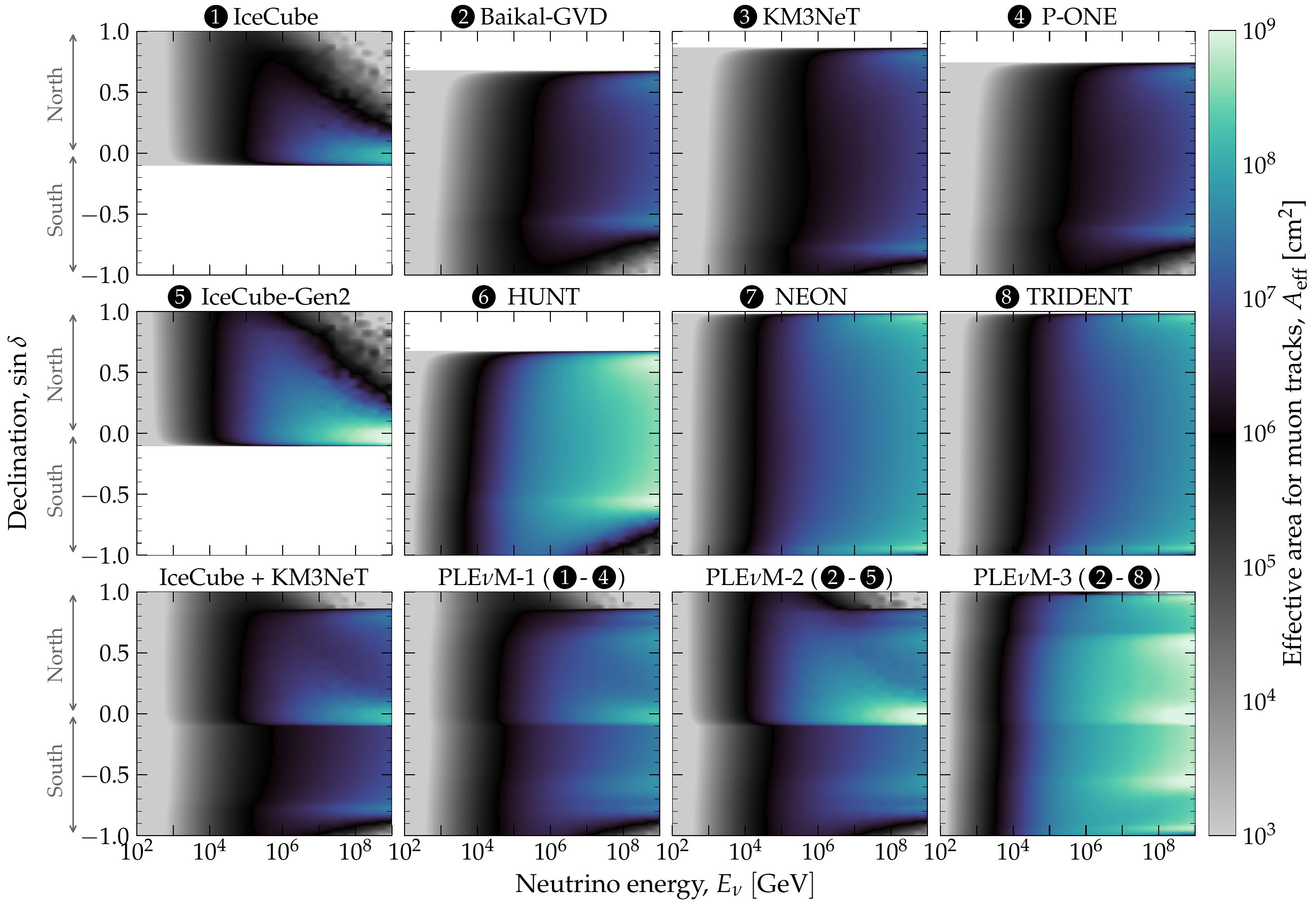} 
 \caption{\textbf{\textit{Effective area for the detection of $\nu_\mu$ in present and future high-energy optical Cherenkov neutrino telescopes.}}  For IceCube, the effective area is extracted from its public 10-year data release~\cite{ICdataRelease2021web, IceCube:2021xar}. 
 For all detectors, we mask down-going directions, i.e., zenith angles between $\ang{0} \leq \theta_z \leq \ang{85}$, to remove the background of atmospheric muons (Sec.~\ref{sec:method_background}). For IceCube at the South Pole, this translates to masking the Southern Sky, i.e., the declination between $\ang{-5} \leq \delta \leq \ang{-90}$ or $-0.1 \leq \sin \delta \leq -1$.
 Thus, depending on the detectors' geographic locations (\figu{skymap_detector_locations}), different declination bands in equatorial coordinates are masked.  We model Baikal-GVD, KM3NeT, and P-ONE as detectors identical to IceCube, but placed elsewhere; IceCube-Gen2, as 7.5~times larger than IceCube, at the same location; and HUNT, NEON, and TRIDENT, as 30, 10, and 7.5 times larger than IceCube, but placed elsewhere.  The effective areas of IceCube + KM3NeT, \plenum-1, \plenum-2, and \plenum-3 are the sum of the effective areas of their constituent detectors.  See Sec.~\ref{sec:method_effective_area} for details.}
 \label{fig:aeff_upgoing}
\end{figure*}

As pointed out in Sec.~\ref{sec:method_what_is_plenum}, to obtain the results in this paper, we assume that all the detectors modeled in \plenum~have the same effective area, only scaled by the size of each detector relative to IceCube.  As a baseline, we use the IceCube effective area for muon tracks of the completed 86-string detector, published by the IceCube Collaboration as part of a recent 10-year public data release~\cite{IceCube:2021xar, ICdataRelease2021web}. 
This data set is selected for $\nu_\mu + \bar{\nu}_\mu$ events and optimized for point-source searches.
Specifically, we adopt the effective area valid for the IC-86-II observation period, called \textit{IC-86} for brevity below.  
[This effective area does not include the subdominant contribution of muon tracks that are made by $\nu_\tau$ (Sec.~\ref{sec:event_rate-how_detected}), though it was included in the dedicated IceCube search that discovered  neutrinos from NGC 1068~\cite{IceCube:2022der}.]

\smallskip

\textbf{\textit{Individual neutrino telescopes.---}}Figure~\ref{fig:aeff_upgoing} shows the effective areas of the neutrino telescopes that we consider and of their combinations.  In each detector, we set the effective area to zero for values of the zenith angle $0 \leq \theta_z \leq 85^\circ$ in order to mask out the otherwise dominant background of down-going atmospheric muons; see Sec.~\ref{sec:method_background} for details.   For IceCube, located at the South Pole, this masks out the declination band  $-\sin (85^\circ) \approx -0.1 \leq \sin \delta \leq -1$; 
for comparison, the unmasked effective area is shown in Fig.~2 of \Refe~\cite{IceCube:2021xar}.
For the other detectors, we apply the same mask on $\theta_z$ at each location.
This reduces the acceptance in different declination bands in each detector when averaging over the daily rotation of the Earth (\figu{skymaps_event_rate}).
(Future revisions of our analysis may unmask down-going directions to include the smaller, but not insignificant contributions from these directions.)

To generate the effective area of each detector, we scale up the IceCube effective area by a factor equal to the volume of the detector relative to IceCube, and rotate the result to the position of the detector (\figu{skymap_detector_locations})~\cite{Schumacher:2021hhm, github-plenum}. 
As mentioned in Sec.~\ref{sec:event_rate-calculation}, we integrate the effective area of each detector over the daily rotation of the Earth.
This is possible because we focus here on steady-state neutrino sources. For transient sources on the timescale of a day or less, this may no longer be justified~\cite{Schumacher:2023mxz}. 

The IceCube effective area in \figu{aeff_upgoing} shows that neutrinos coming from near-horizontal directions, \ie, $\delta \approx 0$, can be detected up to the highest energies.
In contrast, high-energy ($E_\nu \gtrsim 10^6$~GeV) \textit{up-going} ($\sin \delta \gtrsim 0.5$) neutrinos are more likely to be absorbed in the Earth.
As the neutrino cross section falls at lower energies, so does the effective area.
At the lowest energies ($E_\nu \lesssim 10^4$~GeV), the dim muon tracks are detected less efficiently, which reduces the effective area even further.
The features above also appear in the effective areas of the other detectors in \figu{aeff_upgoing} but shifted in declination and smeared out after translating from local zenith to declination and averaging over the daily rotation of the Earth.
They reflect essential limitations of individual detectors that would be broadly present even in more detailed treatments.

\smallskip

\textbf{\textit{Combining neutrino telescopes.---}}Figure~\ref{fig:aeff_upgoing} also illustrates how the above limitations are mitigated by combining neutrino telescopes at different locations.  First, their combined effective area is larger.  However, this, by itself, could be arguably achieved alternatively by building a larger detector at a single location.  Second, their combined effective area covers more of the sky.  Any detector located in the Southern Hemisphere (IceCube, IceCube-Gen2) or Northern Hemisphere (Baikal-GVD, KM3NeT, P-ONE, HUNT, NEON, TRIDENT) has part of its field of view in the opposite hemisphere masked.  
For NEON and TRIDENT, their locations only \ang{17.4} north of the equator are privileged, and only neutrinos with $\sin \delta \gtrsim 0.95 $ are masked out. Using showers in addition to muon tracks---which we do not explore here---could mitigate this, but at the cost of poorer angular resolution; see Sec.~\ref{sec:event_rate-how_detected}. 
Similarly, IceCube uses different approaches to reducing the contribution of the background of atmospheric muons in the samples of detected $\nu_\mu$, but they come at the cost of reduced effective area and energy range~\cite{IceCube:2019cia, IceCube:2020wum, IceCube:2024fxo}.

\textbf{\textit{The capacity to look for neutrino sources across the full sky is the key improvement made possible by a distributed network of neutrino telescopes.}}  For steady-state sources, like the ones we consider here, the improvement is significant.  Later, we show this explicitly via sky maps of expected event rates (Figs.~\ref{fig:skymaps_event_rate} and \ref{fig:skymaps_event_rate_4fgl}).  Even so, one could argue that, for steady-state sources, a single IceCube-sized detector in the Southern Hemisphere and a single detector in the Northern Hemisphere could be sufficient, with the important caveat of needing them to run for longer---in some cases, for decades longer---to achieve what combinations of more detectors would achieve in less time. 

For short-duration transient sources, however, a distributed network of neutrino telescopes is not just desirable, but essential, since only with it can we achieve \textit{instantaneous} full-sky coverage.  An exploration of transient sources lies beyond the scope of this paper and will be presented elsewhere; \Refe~\cite{Schumacher:2023mxz} shows preliminary work.

In this paper, we omit the dependence of the effective area on local detector coordinates because we integrate the observations over year-long timescales so that the daily rotation of the Earth averages out the detector acceptance over right ascension.
In addition, the approximately cylindrical geometry of the neutrino telescopes induces only a mild variance of event rates in local azimuth.
Still, when observing any source in the sky at fixed equatorial coordinates, this source will have time-dependent local zenith and azimuth coordinates.
Especially for detectors not located at the North or South Pole, the time-dependent local zenith angle affects the ratio of signal to background events, which then varies with time.  
Thus, accounting for the time-dependent translation between equatorial and local coordinates can improve the analysis performance -- this will be explored in future work.

(In contrast, accounting for the variation of the effective area on local zenith and azimuth angles is inescapable when searching for signals on short time scales, where the detector acceptance is not averaged due to the rotation of the Earth, even in the azimuth direction.)


\subsubsection{The neutrino event rate}
\label{sec:event_rate-calculation}

Given a flux of high-energy $\nu_\mu + \bar{\nu}_\mu$ arriving at a neutrino telescope from declination $\delta$ and right ascension $\alpha$, $\drv \Phi_\nu/(\drv E_\nu \drv \Omega)$, where $\drv \Omega \equiv \sin \delta \, \drv \delta \, \drv \alpha$ is the differential element of solid angle, the differential number of detected neutrinos is
\begin{equation}    
 \label{equ:diff_number_nu}
 \frac{\drv N_\nu}{\drv E_\nu \drv \Omega}
 =
 T \cdot \aeff \left( E_\nu, \delta \right) \cdot
 \frac{\drv \Phi_\nu(E_\nu, \delta, \alpha)}{\drv E_\nu \drv \Omega} \;,
\end{equation}
where $T$ is the detector live time and $\aeff$ is the energy- and declination-dependent effective area of the telescope for neutrino detection via muon tracks (\figu{aeff_upgoing}).  Because we focus on the detection of point sources, we only consider $\nu_\mu + \bar{\nu}_\mu$ detection via muon tracks.  

The energy and angular distribution of the detected events are affected by the limited energy and angular resolution of the neutrino telescope (Sec.~\ref{sec:event_rate-how_detected}).  
To account for this uncertainty, we compute the differential number of events as a function of reconstructed muon energy, $E_\mu^{\rm rec}$, and reconstructed direction in equatorial coordinates, $\Omega^{\rm rec} = (\delta^{\rm rec}, \alpha^{\rm rec})$, \ie,
\begin{eqnarray}    
 \label{equ:diff_number_nu_rec_muon_energy_general}
 \frac{\drv N_\nu}{\drv E_\mu^{\rm rec} \drv \Omega^{\rm rec}}
 &=& 
 \int_0^\infty \drv E_\nu
 \int \drv \Omega
 \;
 \frac{\drv N_\nu}{\drv E_\nu \drv \Omega} 
 \\
 &&
 \times~
 R_{E_\mu}(E_\mu^{\rm rec}, E_\nu)
 \,
 \times~ R_{\Omega}(\Omega^\nu, \Omega^{\rm rec}, E_\nu) 
 \nonumber
 \;,
\end{eqnarray}
where $R_{E_\mu}$ and $R_{\Omega}$ are, respectively, resolution functions in energy and direction. 
Sections~\ref{sec:method_effective_area} and~\ref{sec:method_resolution} elaborate on our choices for them.

\begin{figure*}[t!]
 \centering
 \includegraphics[width=\textwidth, trim={0.3cm 0 0 0.1cm 0.5cm}, clip]{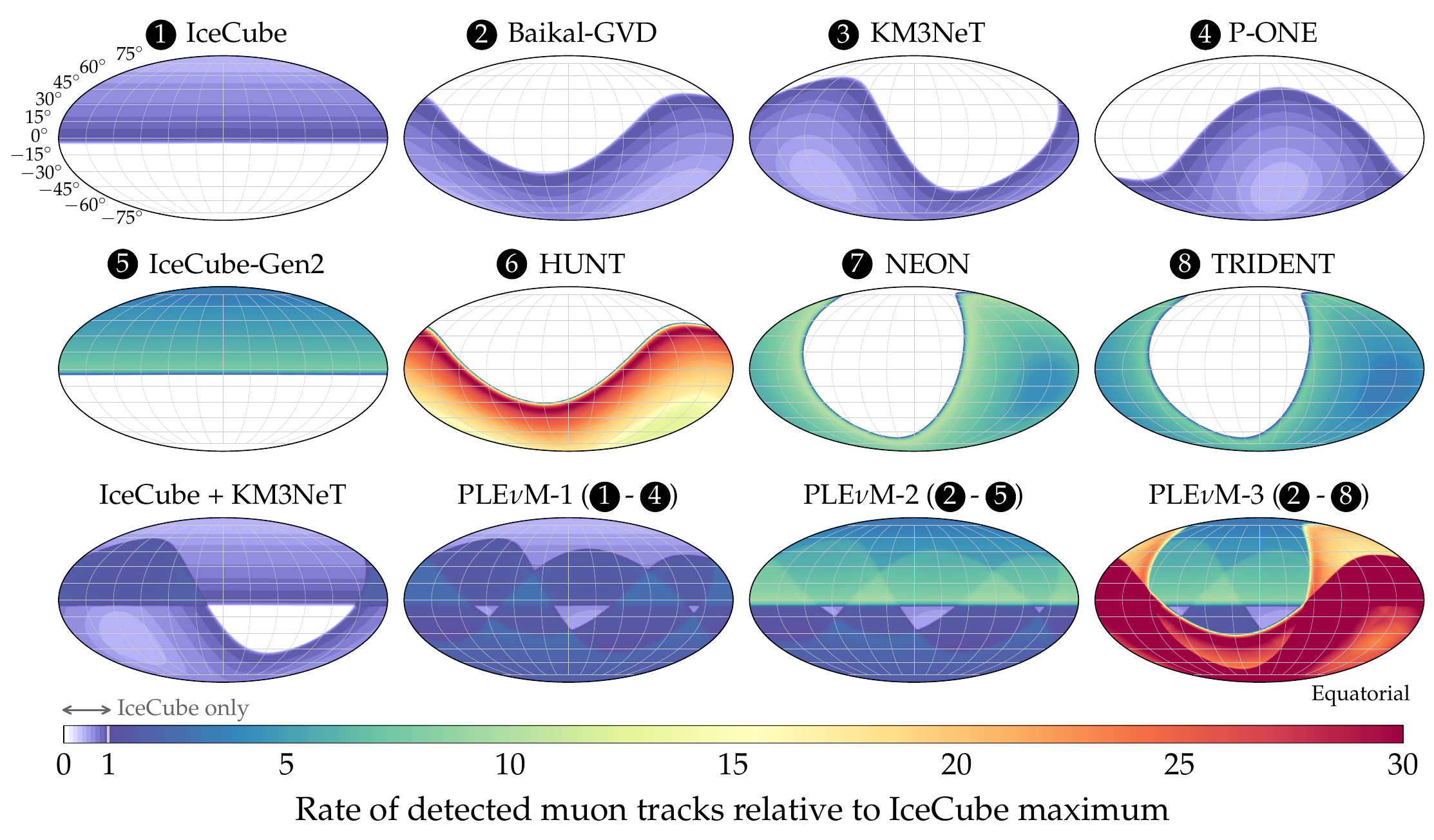}
 \caption{\textbf{\textit{Expected rate of muon tracks detected in present and future high-energy optical Cherenkov neutrino telescopes.}}  For each detector, the expected event rate is computed using \equ{diff_number_nu_rec_muon_energy_general}, integrated over reconstructed muon energy above \SI{100}{GeV} and above zenith angles of $\theta_z\geq -5^\circ$, with its corresponding effective area (\figu{aeff_upgoing}).  Event rates are computed assuming an illustrative $\propto E_\nu^{-2}$ neutrino energy spectrum, are instantaneous at an arbitrary date and time (2025/01/01-00:00 UTC), \ie, not averaged over the rotation of the Earth), and are expressed relative to the maximum event rate achievable in IceCube.  For detector combinations, the event rate is the sum of the contributions of its constituent detectors.
 See Sec.~\ref{sec:event_rate-calculation} for details and \figu{skymaps_event_rate_4fgl} for a comparison to the position of confirmed and potential high-energy astrophysical neutrino sources.}\label{fig:skymaps_event_rate}
\end{figure*}

Figure~\ref{fig:skymaps_event_rate} shows sky maps of the expected rate of detected muon tracks in each neutrino telescope that we consider, and in their combinations, relative to the rate in IceCube.  The event rates are computed using \equ{diff_number_nu_rec_muon_energy_general}, assuming a diffuse energy spectrum $\propto E_\nu^{-2}$, integrated for $E_\mu^{\rm rec} \geq \SI{100}{GeV}$.  The event rates are instantaneous, \ie, \textit{not} averaged by the rotation of the Earth, in order to showcase more clearly the differences in field of view.  The sky maps show how the variation of the effective areas with declination in \figu{aeff_upgoing} translates into its dependence on right ascension and declination, depending on the detector location.

\begin{figure*}[t!]
 \centering
 \includegraphics[width=\textwidth, trim={0.3cm 0 0 0.1cm 0.5cm}, clip]{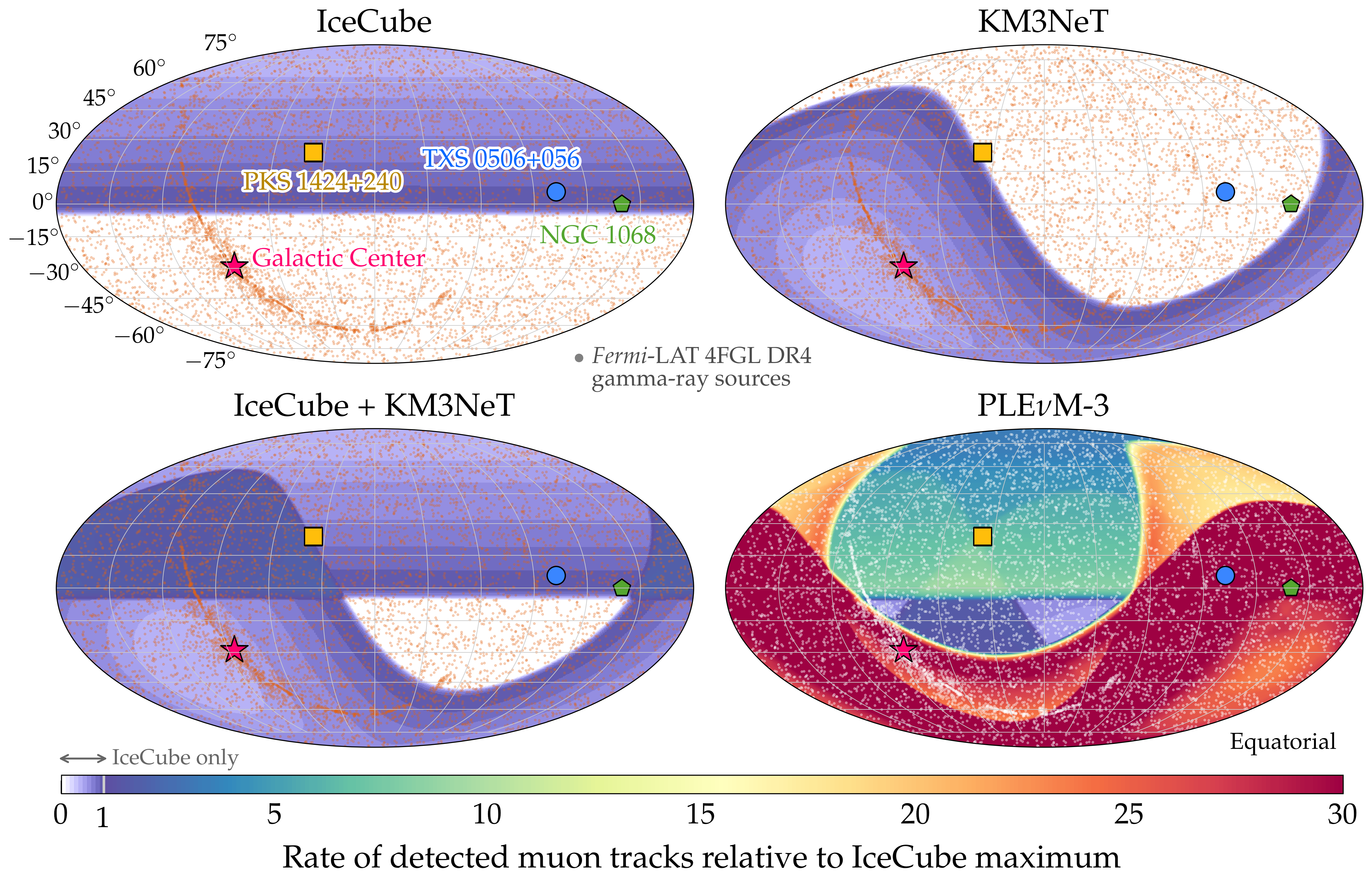}
 \caption{\textbf{\textit{Expected rate of muon tracks detected in present and future neutrino telescopes, compared to the positions of known high-energy astrophysical sources.}} 
 Instantaneous event rate at an arbitrary date and time (2025/01/01-00:00 UTC), same as \figu{skymaps_event_rate}, but shown only for a selection of detectors.  We overlay the position of known high-energy neutrino emitters: the extragalactic steady-state source NGC~1068~\cite{IceCube:2022der} and transient source TXS 0506+056~\cite{IceCube:2018cha} (for source PKS 1424+240 IceCube sees hints of neutrino emission~\cite{IceCube:2019cia}), and the Galactic Plane~\cite{IceCube:2023ame}.  In addition, we overlay the position of gamma-ray sources from the 14-year \textit{Fermi}-LAT 4FGL DR4 catalog~\cite{Ballet:2023qzs}, some of which are well-motivated candidate high-energy neutrino sources.  See Sec.~\ref{sec:event_rate-calculation} for details and \figu{integrated_skymaps_event_rate_4fgl} for daily averaged event rates.}
 \label{fig:skymaps_event_rate_4fgl}
\end{figure*}

Figure~\ref{fig:skymaps_event_rate_4fgl} shows how the larger field of view obtained by combining neutrino telescopes enhances the number of high-energy neutrino sources that are observable.  This includes known candidate sources---active galactic nuclei NGC 1068~\cite{IceCube:2022der}, TXS 0506+056~\cite{IceCube:2018cha}, PKS 1424+240~\cite{IceCube:2019cia}, the Galactic Plane~\cite{IceCube:2023ame}---and hundreds of gamma-ray sources from the \textit{Fermi}-LAT 4FGL DR4~\cite{Ballet:2023qzs} catalog that are well-motivated candidate neutrino sources.  The location of NGC 1068 makes it especially well-suited to benefit from the combination of multiple telescopes.
In Appendix~\ref{app:averaged_rates}, \figu{integrated_skymaps_event_rate_4fgl} shows the equivalent picture based on daily-integrated event rates.

\smallskip

\textbf{\textit{Neutrino point source.---}}Given the angular uncertainty of  track events of about $0.1^\circ$ at best, extragalactic objects can be considered point-like. Thus, the neutrino flux is a delta function in the direction of the source, given by declination $\delta_{\rm src}$ and right ascension $\alpha_{\rm src}$, \ie,
\begin{equation}
 \label{equ:flux_point_source}
 \frac{\drv \Phi_\nu^{\rm ast}}{\drv E_\nu \drv \Omega}
 =
 \frac{\drv \Phi_\nu^{\rm src}}{\drv E_\nu}
 \,
 \delta(\cos \delta^{\rm src}-\cos\delta)
 \,
 \delta(\alpha^{\rm src}-\alpha) 
 \;,
\end{equation}
where $\drv \Phi_\nu^{\rm src} / \drv E_\nu$ is the flux from the source. 
Later (Sec.~\ref{sec:statistics}), we explore different possible forms for the energy spectrum. With this, the differential event rate, \equ{diff_number_nu_rec_muon_energy_general}, simplifies to
\begin{eqnarray}
 \label{equ:diff_number_nu_rec_muon_energy_point_source}
 \frac{\drv N_\nu^{\rm ast}}{\drv E_\mu^{\rm rec} \drv \Omega^{\rm rec}}
 &=&
 T \int_0^\infty \drv E_\nu
 \;
 \frac{\drv \Phi^{\rm src}(E_\nu)}{\drv E_\nu} 
 \,
 \aeff(E_\nu, \delta^{\rm src})
 \nonumber
 \\
 &&
 \times~
 R_{E_\mu}(E_\mu^{\rm rec}, E_\nu)
 \nonumber
 \\
 &&
 \times~
 R_{\Omega}(\Omega^\nu, \Omega^{\rm rec}, E_\nu) \;.
\end{eqnarray}
Equation \ref{equ:diff_number_nu_rec_muon_energy_point_source} shows that a larger lifetime and an overall larger effective area yield the same linear increase in the expected differential event rate.

\subsubsection{Energy and angular resolution}
\label{sec:method_resolution}

The energy and angular resolution functions of a neutrino telescope, $R_{E_\mu}$ and $R_\Omega$ in \equ{diff_number_nu_rec_muon_energy_general}, influence how well it can characterize an astrophysical neutrino source. 
To produce our results below, we adopt the functions published in the IceCube 10-year data release~\cite{ICdataRelease2021web, IceCube:2021xar}, the same one from which we adopt our baseline effective area (Sec.~\ref{sec:method_effective_area}). 

In the data release, the resolution functions are given for three declination bands, corresponding to events with \textit{up-going} ($-90^\circ \leq \delta \leq -10^\circ$), \textit{horizontal} ($-10^\circ \leq \delta \leq 10^\circ$), and \textit{down-going} ($10^\circ \leq \delta \leq 90^\circ$) directions. 
To produce our forecasts, we average the energy and angular resolution functions over their horizontal and up-going bands, as they are similar.  To reproduce results for \ngc based on actual experimental data, we use instead the resolution functions for the horizontal declination range, since \ngc is at $\delta \approx 0^\circ$.  We do not need to consider the resolution function for down-going directions because these are masked out (Sec.~\ref{sec:method_effective_area}).

[The largest differences between the resolution functions in the horizontal and up-going bands occur above \SI{100}{TeV}.  However, many of our results (Sec.~\ref{sec:statistics}) are for sources with soft neutrino spectra, \ie, spectra that decrease strongly with energy, that are dominated by neutrinos with energies of up to tens of TeV.  For these, averaging the resolution functions between the horizontal and up-going bands, rather than using them separately, does not affect our results significantly.  We also show results for hard spectra in \figu{source_discovery_vs_declination} and Appendix~\ref{app:discovery_hard}, where the above averaging is still an acceptable approximation.]

\begin{figure}[t!]
 \centering
 \includegraphics[width=\columnwidth]{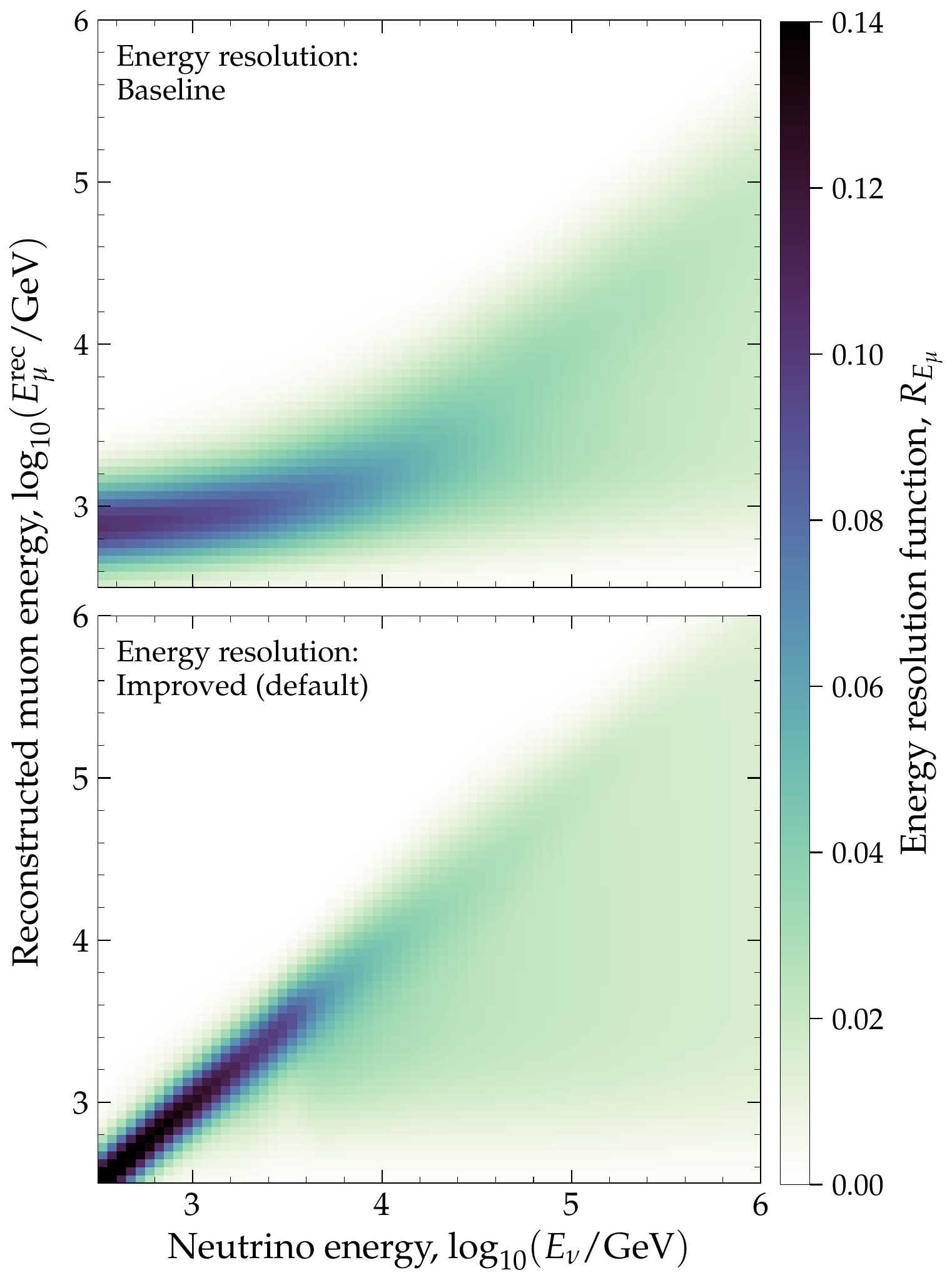}
 \caption{\textbf{\textit{Detector energy resolution function.}} The energy resolution function, $R_{E_\mu}$ in \equ{diff_number_nu_rec_muon_energy_general}, maps the relation between the reconstructed neutrino energy of a detected muon track, $E_\mu^{\rm rec}$---an experimentally measured quantity--and the energy of the parent neutrino that created the track.  \textit{Top:} Baseline resolution function, adopted from the public IceCube 10-year data release~\cite{IceCube:2021xar, ICdataRelease2021web}.  \textit{Bottom:} Artificially improved resolution function built to approximate that of \Refe~\cite{IceCube:2021wbi}.  Figure~\ref{fig:signal_model} shows the impact of using the improved \textit{vs.}~baseline resolution function.  See Sec.~\ref{sec:method_resolution} for details.}
 \label{fig:eres}
\end{figure}

\smallskip

\textbf{\textit{Energy resolution function.---}}The energy resolution function maps the relation between the neutrino energy, $E_\nu$, and the reconstructed muon energy, $E_\mu^{\rm rec}$. 

Figure \ref{fig:eres} shows the two models of energy resolution that we use in this paper: baseline and improved. 
The baseline resolution function is directly taken from the 10-year IceCube data release~\cite{IceCube:2021xar, ICdataRelease2021web}. 
Ideally, these quantities would be tightly correlated, allowing the energy distribution of detected events to reflect the neutrino energy spectrum from a point source and easing the separation between it and the flux of atmospheric neutrinos.  In \figu{eres}, this means that the energy resolution would ideally be a narrow diagonal band along $E_\mu^{\rm rec} \propto E_\nu$.

In reality, this correlation is weaker, and the energy resolution is wider, predominantly for two reasons.  
First, most neutrinos interact outside the detection volume, such that the secondary muons already lose an unknown amount of energy before they reach the detector.
This effect becomes visible in Fig.~\ref{fig:eres} above $E_\nu \gtrsim 100$~TeV, where the energy resolution spreads out such that 
detected low-energy muons may have been made by significantly higher-energy neutrinos. 
Second, the energy of the muon is inferred from its energy loss inside the detector, which is subject to stochastic variations and thus inherently uncertain~\cite{IceCube:2013dkx}.
Lastly, at low energies, the muons become minimally ionizing such that the correlation between muon energy and energy loss is washed out.
This effect is visible in the baseline model, where $E_\mu^{\rm rec}$ and $E_\nu$ become nearly degenerate below $E_\nu \approx \SI{10}{TeV}$.  This limitation is especially detrimental to the characterization of neutrino spectra of soft-spectrum sources, where most of the events come from the lower energy range.

This limitation is no longer present in the improved model of energy resolution that we use as the default to produce our main results.  This model is motivated by the new energy reconstruction method based on deep neural networks presented in \Refe~\cite{IceCube:2021wbi} for the detection of neutrinos from NGC~1068.
Since this new resolution function is not yet publicly available for the full sky, we build one ourselves that captures the features of the improved function from \Refe~\cite{IceCube:2021wbi}. 

The energy resolution in the IceCube public data release is \textit{not} split into the bare muon energy resolution (improvable) and the propagation and kinematic effects (not improvable), as described above, so we must take an indirect approach for building our resolution function.
To estimate the bare muon resolution, we parameterize the baseline energy resolution function---confirming that it reproduces the original function from \Refes~\cite{IceCube:2021xar, ICdataRelease2021web}---and then we tighten the relation between $E_\nu$ and $E_\mu^{\rm rec}$. 
First, we resolve the degeneracy that exists between them in the baseline function at energies of 0.1--10~TeV (see \figu{eres}) by positing a one-to-one relation between them.
Second, we tighten the relation between them by imposing a 50\% reduction in the spread of the function.  Figure~\ref{fig:signal_model} illustrates how the switch from the baseline to the improved energy resolution significantly shifts the position of the energy distribution of detected events and changes its shape.

\smallskip

\textbf{\textit{Angular resolution function.---}}The angular resolution, \ie, the point-spread function (PSF), is central to the discovery of neutrino sources. When searching for astrophysical point sources across the sky, a tighter PSF reduces the contribution of atmospheric neutrinos on the scale of the PSF. This, in turn, improves the signal-to-background ratio in the direction of neutrino sources, thus improving the discovery potential.

To produce the results in this paper, we use the angular resolution function from the 10-year IceCube data release~\cite{IceCube:2021xar, ICdataRelease2021web}.
This is a departure from a simple Gaussian approximation based on per-event estimators of the angular resolution as used in previous analyses, \eg, \Refe~\cite{IceCube:2019cia}.
Instead, it is closer to---but simpler than---the modeling of the angular resolution based on Monte-Carlo simulations used in \Refe~\cite{IceCube:2022der}.

\begin{figure}[t!]
 \centering
 \includegraphics[width=\columnwidth]{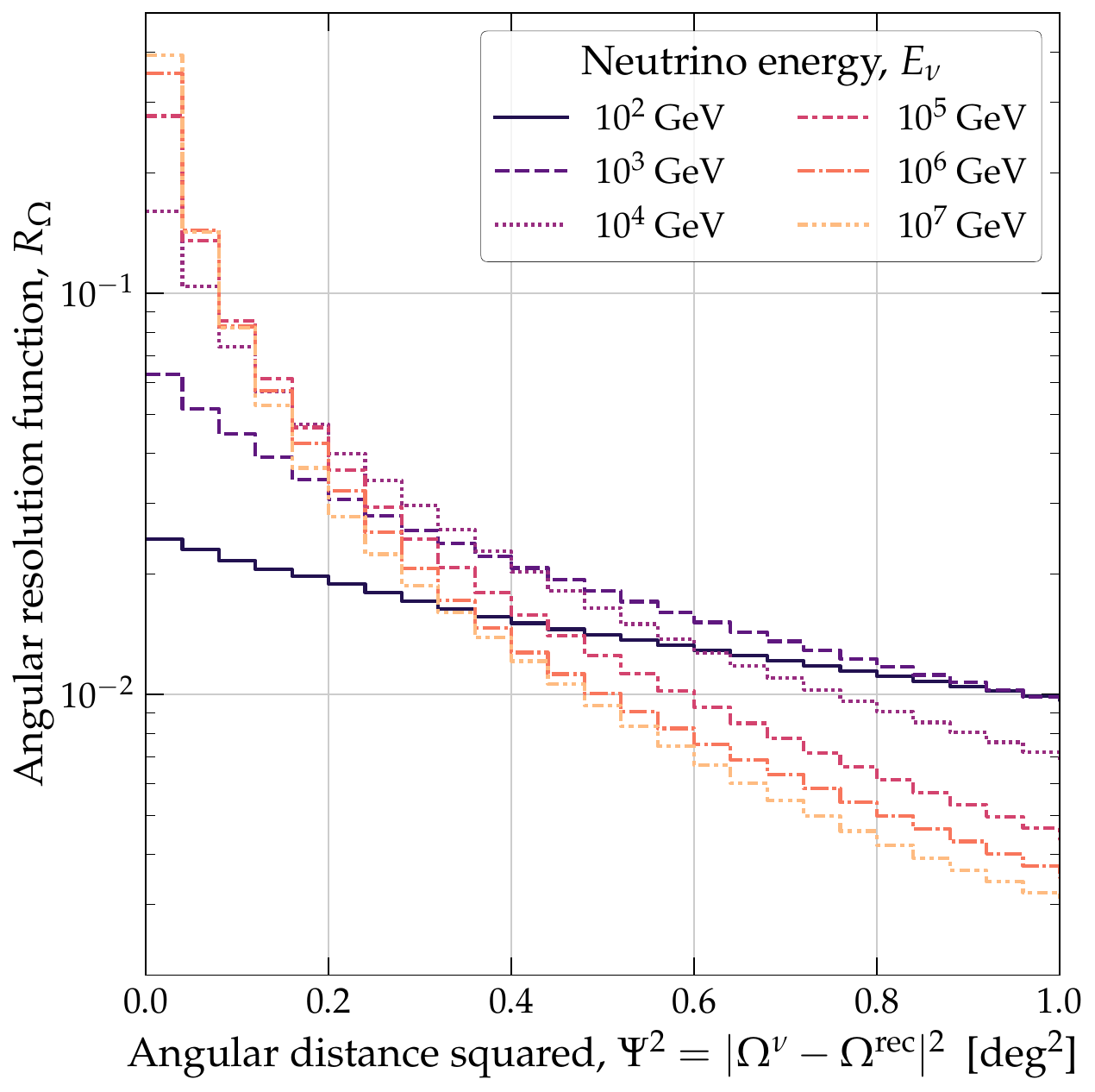}
 \caption{\textbf{\textit{Detector angular resolution function.}} The angular resolution function, $ R_{\Omega}(\Omega^\nu, \Omega^{\rm rec}, E_\nu)$ in \equ{diff_number_nu_rec_muon_energy_general}, maps the relation between the squared angular distance from the true direction of the neutrino to the reconstructed direction of the detected muon track, $\Psi^2 = |\Omega^\nu -\Omega^{\rm rec}|^2$ and the energy of the neutrino, $E_\nu$. Directly adopted from the public IceCube 10-year data release~\cite{IceCube:2021xar, ICdataRelease2021web}. See Sec.~\ref{sec:method_resolution} for details.}
 \label{fig:psi_res}
\end{figure}

Figure~\ref{fig:psi_res} shows the square of the angular distance of neutrino events to a point neutrino source, $\Psi^2 = \left|\Omega^{\rm src} - \Omega^{\rm rec} \right|^2$, for different neutrino energies.
On average, muons produced by high-energy neutrinos are reconstructed closer to the neutrino direction than muons produced by lower-energy neutrinos, which is reflected in their PSF being more peaked towards $\Psi^2 = 0$.
Muons produced by lower-energy neutrinos have an extra angular deviation due to the non-negligible kinematic angle between the muon and its parent neutrino~\cite{Cooper-Sarkar:2011jtt}. 

We expect water-Cherenkov neutrino telescopes, like Baikal-GVD, HUNT, KM3NeT, NEON, P-ONE, and TRIDENT, to have a better angular resolution than IceCube~\cite{Twagirayezu:2023cpv, KM3NeT:2018wnd}, on account of the scattering length of light in water being longer than in ice. 
However, we do not account for this in the present results and instead leave this improvement for future work.


\subsubsection{Background of atmospheric neutrinos and muons}
\label{sec:method_background}

In searches for high-energy astrophysical neutrinos, the main background is the large flux of atmospheric neutrinos and muons produced in the interaction of high-energy cosmic rays in the atmosphere of the Earth.  

\smallskip

\textbf{\textit{Atmospheric neutrino flux.---}}In a detector, along the horizontal and up-going directions of the sky, the flux of atmospheric neutrinos is the main background to searches for astrophysical point sources. 
Unlike the flux of neutrinos from a point source, the atmospheric neutrino flux, $\drv \Phi_\nu^{\rm atm}/(\drv E_\nu \drv \Omega)$, arrives from all directions.
It is essentially isotropic in azimuth angle but not so in zenith angle---it is higher around the horizon, where the column depth in the atmosphere is thicker than along vertical directions. 
Because the effective area varies slowly on angular distances comparable to the scale of the angular resolution, we assume that the background rate of atmospheric neutrinos is constant in the vicinity of an astrophysical source. 
We compute the differential atmospheric event rate, $\drv N_\nu^{\rm atm}/(\drv E_\mu^{\rm rec} \drv \Omega^{\rm rec})$, using \equ{diff_number_nu_rec_muon_energy_general} and, later, the binned event rates, $\mu_{ij}^{\rm atm}$, using \equ{event_rate_bin}.

\begin{figure}[t!]
 \centering
 \includegraphics[width=\columnwidth]{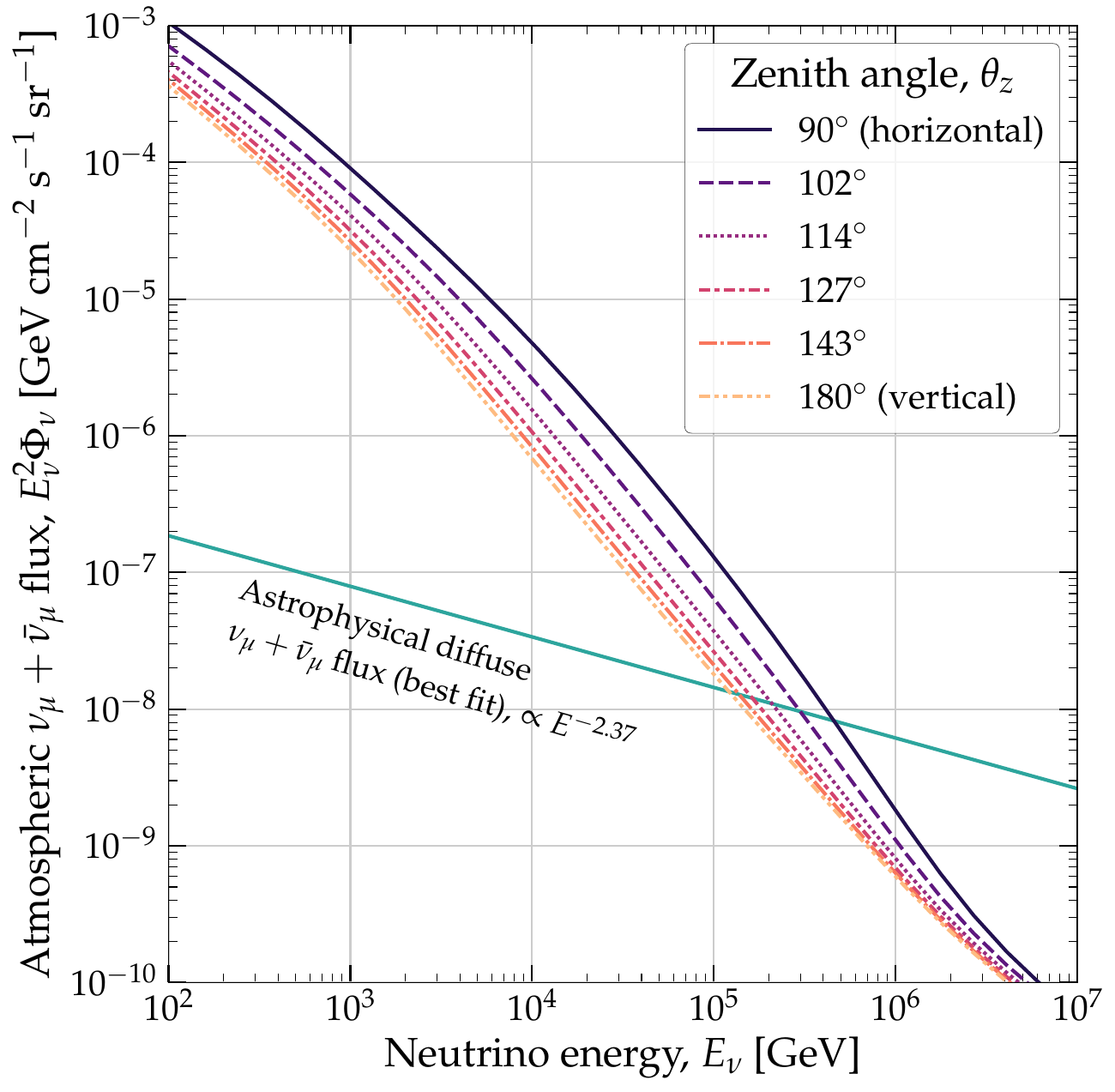}
 \caption{\textbf{\textit{Energy spectrum of atmospheric $\nu_\mu$.}}  The spectrum is from the \textsc{Daemonflux}~\cite{Yanez:2023lsy} prescription, shown here for a few representative choices of incoming neutrino direction, parametrized by the zenith angle measured in local detector coordinates.  For comparison, we show the diffuse spectrum of astrophysical $\nu_\mu$~\cite{IceCube:2021uhz}.}
 \label{fig:atm_flux}
\end{figure}

Figure~\ref{fig:atm_flux} shows the spectrum of atmospheric neutrinos for different zenith angles measured in the local coordinate system of IceCube. 
We account for the energy- and zenith-dependence of the atmospheric neutrino flux by adopting the \textsc{Daemonflux}~\cite{Yanez:2023lsy, daemonflux} flux prescription. This is a state-of-the-art data-driven computation of the atmospheric neutrino flux via \textsc{MCEq}~\cite{Fedynitch:2017trv, MCEq}, the same computational tool used by the IceCube Collaboration. 

We omit the relatively small differences in the atmospheric neutrino background that exist between the different telescope locations, using instead the background in \figu{atm_flux} for all telescopes.
We keep the shape of the \textsc{Daemonflux} neutrino energy spectrum fixed, but allow its normalization, \phiB, to float freely as our single free parameter for the background model. (Full analyses by experimental collaborations, like \Refes~\cite{IceCube:2020wum, Abbasi:2021qfz}, additionally vary the shape of the neutrino energy spectrum.)

When analyzing real IceCube experimental rather than simulated data, we calculate the background expectation directly from the data, after randomizing the right ascension of the detected events, making the analysis less reliant on having an accurate description of the background via simulation (Sec.~\ref{sec:results-exp_data}).
As for the effective areas, we average the background flux of atmospheric neutrinos at the declination of the analyzed sources over a full daily rotation of the Earth.

\smallskip

\textbf{\textit{Atmospheric muons.---}}Along the locally horizontal and up-going directions of the sky, atmospheric muons are quickly absorbed during their propagation inside the Earth and through ice or water, leaving only atmospheric and astrophysical neutrinos to reach IceCube. Because of this, most searches for astrophysical sources of high-energy neutrinos use up-going tracks.  

In contrast, from the Southern Hemisphere, detected muon tracks from down-going atmospheric muons vastly outnumber those from astrophysical neutrinos.  This is why we mask out down-going directions in the effective area (Sec.~\ref{sec:method_effective_area}).  For IceCube, this means setting the effective area to zero for zenith angles between \ang{0} (Zenith) and \ang{85} (\ang{5} above local Horizon; see Fig.~\ref{fig:aeff_upgoing}).  After this, our samples of muon tracks have a neutrino purity of 99.9\%, with only the remaining 0.1\% of tracks due to atmospheric muons~\cite{IceCube:2019cia}. 

We do not model the contamination from atmospheric muons because there is no public IceCube effective area for them that we can use. However, their contribution is negligible in the up-going and horizontal directions~\cite{IceCube:2019cia}, so it is safe to ignore it in our work.

\section{Models of astrophysical high-energy neutrino emission}
\label{sec:flux_models}

\begin{figure}[t!]
 \centering
 \includegraphics[width=\columnwidth]{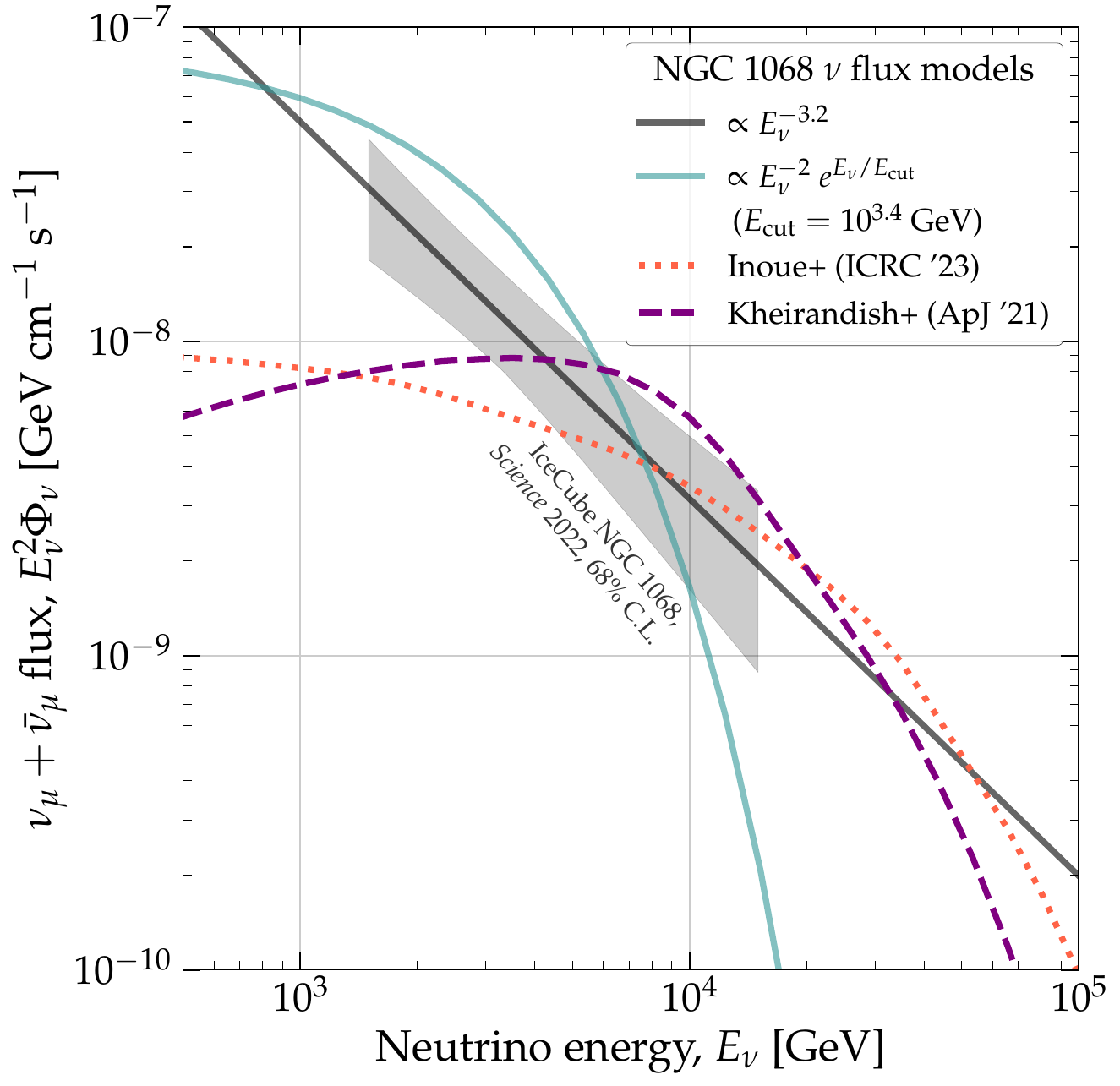}
 \caption{\label{fig:flux_models}
 \textbf{\textit{Models of high-energy neutrino flux from the source NGC 1068.}}  We consider two generic benchmark models---a power law (PL) and a power law with cut-off (PLC)---and two dedicated models by Kheirandish \textit{et al.}~\cite{Kheirandish:2021wkm} and by Inoue \textit{et al.}~\cite{Inoue:2022yak}.
 The IceCube 68\%~C.L. (confidence level)~allowed flux region, obtained assuming PL, is from \Refe~\cite{IceCube:2022der}.}
\end{figure}

In an astrophysical source, high-energy neutrinos are believed to be produced by the interaction of high-energy protons and nuclei with surrounding matter~\cite{Kelner:2006tc} and radiation~\cite{Kelner:2008ke}.  The amount of produced neutrinos and the shape of their energy spectrum depends on those of the parent protons, the geometry of the production region, and the physical conditions present in it.  Dedicated theory models use the above ingredients to make detailed predictions of the neutrino spectrum emitted by different candidate sources.  See, \eg, \figu{flux_models} below and Fig.~2 in \Refe~\cite{Fiorillo:2022rft} for an overview of the variety in the theoretical predictions of high-energy neutrino spectra.

In this paper, our goal is to showcase the future capabilities of neutrino telescopes rather than to perform detailed analyses.  Therefore, in lieu of exploring different sophisticated models of the neutrino spectrum from different candidate sources, we adopt two generic benchmark choices: a power law in neutrino energy (PL) and a power law with an exponential cut-off (PLC). 
Both spectra are frequently considered in the literature, especially in fits to observations; see, \eg, \Refes~\cite{IceCube:2020wum, Abbasi:2021qfz, IceCube:2024fxo}.  Later, we assess the power to experimentally distinguish between the PL and PLC models, \ie, the prospects for identifying a cut-off in the spectrum.  The spectra below are implicitly assumed to be for $\nu_\mu + \bar{\nu}_\mu$.

\begin{table*}[t!]
 \begin{ruledtabular}
  \caption{\label{tab:params}\textbf{\textit{Parameters and baseline values of our benchmark neutrino energy spectra.}} 
  We explore two alternative, generic neutrino energy spectra: a power law (PL) and a power law with a high-energy cut-off (PLC).  Their baseline values are (PL) or approximate (PLC) the IceCube best-fit values for the neutrino observations from the steady-state source NGC 1068~\cite{IceCube:2022der}.  We adopt these values for our calculations (Sec.~\ref{sec:statistics}).
  For the atmospheric neutrino flux, we use the state-of-the-art prediction from \textsc{Daemonflux}~\cite{Yanez:2023lsy}, keeping the shape of the energy spectrum fixed and varying only its normalization.
  See Sec.~\ref{sec:flux_models} for details.}
  \centering
  \renewcommand{\arraystretch}{1.3}
  \begin{tabular}{lllcl}
   Source type & Spectral shape & Parameter & Symbol & Baseline value \\
   \hline
   \multirow{2}{*}{\makecell[l]{Soft-spectrum source: \\ \textbf{\ngc} (Figs.~\ref{fig:flux_models}--\ref{fig:time_evolution_discovery_soft_spectrum}, \ref{fig:ngc_contour}, \ref{fig:time_evolution_discrimination_pl_vs_plc})}} &
   Power law (PL), \equ{spl} &
   Normalization & $\phizero$ & $5.0 \cdot 10^{-14}$~GeV$^{-1}$~cm$^{-2}$~s$^{-1}$ \\
    & & Spectral index & $\gamma$ & 3.2\\
   \hline
   \multirow{2}{*}{\makecell[l]{Hard-spectrum source: \\
   \textbf{\txs} (Figs.~\ref{fig:source_discovery_vs_declination}, \ref{fig:time_evolution_discovery_hard_spectrum})}} &
   Power law (PL), \equ{spl} &
   Normalization & $\phizero$ & $2.7 \cdot 10^{-16}$~GeV$^{-1}$~cm$^{-2}$~s$^{-1}$ \\
    & & Spectral index & $\gamma$ & 2.0 \\
   \hline
   \multirow{2}{*}{\makecell[l]{NGC 1068-like source \\
   (Figs.~\ref{fig:flux_models}, \ref{fig:signal_model}, \ref{fig:time_evolution_discrimination_pl_vs_plc})}}
   &
   \multirow{2}{*}{\makecell[l]{Power law with cut-off \\ (PLC),
   \equ{cut} }} &
   Normalization & $\phizero$ & $8.9 \cdot 10^{-14}$~GeV$^{-1}$~cm$^{-2}$~s$^{-1}$  \\
   & & Spectral index & $\gamma$ & 2\\
   & & Cut-off energy & $\ecut$ & $10^{3.4}~{\rm GeV} \approx 2.5 \cdot 10^3$~GeV \\
   \hline
   \multirow{2}{*}{\makecell[l]{Atmospheric neutrino flux \\(Figs.~\ref{fig:atm_flux}, \ref{fig:signal_model})}} &
   \multirow{2}{*}{\makecell[l]{Parametric \\ (\textsc{Daemonflux}) }}&
   Normalization & $\Phi_0^{\rm atm}$ & 1  \\
   & & & & \\
  \end{tabular}
 \end{ruledtabular}
\end{table*}

\smallskip

\textbf{\textit{Power law (PL).---}}The PL spectrum is often the baseline choice in analyses of the diffuse flux of high-energy neutrinos and in searches for point neutrino sources.  The PL spectrum is
\begin{equation}
 \label{equ:spl}
 \frac{\drv \rm \Phi_{\nu, \plw}^{\rm src}}{\drv E_\nu} = 
 \phizero \, \left( \frac{E_\nu}{\SI{1}{TeV}} \right) ^ {-\gamma} \;,
\end{equation}
where the model parameters are the flux normalization, $\phizero$, and the spectral index, $\gamma$.  The PL spectrum is motivated by the possibility of neutrino production via interactions of high-energy protons with surrounding matter, where the daughter neutrinos inherit the power-law spectrum from their parent protons~\cite{Kelner:2006tc}.  This could happen, \eg, in starburst galaxies~\cite{Loeb:2006tw, Thompson:2006qd, Stecker:2006vz, Tamborra:2014xia, Palladino:2018bqf, Peretti:2018tmo, Peretti:2019vsj, Ambrosone:2020evo}, galaxy clusters~\cite{Berezinsky:1996wx, Murase:2008yt, Kotera:2009ms, Murase:2013rfa}, and low-luminosity active galactic nuclei~\cite{Kimura:2014jba, Kimura:2020thg}. 

\smallskip

\textbf{\textit{Power law with a cut-off (PLC).---}}The PLC spectrum conveys the fact that astrophysical sources are expected to accelerate protons and nuclei only up to a maximum energy.  The value of their maximum attainable energy is source- and model-specific and depends on conditions such as the bulk speed of the acceleration region, its size, and the intensity of the magnetic field it contains; see, \eg, \Refes~\cite{Anchordoqui:2018qom, AlvesBatista:2019tlv}.  As a result, daughter neutrinos are scarcer above a cut-off neutrino energy that reflects the maximum energy of the parent protons.  The PLC spectrum captures this by augmenting the PL spectrum with a high-energy exponential cut-off, \ie,
\begin{equation}
 \label{equ:cut}
  \frac{\drv \rm \Phi_{\nu, \ec}^{\rm src}}{\drv E_\nu} = 
  \phizero \, \left( \frac{E_\nu}{\SI{1}{TeV}} \right) ^ {-\gamma} \, \exp\left( - \frac{E_\nu}{\ecut}\right) \;,
\end{equation}
where the cut-off energy, \ecut, is an additional parameter.  In our projections, we do not compute the value of \ecut~using models of neutrino production; instead, we fix its value to fit present-day observations (Appendix ~\ref{app:plc_model_choice}) when generating mock data. When $E_{\rm cut}$ is much higher than the energies observed by IceCube, the PLC model effectively reduces to the PL model.  Because the deviations of the PLC spectrum relative to the PL spectrum become evident at high energies---where the IceCube event rate is scant but the \plenum event rate is much higher---it is interesting to assess the capability \plenum to distinguish between PLC and PL spectra assuming the source is already known.

\smallskip

Table~\ref{tab:params} lists the PL and PLC model parameters and their baseline values.  For the PL model, the baseline parameter values are the best-fit values from the IceCube analysis of high-energy neutrinos from NGC 1068 and \txs~\cite{IceCube:2022der}.  For the PLC model, we choose baseline parameter values that are compatible with the PL fit to the NGC 1068 observations; Appendix~\ref{app:plc_model_choice} details how.  Later, as part of our statistical methods (Sec.~\ref{sec:statistics}), we fix the model parameters to their baseline values to produce mock samples of observed events, and let the parameter values float freely when, in comparison to the observations, we assess how well we can measure them.  

\begin{figure*}
 \centering
 \includegraphics[width=\textwidth]{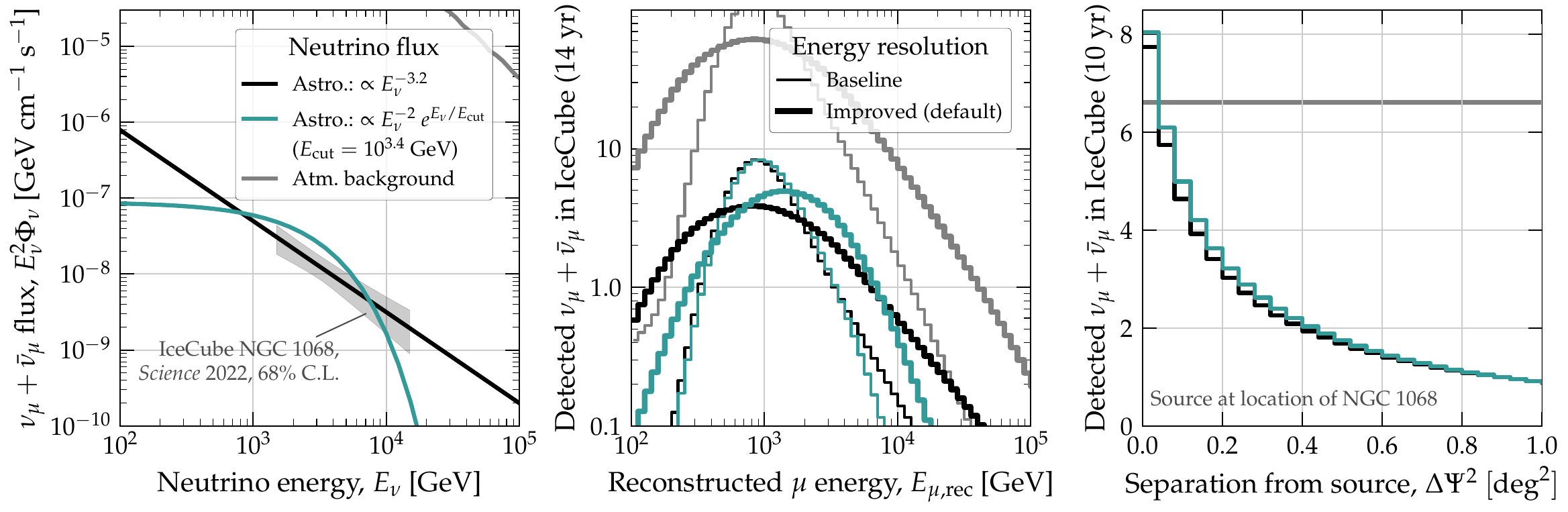}
 \caption{\label{fig:signal_model}\textbf{\textit{Benchmark high-energy neutrino flux models from an astrophysical point source and event distributions at IceCube.}}  For this plot, we assume that the source is located at the position of NGC 1068.  \textit{Left:} Our astrophysical flux models are a power law (PL) and a power law with a cut-off (PLC), here computed using the baseline values of the model parameters (Table~\ref{tab:params}); see Sec.~\ref{sec:flux_models}.  The atmospheric neutrino background is from \textsc{Daemonflux}~\cite{Yanez:2023lsy}.  See also \figu{flux_models}. 
 \textit{Center:} Associated distribution of detected muon tracks by IceCube in reconstructed muon energy, after 10 years of live time, computed using the methods in Sec.~\ref{sec:method}.  Our main results are obtained using the improved energy resolution; see Sec.~\ref{sec:method_resolution}. 
 \textit{Right:} Distribution of detected muon tracks in squared angular separation from the position of the astrophysical source.}
\end{figure*}

Figure \ref{fig:flux_models} (also \figu{signal_model}) shows our baseline PL and the PLC spectra.  In addition, \figu{flux_models} shows two detailed models of neutrino emission from \ngc: the disk-corona (DC) model by Kheirandish \textit{et al.}~\cite{Kheirandish:2021wkm} and the torus-wind (TW) model by Inoue \textit{et al.}~\cite{Inoue:2022yak} that we investigate here.
We consider them, first, as-is and, second, with a free normalization parameter in fits to data.

\section{Statistical methods}
\label{sec:statistics}

In our projections, we address three questions: the measurement of the parameters describing the high-energy neutrino flux from a source, the discovery of neutrino point sources, and the discrimination between the PL and PLC models of the neutrino energy spectrum.  To do so, we perform different, but related statistical analyses on mock data samples detected by the different \plenum~configurations.  When computing and reporting statistical significance, we adopt a frequentist approach.

As established earlier, we treat only steady-state 
or long-duration sources; see \Refe~\cite{Schumacher:2023mxz} for preliminary work on discovering transient sources with \plenum.

\smallskip

\textbf{\textit{Binned event rates.---}}The state-of-the-art method to discover point neutrino sources uses an unbinned likelihood analysis that assesses the chance that each event detected by a neutrino telescope comes from an astrophysical source or the atmospheric background~\cite{IceCube:2019cia}.
While this method produces excellent results, it is computationally expensive: the likelihood evaluation itself is more expensive, and numerous mock experiments are required to estimate the source discovery potential at the target $5\sigma$ statistical significance.

Since our goal is to forecast the future capabilities of neutrino telescopes rather than perform detailed analyses, we adopt instead a binned likelihood analysis, which is computationally less demanding.  We bin events across two dimensions: the reconstructed muon energy, $E_\mu^{\rm rec}$, and the squared reconstructed angular separation of the event direction, $\Omega^{\rm rec}$, relative to the true source direction, $\Omega^{\rm src}$, \ie, $\Psi^2 = |\Omega^{\rm src} - \Omega^{\rm rec}|^2$.  We choose to bin in $\log_{10}(E_\mu^{\rm rec}/{\rm GeV})$ due to the power-law or nearly power-law shape of the energy spectra we consider, and in $\Psi^2$, not $\Psi$, because the background of atmospheric neutrinos around the source is flat in this observable (Sec.~\ref{sec:method_background}).

We compute the mean expected event rate in each bin by integrating Eq.~(\ref{equ:diff_number_nu_rec_muon_energy_general}), \ie,
\begin{equation}
 \label{equ:event_rate_bin} 
 \mu_{ij}
 =
 \int_{\textrm{Bin}~i} \drv \log_{10} E_\mu^{\rm rec}
 \int_{\textrm{Bin}~j} \drv \Psi^2
 \frac{\drv N_{\nu}}{\drv E_\mu^{\rm rec} \drv \Omega^{\rm rec}} \;.
\end{equation}
We use $N_{E_\mu^{\rm rec}} = 139$ bins in $\log_{10}(E_\mu^{\rm rec}/{\rm GeV})$, evenly spaced from $E_\mu^{\rm rec} = 10^2$ to $10^{8.95}$  \si{GeV}, and $N_{\Psi}=225$ bins in $\Psi^2$, evenly spaced from 0 to \SI{9}{deg^2}, corresponding to a maximum angular distance of $\Psi = 3^\circ$.

For a given neutrino detector, we use the methods in Sec.~\ref{sec:event_rate} to produce mock binned samples of the mean number of expected events in the $i$-th bin of $E_\mu^{\rm rec}$ and the $j$-th bin of $\Psi^2$, including the contributions of astrophysical and atmospheric neutrinos, \ie,
\begin{equation}
 \label{equ:event_rate_model_parameters}
 \mu_{ij}(\boldsymbol{\theta})
 =
 \mu_{ij}^{\rm ast} (\boldsymbol{\theta}^{\rm ast})
 +
 \mu_{ij}^{\rm atm} (\Phi_0^{\rm atm}) \;,
\end{equation}
where $\boldsymbol{\theta} \equiv (\boldsymbol{\theta}^{\rm ast}, \Phi_0^{\rm atm})$ are the free model parameters (Table~\ref{tab:params}).
Specifically, $\boldsymbol{\theta}^{\rm ast}$ are the parameters of the PL or PLC model, or, in the case of the DC and TW models of NGC 1068, the flux normalization only. 

Figure~\ref{fig:signal_model} shows the distributions of mean event rates in energy and direction of the baseline PL and PLC models in IceCube, separately for the astrophysical and atmospheric contributions.  The baseline PL model yields about 98 events in 10 years of observation with IceCube, integrated across all energies; the baseline PLC model yields comparable numbers by design (Appendix~\ref{app:plc_model_choice}).  In contrast, the atmospheric neutrino flux yields about 1490 events within a radius of $3^\circ$ from the source, dwarfing the astrophysical contribution.  

Thus, \figu{signal_model} reveals that source discovery stems not from the total event rate of the astrophysical flux being higher than that of the atmospheric flux. Rather, it stems predominantly from the angular distribution of the detected events: close to the position of the source, the observed event rate grows, an indication of its presence.  We quantify the significance of this in Sec.~\ref{sec:results-ps_discovery}.

\smallskip

\textbf{\textit{Likelihood function.---}}We compare the predictions made with \equ{event_rate_model_parameters}, $\mu_{ij}(\boldsymbol{\theta})$, for varying \textit{test} values of $\boldsymbol{\theta}$, against present-day real observations or projected mock \textit{observations}, $n_{ij}$.  For the former, we use the public IceCube observations of NGC 1068.  For the latter, we use an Asimov data set~\cite{Cowan:2010js} computed with \equ{event_rate_model_parameters}, using the baseline values of the model parameters (Table~\ref{tab:params}), which we take as their true values in our calculations; \figu{signal_model} shows this for the PL and PLC models.

In each bin, we compare test \textit{vs.}~observed event rates via a Poisson distribution.  Thus, the likelihood function for a given neutrino detector is
\begin{equation}
 \label{equ:likelihood_det}
 \mathcal{L}_{\rm det}(\boldsymbol{\theta}; n_{ij} )
 = 
 \prod_{i=1}^{N_{E_\mu^{\rm rec}}} 
 \prod_{j=1}^{N_{\Psi}}
 \frac{ [\mu_{ij}(\boldsymbol{\theta})]^{n_{ij}}}{n_{ij}!}
 \,
 e^{-\mu_{ij}(\boldsymbol{\theta})}.
\end{equation}
Our likelihood function and the statistical procedure that we introduce below are similar to those used in analyses performed by the IceCube Collaboration~\cite{IceCube:2019cia, IceCube:2022der, IceCube:2023ame}.

In \plenum, we treat the detection of neutrinos from a point source in each detector as an independent observation of the source from the vantage point of the detector (Sec.~\ref{sec:event_rate}).  Therefore, in each of the future possible detector scenarios that we consider (Sec.~\ref{sec:method_what_is_plenum}), the total likelihood function is simply the product of the likelihood functions of the detectors envisioned for the scenario, \ie,
\begin{equation}
 \label{equ:likelihood}
 \mathcal{L}(\boldsymbol{\theta}; n_{ij})
 = 
 \prod_{\rm det} 
 \mathcal{L}_{\rm det}(\boldsymbol{\theta}; n_{ij}) \;,
\end{equation}
where $\mathcal{L}_{\rm det}$ is computed using \equ{likelihood_det}.  Using this likelihood function, we perform three statistical tests, as described next.

\smallskip

\textbf{\textit{Measuring astrophysical flux parameters.---}}For a given observed event rate, $n_{ij}$, we find the best-fit values of the model parameters, $\hat{\boldsymbol{\theta}} \equiv \left( \hat{\boldsymbol{\theta}}^{\rm ast}, \hat{\Phi}_0^{\rm atm} \right)$, by maximizing the likelihood function, \equ{likelihood}.  We have verified that these values match the true values that we assume to construct the Asimov data for our forecasts.  We report (Figs.~\ref{fig:ngc_contour} and \ref{fig:exp_2Dscan}) mainly on the best-fit values and allowed ranges of the astrophysical flux normalization, $\Phi_0$, and the spectral index, $\gamma$, and treat the remaining parameters as nuisance.

\smallskip

\textbf{\textit{Point-source discovery potential.---}}Further, we compute the point-source discovery potential---\ie, the flux a point source needs to have, on average, to be discovered in the background of atmospheric neutrinos.  To do so, we use a likelihood-ratio test that calculates the discrimination power between the signal hypothesis---where there is an astrophysical neutrino signal plus the atmospheric background---and the background hypothesis---where there is only the atmospheric background. 
We use the conventional test statistic
\begin{equation}
 \label{equ:ts_src}
 \Lambda_{\rm src}(n_{ij})
 = 
 -2 \ln
 \frac
 {\mathcal{L}\left( \hat{\Phi}_0^{\rm atm}, \Phi_0^{\rm ast} = 0; n_{ij}\right)}
 {\mathcal{L}\left( \hat{\Phi}_0^{\rm atm}, \hat{\boldsymbol{\theta}}^{\rm ast}; n_{ij} \right)}
 \;.
\end{equation}
We report (Figs.~\ref{fig:time_evolution_discovery_soft_spectrum} and \ref{fig:source_discovery_vs_declination}) source discovery potential at $5\sigma$ (\textit{p}-value of \num{1.43E-7}) by requiring $\Lambda_{\rm src}^{5\sigma} = 31.5$ and adjusting $\Phi_0$ of the source flux accordingly.
These values are based on Wilks' theorem~\cite{wilks1938}, which we use because the null hypothesis (only atmospheric neutrinos) is a subset of the signal hypothesis (atmospheric plus source neutrinos).
With two degrees of freedom, we require $\int_{\Lambda_{\rm src}^{5\sigma}}^\infty \chi^2 (\Lambda,~2~{\rm d.o.f.})~\drv \Lambda = p$, where $\chi^2$ is the chi-squared distribution with two degrees of freedom.

\smallskip

\textbf{\textit{Discriminating between PL and PLC spectra.---}}Finally, we compute the discrimination power, given an observation, between the neutrino spectrum being one or the other of our benchmarks, PL or PLC (Sec.~\ref{sec:flux_models}).  We use the test statistic
\begin{equation}
 \label{equ:ts_discrimination_pl_vs_plc}
 \Lambda_{\rm dis}(n_{ij})
 = 
 -2 \ln
 \frac
 {\mathcal{L}\left( \hat{\Phi}_0^{\rm atm}, \hat{\boldsymbol{\theta}}^{\rm ast}_{\rm PL}; n_{ij}\right)}
 {\mathcal{L}\left( \hat{\Phi}_0^{\rm atm}, \hat{\boldsymbol{\theta}}^{\rm ast}_{\rm PLC}; n_{ij}\right)}
 \;,
\end{equation}
where the numerator is computed assuming the PL model to fit observations, where $ \hat{\boldsymbol{\theta}}^{\rm ast}_{\rm PL}$ are the best-fit values of the PL parameters, and the denominator is computed assuming the PLC model to fit observations, where $ \hat{\boldsymbol{\theta}}^{\rm ast}_{\rm PLC}$ are the best-fit values of the PLC parameters. 
We report (Figs.~\ref{fig:discrimination_soft_spectrum_vs_declination} and \ref{fig:time_evolution_discrimination_pl_vs_plc}) discrimination at $3\sigma$ (\textit{p}-value of \num{6.75E-4}) by setting $\Lambda_{\rm dis} = 11.6$, using Wilks' theorem~\cite{wilks1938} and the $\chi^2$ distribution with one degree of freedom.  

\smallskip

\textbf{\textit{Corrections for multiple testing.---}} When calculating the p-value of a source, it has to be corrected once multiple testing occurs.
In the case of \ngc, two corrections are noted in~\cite{IceCube:2022der}: the location with the smallest p-value in the Northern Hemisphere, which is \ang{0.11} away from \ngc, has a significance of $5.3\sigma$. After correcting for all tested source locations on a $(\ang{0.2} \times \ang{0.2})$ grid in the Northern Sky, the significance reduces to $2\sigma$, \ie, by a factor of about \num{4E5}\footnote{The correction factor roughly matches the number of scanned pixels, however, the IceCube result accounts for local pixel-to-pixel correlations as well.}.
However, when pre-selecting a list of 110 source candidates, the significance at the location of \ngc only reduces by this factor, \ie, from $5.2\sigma$ to $4.2\sigma$.
Given that the correction factor changes significantly depending on the pre-selection of source candidates, it is a convention to quote only the uncorrected (\ie, local) p-value when comparing analyses, e.g.~with discovery potentials.
In the following, we do not apply any corrections that may be needed due to multiple testing.
Specifically, our predictions on discriminating spectra assume that we are analyzing a known source such that no correction is needed.


\section{Results}
\label{sec:results}

Using the methods from Sec.~\ref{sec:statistics}, we report forecasts for the discovery of steady-state neutrino sources, the measurement of their flux parameters, and the distinction between a PL and PLC neutrino spectrum. We show most of our results for soft-spectrum sources with energy spectra that resemble that of \ngc (\ie, $\propto E_\nu^{-3.2}$), and some results for hard-spectrum sources with spectra that resemble that of \txs (\ie, $\propto E_\nu^{-2}$).  Additionally, we show the application of the \plenum tools to real, present-day public IceCube data on \ngc. 

Because our forecasts do not contemplate improvements other than an increase in the cumulative detector exposure, they are conservative.  Likely future improvements in energy and directional resolution and in background rejection, and the combination of muon tracks with other detection channels, like cascades---not included here---would only improve our forecasts.


\subsection{Point-source discovery potential}
\label{sec:results-ps_discovery}

\begin{figure}[t!]
 \centering 
 \includegraphics[width=\columnwidth]{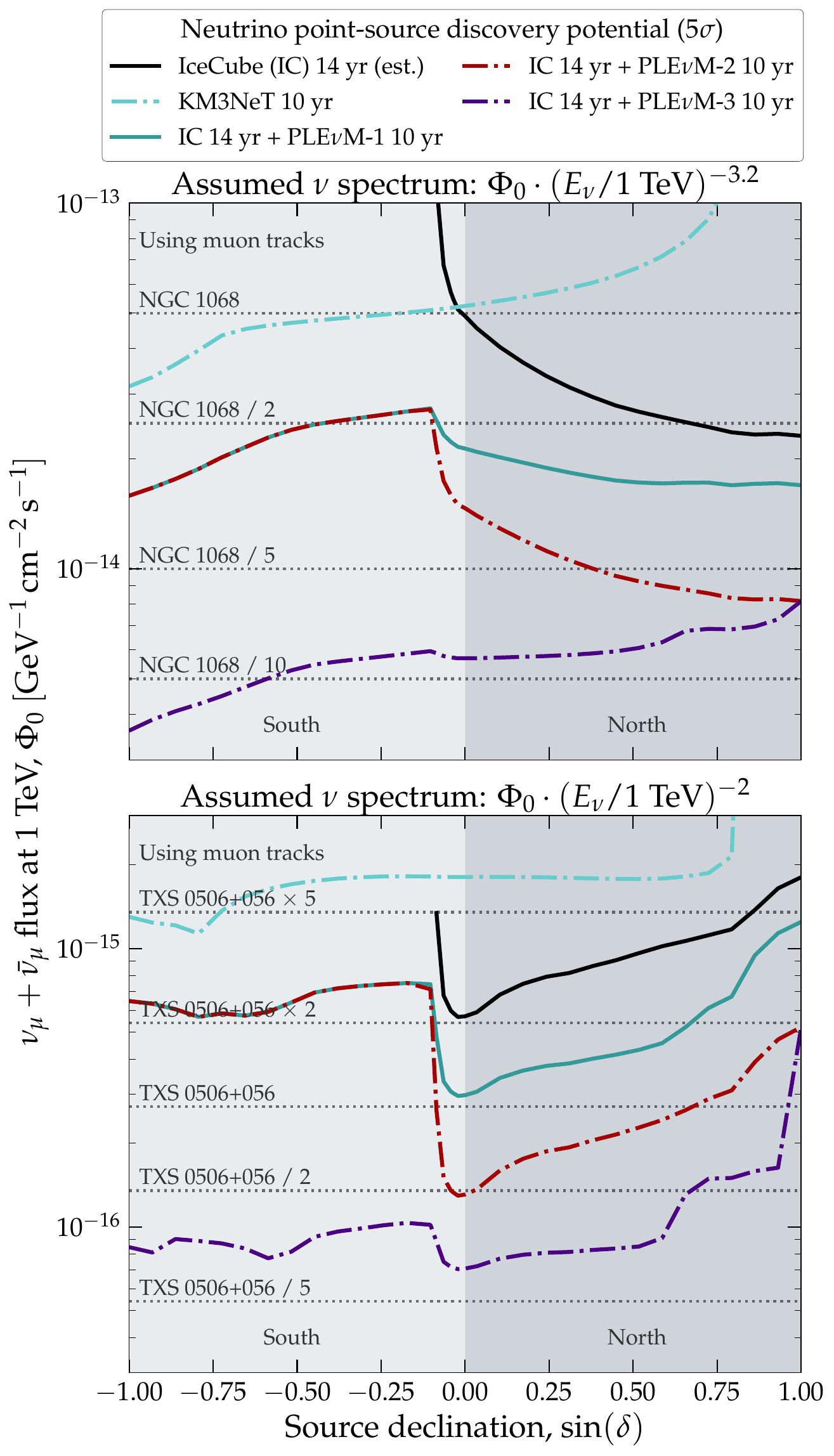}
 \caption{\textbf{\textit{Projected discovery potential ($5\sigma$) of a steady-state point source of high-energy neutrinos.}} \textit{Top: }The source has a power-law spectrum, $\Phi_0 \, (E_\nu / 1~{\rm TeV})^{-\gamma}$, with $\gamma = 3.2$, as measured for NGC 1068~\cite{IceCube:2022der}. \textit{Bottom:} The source has a power-law spectrum with $\gamma = 2$, as measured for \txs~\cite{IceCube:2022der}. We find the value of $\Phi_0$ that would yield discovery with a statistical significance of $5\sigma$, employing muon tracks observed by one or more neutrino telescopes, using the methods in Sec.~\ref{sec:statistics}. As benchmarks, we show the baseline flux level measured for \ngc and \txs (Table~\ref{tab:params}), and 50\% and 20\% of it. Table~\ref{tab:detectors} defines the detector combinations \plenum-1, \plenum-2, and \plenum-3.  The 14-year IceCube discovery potential is estimated using the same methods used for the forecasts; it stops just below the horizon, where the IceCube effective area becomes null in our analysis (\figu{aeff_upgoing}).  See Sec.~\ref{sec:results-ps_discovery} for details.  \textbf{\textit{A globally distributed network of neutrino telescopes would enable the discovery of dim sources anywhere in the sky.}}
 \vspace{2em}
 }
 \label{fig:source_discovery_vs_declination}
\end{figure}

\begin{figure*}
 \centering\includegraphics[width=\textwidth]{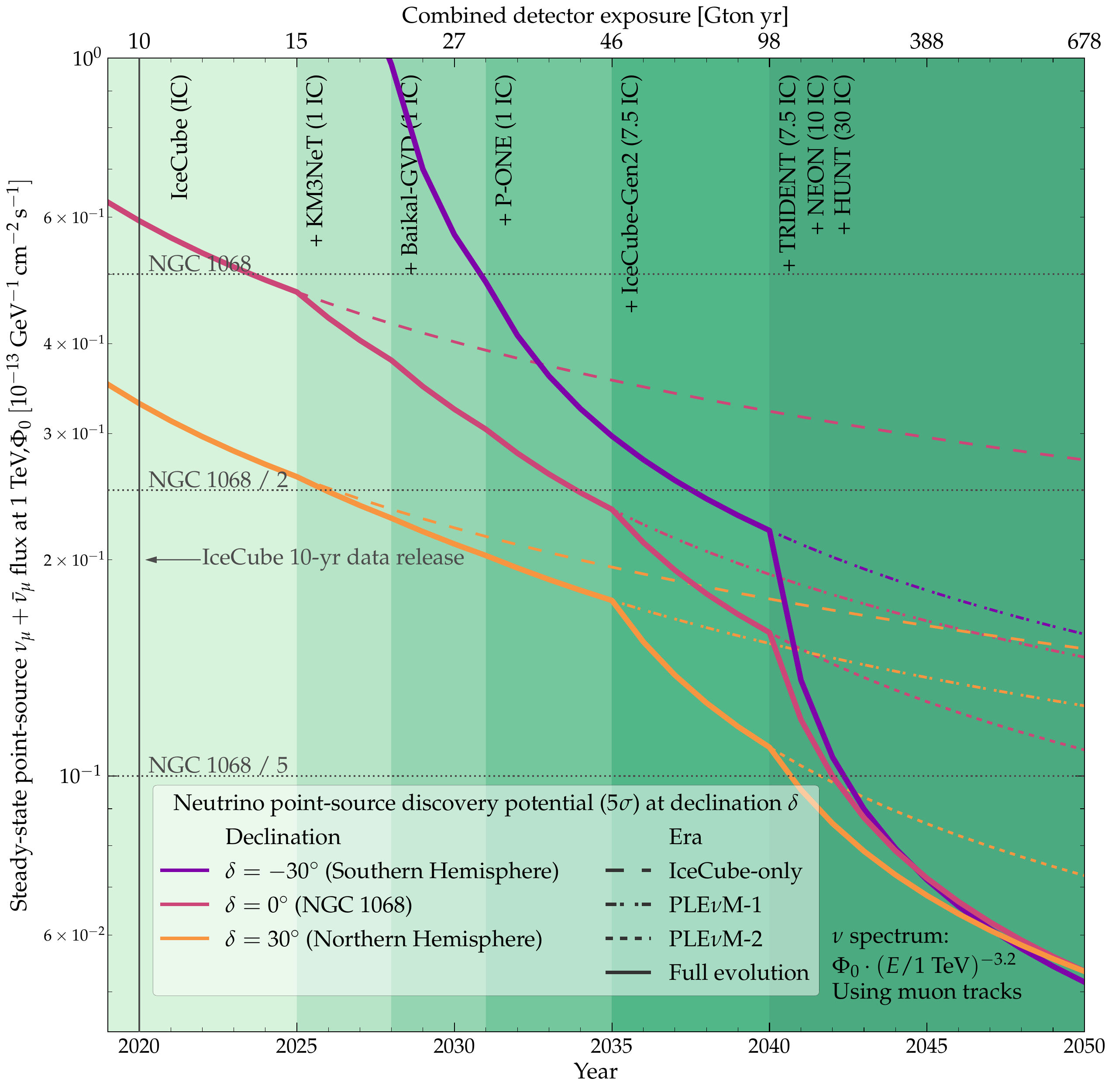}
 \caption{\textbf{\textit{Projected evolution of the discovery potential ($5\sigma$) of a steady-state point source of high-energy neutrinos with a soft energy spectrum.}}
 The source has a power-law spectrum, $\Phi_0 (E_\nu / 1~{\rm TeV})^{-\gamma}$, with $\gamma = 3.2$, as measured for NGC 1068~\cite{IceCube:2022der}, and we find the value of $\Phi_0$ that would yield discovery with a statistical significance of $5\sigma$, employing muon tracks observed by one or more neutrino telescopes, using the methods in Sec.~\ref{sec:statistics}.  We show results for sources at three illustrative declinations: $\delta = -30^\circ$, 0, and $30^\circ$; \figu{source_discovery_vs_declination} shows results for other choices.  As benchmarks, we show the baseline flux level measured for NGC 1068 (Table~\ref{tab:params}), and 50\% and 20\% of it.   The start date of future detectors is staggered and follows the tentative timeline in Table~\ref{tab:detectors}.   Each detector is a scaled version of IceCube (Table~\ref{tab:detectors}), translated to the detector location (\figu{skymap_detector_locations}); see Sec.~\ref{sec:method}.  The top x-axis shows the accumulated exposure of \textit{all} available detectors up to the year of the bottom x-axis; it is only valid for the solid lines, \ie, ``Full evolution''. Table~\ref{tab:exposure} in Appendix~\ref{app:exposure} shows all summed exposures for the different detector eras. See Sec.~\ref{sec:results-ps_discovery} for details.  
 \textbf{\textit{By the mid 2040s, the cumulative exposure of a global network of neutrino telescopes could enable the discovery of neutrino sources five times dimmer than \ngc and half as dim as \txs anywhere in the sky.}}}
 \label{fig:time_evolution_discovery_soft_spectrum}
 \end{figure*}

\textbf{\textit{Rediscovering NGC 1068.---}}First, as a test of our methods, we use them to estimate the discovery significance of NGC 1068 by applying them to Asimov data generated using the best-fit values of the neutrino spectrum reported by IceCube (Table~\ref{tab:params}) and analyzing them via the test statistic in \equ{ts_src}. 
With a 10-year live time, this Asimov data set yields a source discovery significance of $4.3\sigma$ ($p$-value of \num{9e-6}) using our improved energy resolution, or $4.2\sigma$ ($p$-value of \num{1.5e-5}) using the baseline resolution.
Note that this significance is not directly comparable to the IceCube results on real data, since the real data consists of several detector configurations in addition to the IC-86 configuration we use by itself.
[Later (Sec.~\ref{sec:results-exp_data}), we revisit this  using real IceCube data.]

To enable a more direct comparison, we have evaluated simulated pseudo data to find that our simplified analysis approximates the analysis performance obtained by the IceCube search for sources in \Refe~\cite{IceCube:2019cia} and performs 30--50\% worse than the state-of-the-art analysis used to discover \ngc in \Refe~\cite{IceCube:2022der}.
This shows that the results generated using our methods are, if anything, conservative. 
Even so, combining 14 years of IceCube data with 10 years of \plenum-1, the discovery significance of \ngc based on Asimov data grows to over $11\sigma$.

\smallskip

\textbf{\textit{Discovering sources across the sky.---}}Figure~\ref{fig:source_discovery_vs_declination} shows the $5\sigma$ discovery potential of a soft-spectrum source---PL with $\gamma = 3.2$, motivated by \ngc---and a hard-spectrum source---PL with $\gamma = 2$, motivated by \txs---depending on its declination.  The dependence reflects the angular distribution of the expected event rates of the different detectors and their combinations in \plenum (\figu{skymaps_event_rate}).  

Regarding soft-spectrum sources, their discovery relies on detecting mainly neutrinos below 10~TeV, where they are more abundant.
IceCube alone can discover sources half as bright as \ngc only in the Northern Hemisphere due to its location at the South Pole.
[Due to the cut we apply, the effective area for muon neutrinos is null in most of the Southern Hemisphere (\figu{aeff_upgoing}); however, the discovery potential would deteriorate significantly in the Southern Hemisphere even without this cut (see, \eg, Fig.~3 in \Refe~\cite{IceCube:2019cia}).]

Conversely, KM3NeT alone can discover sources slightly dimmer than NGC 1068 only in the Southern Hemisphere, due to its location in the Northern Hemisphere.  IceCube or KM3NeT alone can discover sources located above the North Pole ($\sin \delta = 1$) or South Pole ($\sin \delta = -1$), respectively, even if they are about half as bright as \ngc, since in these directions the event rates are highest (\figu{skymaps_event_rate}) and the signal-to-background ratio is most favorable.

Because \plenum has a larger sky coverage than any individual detector (\figu{skymaps_event_rate}), it would allow us to discover sources regardless of their declination, even if they are significantly dimmer than NGC 1068.  Figure~\ref{fig:source_discovery_vs_declination} shows that \plenum-1 (IceCube + Baikal-GVD + KM3NeT + P-ONE) could discover sources roughly half as bright as NGC 1068 anywhere.  \plenum-2 (IceCube-Gen2 + Baikal-GVD + KM3NeT + P-ONE) could discover Northern-Hemisphere sources only 20\% as bright as NGC 1068, thanks to the large size of IceCube-Gen2 (Table~\ref{tab:detectors}).  And \plenum-3 (IceCube-Gen2 + Baikal-GVD + KM3NeT + P-ONE + NEON + HUNT + TRIDENT) could discover even dimmer sources also in the Southern Hemisphere, thanks mainly to HUNT.

Regarding hard-spectrum sources, similar trends are apparent, with key differences.
In contrast to soft-spectrum sources, the discovery of hard-spectrum sources relies more on detecting high-energy neutrinos between 10~TeV and 10~PeV. While astrophysical neutrinos at these energies are scarcer, so are atmospheric neutrinos, especially above 100~TeV, which recovers the source discovery potential. 
Because these high-energy neutrinos are more strongly absorbed while propagating through the Earth, the discovery potential of hard-spectrum sources in individual detectors is markedly better around the horizon (\ie, around $\cos \theta_z = 0$, which, for IceCube only, corresponds to $\sin \delta = 0$ in \figu{source_discovery_vs_declination}).
In these directions, the trajectories of neutrinos underground are shorter, which lessens their absorption.

As a result, discovering hard-spectrum sources is more challenging. At a minimum, we would need \plenum-2 to discover steady-state sources half as bright as \txs located at favorable locations around $\delta = 0$.
\plenum-3 could detect sources half as bright as \txs across most of the sky.

\smallskip

\textbf{\textit{Evolution of the discovery potential.---}}So far, we have demonstrated the power of combining multiple neutrino telescopes to discover sources using data-taking periods in 10-year increments (Sec.~\ref{sec:method_what_is_plenum}).  

Figure~\ref{fig:time_evolution_discovery_soft_spectrum} shows the projected time evolution of the source discovery potential using instead the staged increase of combined detector exposure over time contained in Table~\ref{tab:detectors}, which roughly reflects the current plans of each experiment.  The timeline is unavoidably tentative, based on information that is not final at present and, for simplicity, ignores the contributions from detectors running with partial configurations.

Figure~\ref{fig:time_evolution_discovery_soft_spectrum} shows the evolution of the $5\sigma$ discovery potential of NGC 1068, located at $\delta = 0$, and of a source with a similar soft spectrum, \ie, $\propto E_\nu^{-3.2}$, but located in the Southern Hemisphere, at $\delta = -30^\circ$, or in the Northern Hemisphere, at $\delta = 30^\circ$.  Extrapolating from the 10-year IceCube data release~\cite{IceCube:2019cia}, \figu{time_evolution_discovery_soft_spectrum} shows what level of high-energy neutrino flux can be discovered as global detector exposure grows over time.  
Like for \figu{source_discovery_vs_declination}, the dependence of the source discovery potential on the source location reflects the differences in expected signal and background event rates in \figu{skymaps_event_rate}.  Accordingly, the results in \figu{time_evolution_discovery_soft_spectrum} (and \figu{time_evolution_discovery_hard_spectrum}) agree with those in \figu{source_discovery_vs_declination}.

To discover a source like \ngc, located at $\delta = 0 $ but with half the flux reported by IceCube, we need about 42~Gton~yr of detector exposure to be compared to the 13-14 years that IceCube alone, with about 1~Gton, would need to discover \ngc using the \plenum analysis\footnote{IceCube needed 8.72 years to reach $5.2\sigma$ for \ngc with their state-of-the-art methods and data sets~\cite{IceCube:2022der}}.  According to our detector timeline, this could be achievable by 2034, combining the cumulative data from IceCube, KM3NeT, Baikal-GVD, and P-ONE into \plenum-1.  Otherwise, using IceCube alone would require taking data past 2050. 

In the Northern Hemisphere, a source with half the NGC 1068 flux could be discovered earlier, with about 17~Gton-yr by 2026, even using IceCube alone, since this is where its visibility via muon tracks is best (Figs.~\ref{fig:aeff_upgoing} and \ref{fig:skymaps_event_rate}).  It could even be possible to discover a dimmer source with only 20\% of the NGC 1068 flux, with about 156~Gton~yr by 2041, combining data from IceCube, KM3NeT, Baikal-GVD, P-ONE, and IceCube-Gen2 into \plenum-2 plus 1 year of data from the NEON, TRIDENT, and HUNT.

In the Southern Hemisphere, the improvement achieved by combining multiple neutrino telescopes is more evident. 
Presently, with IceCube alone, discovering a source with a flux like that of \ngc is not possible (with the data set of through-going muons that we use), given the overwhelming background of atmospheric muons from Southern-Hemisphere directions.  Only with the addition of two Northern-Hemisphere detectors, KM3NeT and Baikal-GVD, and a combined detector exposure of about 30~Gton~yr, could discovery become possible by 2031, according to our timeline.  Adding P-ONE would make it possible to discover a source half as bright as \ngc by around 2038; and adding NEON, TRIDENT, and HUNT would make it possible to discover a source 20\% as bright by around 2042 with a combined detector exposure of 215~Gton~yr.

\begin{figure}[t!]
 \centering
 \includegraphics[width=1\columnwidth]{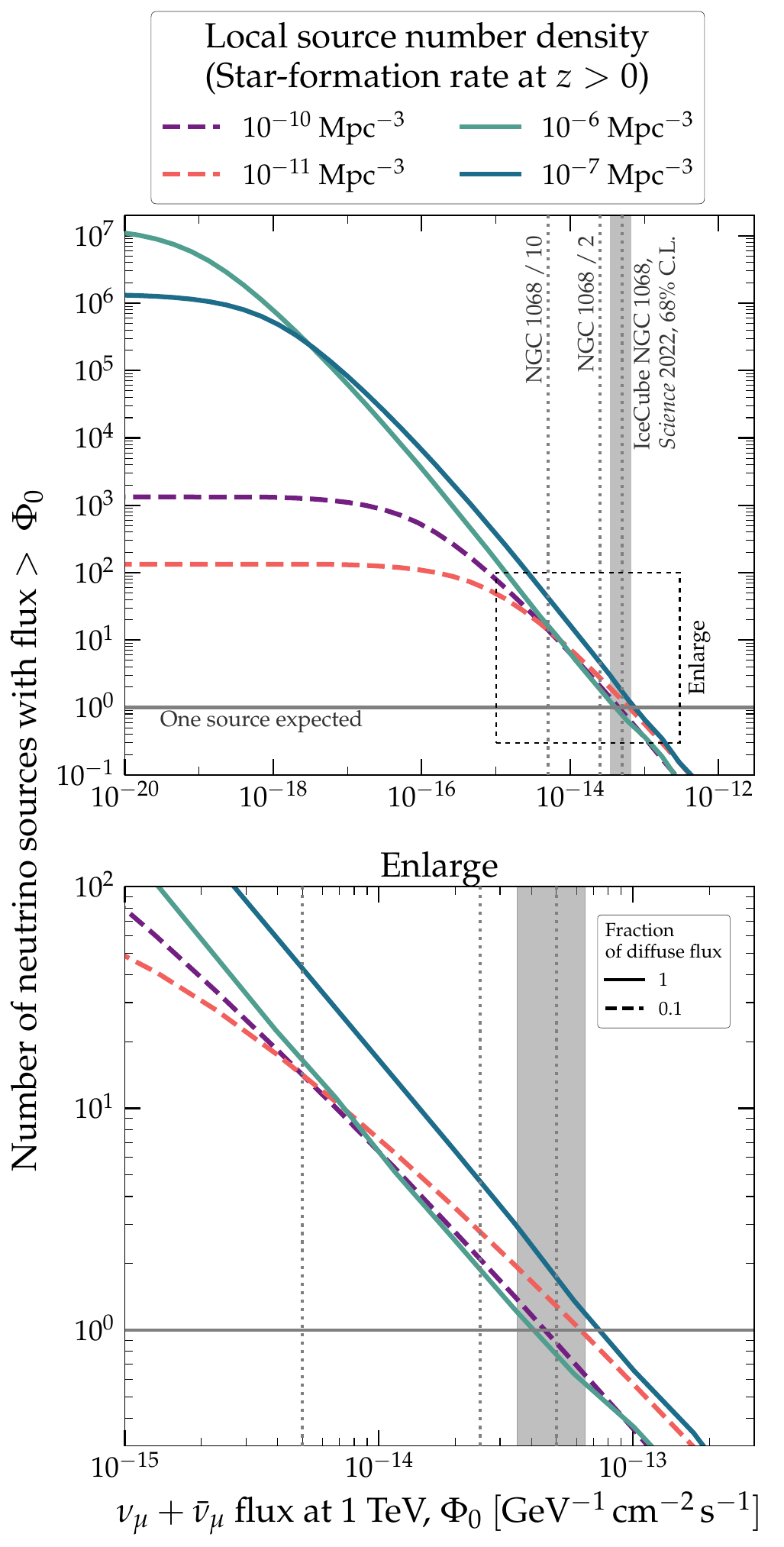}
 \caption{\textbf{\textit{Cumulative distribution of the expected number of neutrino sources above a given flux.}}
 All sources in a source population emit a soft power-law neutrino spectrum, $\Phi_0 \, (E_\nu/1~{\rm TeV})^{-3.2}$, similar to that of \ngc. Their number density follows the star-formation-rate evolution~\cite{Madau:2014bja} with the standard-candle flux per source normalized such that they sum up to 100\% (solid lines), or 10\% (dashed lines), of the diffuse flux measured at \SI{10}{TeV}, \ie, $E_\nu^2 \Phi_{\rm diffuse}\vert_{10~{\rm TeV}} = \SI{3E-8}{GeV~cm^{-2}~s^{-1}~sr^{-1}}$~\cite{Naab:2023xcz}.
 The local source number density (colors, see legend) is chosen such that, on average, about one source in the population produces a flux as bright as \ngc or brighter.
 The gray vertical band and the central gray dotted line indicate the measured flux of \ngc and the corresponding uncertainty~\cite{IceCube:2022der}.} 
 \label{fig:source_evolution_ngc}
\end{figure}

Figure~\ref{fig:time_evolution_discovery_hard_spectrum} in Appendix~\ref{app:discovery_hard} shows similar behavior for the time evolution of the discovery potential of hard-spectrum sources that have a spectrum $\propto E_\nu^{-2}$, like \txs.  Finding hard-spectrum sources over time is more challenging than soft-spectrum sources, for the same reasons outlined above regarding \figu{source_discovery_vs_declination}.

As a sanity check, we evaluate how the discovery potential for hard- and soft-spectrum sources scales with the livetime of IceCube. 
For soft-spectrum sources, we find that the discovery potential scales approximately proportional to the square root of the livetime. This is expected as the signal of soft-spectrum sources is background-dominated.
For hard-spectrum sources, we find that the discovery potential scales better
with a power of approximately 0.7, which is expected due to less background contamination at higher energies. 
Although varying with source declination and detector configuration, we recover similar scaling behaviors of discovery potential with exposure in Figure~\ref{fig:time_evolution_discovery_soft_spectrum} and Figure~\ref{fig:time_evolution_discovery_hard_spectrum}.

\medskip

Figures~\ref{fig:source_discovery_vs_declination} and \ref{fig:time_evolution_discovery_soft_spectrum} (and \ref{fig:time_evolution_discovery_hard_spectrum}) showcase the \textbf{\textit{transformative gain garnered from a distributed network of neutrino telescopes: to enable the discovery of dim neutrino sources anywhere in the sky}}.  According to our tentative detector timeline (Table~\ref{tab:detectors}), by the year 2043, it would be possible to discover a steady-state soft-spectrum neutrino source only 20\% as bright as \ngc or a hard-spectrum source half as bright as \txs, regardless of its position.

\medskip

\textbf{\textit{How many sources are discoverable.---}}Based on the above potential to discover single neutrino sources, we compute how many sources belonging to an underlying source population we could discover across the sky.  

We assume a single population of nondescript sources, distributed isotropically, whose number density evolves with redshift, $z$, following the star-formation rate~\cite{Madau:2014bja}. We use \textsc{Firesong}~\cite{Tung:2021npo} to generate the probability distribution functions of populations of neutrino sources with identical luminosities, \ie, standard-candle sources, but located at different redshifts.  We normalize the flux per source such that the sum of the fluxes from all the sources matches either 100\% or 10\% of the diffuse flux measured by IceCube at \SI{10}{TeV}~\cite{Naab:2023xcz}.  This energy falls within the range with which IceCube observed \ngc, \ie, about 1.5--15~TeV~\cite{IceCube:2022der}.

\begin{figure}[t!]
 \centering
 \includegraphics[width=0.95\columnwidth]{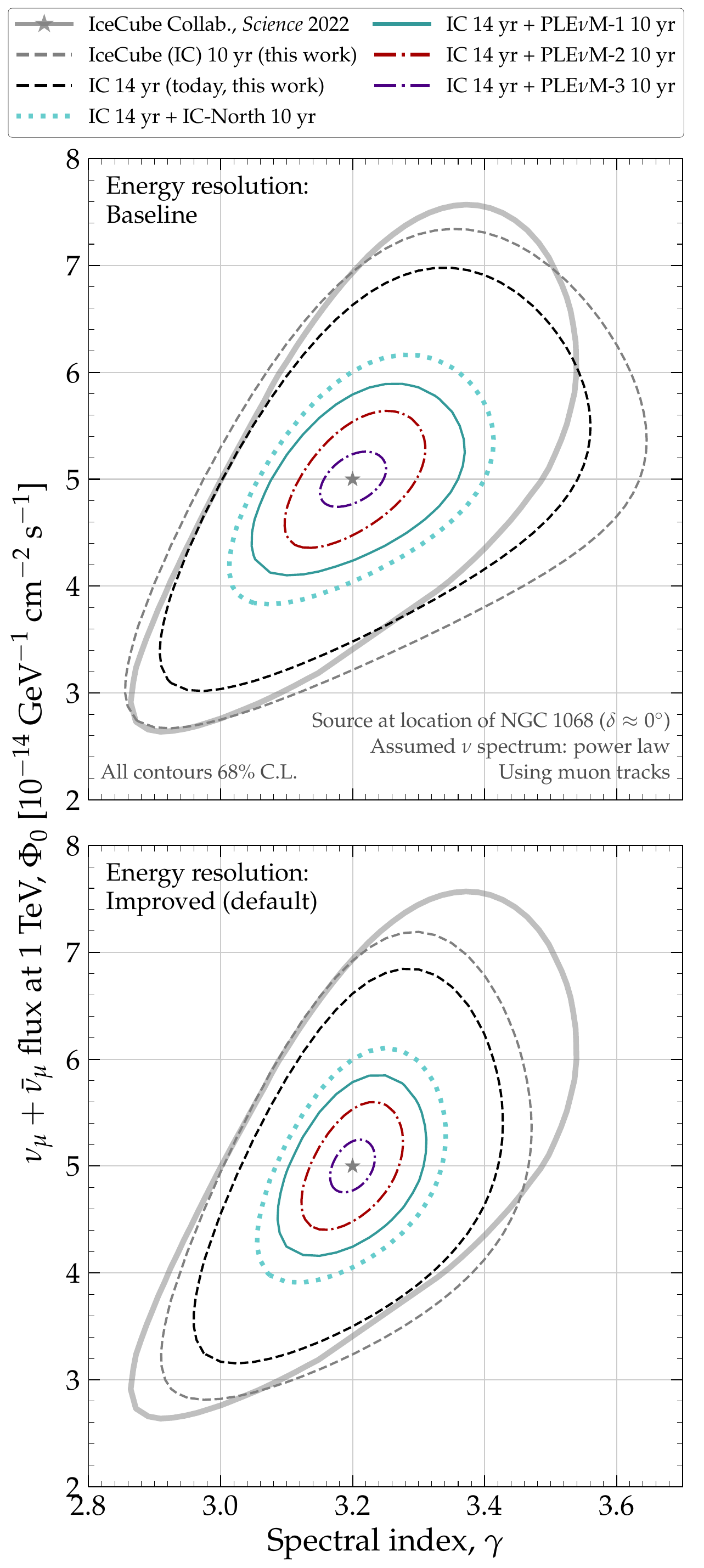}
 \caption{\textbf{\textit{Projected joint measurement of the parameters of the high-energy neutrino flux from NGC 1068.}}  
 The source emits a power-law neutrino spectrum, \ie, $\Phi_0 \, (E_\nu/1~{\rm TeV})^{-\gamma}$.  The contours are 68\%~C.L.~allowed regions of $\Phi_0$ and $\gamma$ (assumed true values in Table~\ref{tab:params}) obtained from the detection of muon tracks (Sec.~\ref{sec:statistics}), using the baseline (\textit{top}) or improved (\textit{bottom}) energy resolution (\figu{eres}).  Table~\ref{tab:param_uncertainties} shows one-dimensional parameter ranges.
 \textbf{\textit{A global network of neutrino telescopes, \plenum-3 (\figu{skymap_detector_locations}), would shrink the allowed region of flux parameters by a factor of 8 in each dimension in the 2040s \textit{vs.}~today.}}
 }
 \label{fig:ngc_contour}
\end{figure}

Figure~\ref{fig:source_evolution_ngc} shows our results.  We study four illustrative scenarios of the local source number density (\ie, at $z = 0$) that ensure that the source population produces, on average, about one source with a flux at least as high as that of \ngc.
This is necessary to fully determine the parameters of the source population.  We do not consider additional constraints on allowed source populations for the sake of simplicity.
For each scenario, we compute the all-sky number of sources in the population that emit neutrinos above a specific minimum flux, which we vary.

From \figu{source_discovery_vs_declination}, we know that \plenum-1 would be able to detect sources about 40\% as bright as \ngc anywhere in the sky with a $5\sigma$ significance; and that \plenum-3 would be able to do so for sources 10\% as bright as \ngc.  Given that, \figu{source_evolution_ngc} reveals that 
\plenum-1 will be able to detect 2--6 sources, and \plenum-3 will be able to detect 10--40 sources, depending on the local source density and on the fraction of the diffuse neutrino flux that the source population is responsible for. 

Weakening the evidence demanded for source discovery to only $3\sigma$ significance (not shown in Figs.~\ref{fig:source_discovery_vs_declination} and \ref{fig:source_evolution_ngc}) improves the threshold for discovery to about 25\% and 8\% of the brightness of \ngc for \plenum-1 and \plenum-3, respectively.
This raises the number of detectable sources to 5--10 and 20--60, respectively, again depending on the local source density and the fraction of the diffuse flux that these sources make up.
These numbers refer to individual source detections on the respective significance level.
Analyses exploiting source catalogs to stack sources, i.e., combine their individual fluxes as e.g.~presented in~\cite{Abbasi_bright_seyfert_2025,Abbasi_x_ray_agn_2025},
will improve the number of discoverable sources even more.

Although our estimates lack the complexity of modeling a specific candidate source class in detail, they show that \textbf{\textit{\plenum will be able to discover tens of new neutrino sources}}. Combining the fluxes of multiple sources from a candidate population, \ie, \textit{stacking} sources, can strongly increase the prospects for discovery compared to the discovery of single sources that we consider here. Results for specific source classes or the co-existence of multiple source populations require detailed study beyond the scope of this paper.


\subsection{Measuring astrophysical flux parameters}
\label{sec:results-astro_parameters}

Figure~\ref{fig:ngc_contour} shows the allowed regions of the PL flux parameters, $\Phi_0$ and $\gamma$, inferred from mock observations of NGC 1068, computed for the same detector configurations used in \figu{source_discovery_vs_declination} (and \figu{discrimination_soft_spectrum_vs_declination}).  Using our improved energy resolution (Sec.~\ref{sec:method_resolution}), we find that the 68\%~C.L.~allowed contour of $\Phi_0$ and $\gamma$ obtained using 10 years of IceCube approximates the result reported by the IceCube Collaboration in \Refe~\cite{IceCube:2022der}, lending credibility to our results, which use the improved resolution by default.

Table~\ref{tab:param_uncertainties} shows the one-dimensional allowed intervals of $\Phi_0$ and $\gamma$.  Using our improved energy resolution, the relative error on $\Phi_0$ shrinks by up to a factor of 8.6, from 29\% using 10 years of IceCube to 3.3\% when combining that with 10 years of \plenum-3. 
Similarly, for $\gamma$, it shrinks by up to a factor of 7.9, from about 5.7\% to 0.72\%.  This level of measurement error would allow precision tests of the shape of the energy spectrum; we explore this further below in 
Secs.~\ref{sec:results-discrimination_pl_vs_plc} and \ref{sec:results_dedicated-models}.

We only investigate how larger detector exposure will shrink statistical uncertainties. 
Given how small they are expected to be in \plenum, systematic uncertainties may surpass them. 
The statistical and systematic errors on the number of events detected by IceCube in the direction of \ngc are  $79^{+22}_{-20}\pm2$ and $3.2 \pm 0.2 \pm 0.07$ on the spectral index~\cite{IceCube:2022der}.
Currently, the main systematic uncertainties considered in IceCube in source searches with tracks are on the detection efficiency and the optical properties of the ice. Uncertainties on the expected flux of atmospheric neutrinos are factored out by using background estimates directly extracted from data.
However, these systematic uncertainties may also shrink over time with a better understanding of the above-mentioned effects, which we do not account for in our forecasts.

\smallskip

\begin{table*}[t!] 
 \begin{ruledtabular}
  \caption{\label{tab:param_uncertainties}\textbf{\textit{Projected measurement of the parameters of the high-energy neutrino flux from NGC 1068.}}  As in \figu{ngc_contour}, the source emits a power-law neutrino spectrum, \ie, $\Phi_0 \, (E_\nu/1~{\rm TeV})^{-\gamma}$.  We show the best-fit values and 68\%~C.L.~allowed intervals of $\Phi_0$ and $\gamma$, \ie, their one-dimensional intervals obtained by profiling \equ{ts_src}.  Their assumed true values are in Table~\ref{tab:params}.
  We show results for the same possible future detector scenarios (Sec.~\ref{sec:method_what_is_plenum}) shown in \figu{ngc_contour} and using the baseline and improved energy resolution (Sec.~\ref{sec:method_resolution}), the latter of which is our default choice.
  Figure~\ref{fig:ngc_contour} shows the joint allowed ranges of $\Phi_0$ and $\gamma$.}
  \centering
  \renewcommand{\arraystretch}{1.3}
  \begin{tabular}{llcc}
   Neutrino telescopes & Energy resolution &
   \multicolumn{2}{c}{Measured $\nu$ flux (PL), b.f.~$\pm 1\sigma$} 
   \\
   \cline{3-4}
   &
   &
   Normalization,\footnote{In units of $5 \cdot 10^{-14}$~GeV$^{-1}$~cm$^{-2}$~s$^{-1}$, the baseline value of $\Phi_0$ for the PL flux benchmark (Table~\ref{tab:params}).} $\Phi_0$ &
   Spectral index, $\gamma$ \\
   \hline
   IceCube 10~yr
   & Baseline & $1 \pm 0.31$ & $3.2 \pm 0.25$ \\
   & Improved\footnote{Our analysis using the improved energy resolution approximates the present-day measurements of $\Phi_0$ and $\gamma$ reported by the IceCube Collaboration using 10 years of data.} & $1 \pm 0.29$ & $3.2 \pm 0.18$ \\
   \hline
   IceCube 14~yr
   & Baseline & $1 \pm 0.26$ & $3.2 \pm 0.21$ \\
   & Improved & $1 \pm 0.24$ & $3.2 \pm 0.15$ \\
   \hline
   IceCube 14 yr + KM3NeT\footnote{IceCube-sized neutrino telescope at the location of KM3NeT.} 10 yr
   & Baseline & $1 \pm 0.15$ & $3.2 \pm 0.13$ \\
   & Improved & $1 \pm 0.14$ & $3.2 \pm 0.09$ \\
   \hline
   IceCube 14 yr + \plenum-1 10 yr
   & Baseline & $1 \pm 0.12$ & $3.2 \pm 0.11$ \\
   & Improved & $1 \pm 0.11$ & $3.2 \pm 0.075$ \\
   \hline
   IceCube 14 yr + \plenum-2 10 yr
   & Baseline & $1 \pm 0.084$ & $3.2 \pm 0.071$ \\
   & Improved & $1 \pm 0.079$ & $3.2 \pm 0.051$ \\
   \hline
   IceCube 14 yr + \plenum-3 10 yr
   & Baseline & $1 \pm 0.034$ & $3.2 \pm 0.033$ \\
   & Improved & $1 \pm 0.033$ & $3.2 \pm 0.023$ 
  \end{tabular}
 \end{ruledtabular}
\end{table*}


\subsection{Identifying a high-energy cut-off}
\label{sec:results-discrimination_pl_vs_plc}

So far, the IceCube Collaboration has reported the measurement of the neutrino energy spectrum of \ngc by determining the values of the free parameters of the power-law spectrum, $\Phi_0$ and $\gamma$ [PL, \equ{spl}]~\cite{IceCube:2022der}---see Table~\ref{tab:params} for their central values---and the normalization of the DC model (close to the normalization it predicts)~\cite{Abbasi_bright_seyfert_2025}. 
However, no further comparison between competing spectra has been performed.
To address this, we use our methods from Sec.~\ref{sec:statistics}---specifically, \equ{ts_discrimination_pl_vs_plc}---to compute the projected potential to discriminate a PL spectrum from a power-law spectrum with a high-energy cut-off [PLC, \equ{cut}].

\smallskip

\begin{figure}
 \centering
 \includegraphics[width=\columnwidth]{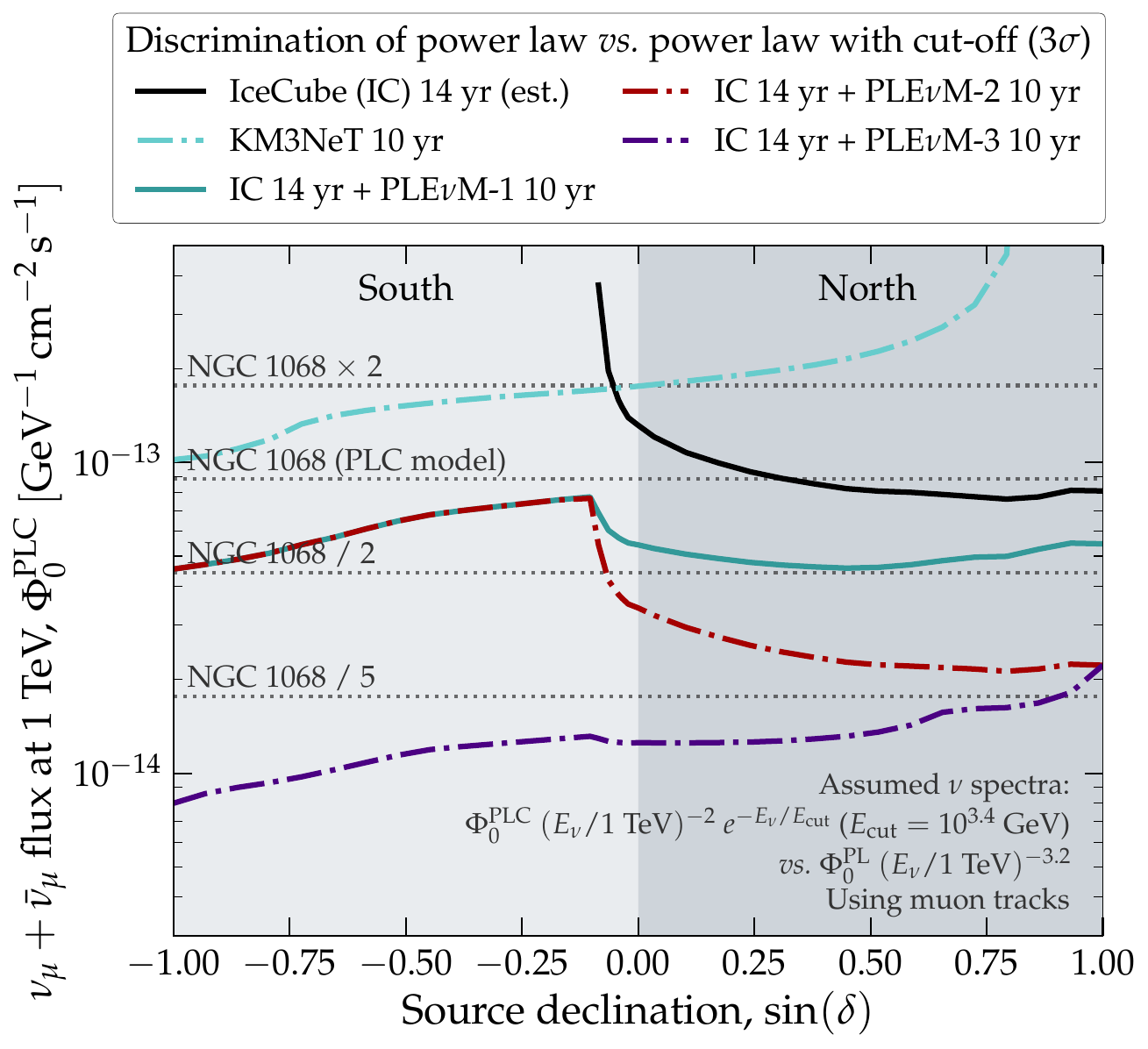}
 \caption{\textbf{\textit{Projected discrimination ($3\sigma$) between a power-law neutrino spectrum \textit{vs.}~one with a high-energy cut-off.}}  
 Results are for a point source located at declination, $\delta$, emitting a power-law-with-cut-off spectrum $\propto \Phi_0^{\rm PLC} \,E_\nu^{-2}\, e^{-E_\nu/E_{\rm cut}}$ [\equ{cut}].
 We find the value of $\Phi_0^{\rm PLC}$ that would yield discrimination between PL and PLC with a statistical significance of $3\sigma$ for a high-energy cut-off, employing muon tracks observed by one or more neutrino telescopes, using the test statistic in \equ{ts_discrimination_pl_vs_plc} and the methods in Sec.~\ref{sec:statistics}.  As benchmarks, we show the baseline flux level of NGC 1068 if it had a PLC spectrum~(Table~\ref{tab:params}), and 50\% and 20\% of it.  The 14-year IceCube result is estimated using the same methods used for the forecasts.  Table~\ref{tab:detectors} defines the detector combinations \plenum-1, \plenum-2, and \plenum-3.  See Sec.~\ref{sec:results-discrimination_pl_vs_plc} for details.
 \textbf{\textit{A globally distributed network of neutrino telescopes would enable the identification of a high-energy neutrino cut-off even for dim sources located anywhere in the sky.}}}
 \label{fig:discrimination_soft_spectrum_vs_declination}
\end{figure}

\textbf{\textit{PL vs.~PLC spectra across the sky.---}}Figure~\ref{fig:discrimination_soft_spectrum_vs_declination} shows the $3\sigma$ discrimination potential between PL and PLC, \ie, evidence for an exponential cut-off in energy, for a neutrino source depending on its declination.  As an illustration, we fix the cut-off energy in \equ{cut} to $E_{\rm cut} = 10^{3.4}$~GeV and $\gamma = 2$ to generate mock observations; these are values that we found to make the PLC flux roughly compatible with present-day IceCube measurements (Appendix~\ref{app:plc_model_choice}). Then, we allow the values of these parameters to float freely in fits to the mock observations.  
Overall, we recover a similar dependence of the discrimination potential on declination as in \figu{source_discovery_vs_declination}, primarily due to the declination-dependent signal-to-background ratio.

IceCube alone can discriminate between the PL and PLC models only for sources in the Northern Hemisphere, for the same reasons that it can only discover sources in that hemisphere when using muon tracks exclusively (Sec.~\ref{sec:results-ps_discovery}).  Yet, \figu{discrimination_soft_spectrum_vs_declination} shows that presently---with 10 years of IceCube data---discrimination at the $3\sigma$ level is only feasible for sources about twice as bright in neutrinos as NGC 1068.  (This is confirmed by our analysis of real IceCube data later, in Sec.~\ref{sec:results-exp_data}.)  Conversely, KM3NeT alone---or Baikal-GVD or P-ONE alone (not shown)---can achieve discrimination primarily for sources in the Southern Hemisphere, and only if they are more than twice as bright than \ngc.  Further, if the cut-off energy were higher, \eg, 100~TeV, the discrimination between PL and PLC would weaken, given that it would have to rely on the detection of higher-energy, scarcer neutrinos.  We leave dedicated studies on the spectra of high-energy sources like \txs for future work.

Discrimination between PL and PLC spectra at the $3\sigma$ level becomes possible only with \plenum and, even with it, remains challenging.  With \plenum-1, discrimination is only possible for sources as bright as NGC 1068 located in the Northern Hemisphere or directly above the South Pole.  With \plenum-2, discrimination in the Northern Hemisphere is possible for sources half as bright as NGC 1068---thanks to IceCube-Gen2---and, in the Southern Hemisphere, for sources as bright as NGC 1068 if they are above the South Pole or for sources up to 40\% brighter if they are elsewhere.  With \plenum-3, discrimination is possible for sources half as bright as NGC 1068 or significantly dimmer regardless of their declination---thanks primarily to HUNT---but especially for sources in the Southern Hemisphere, where it is possible even if they are only 20\% as bright as NGC 1068.

\smallskip

\begin{figure*}
 \centering
 \includegraphics[width=\textwidth]{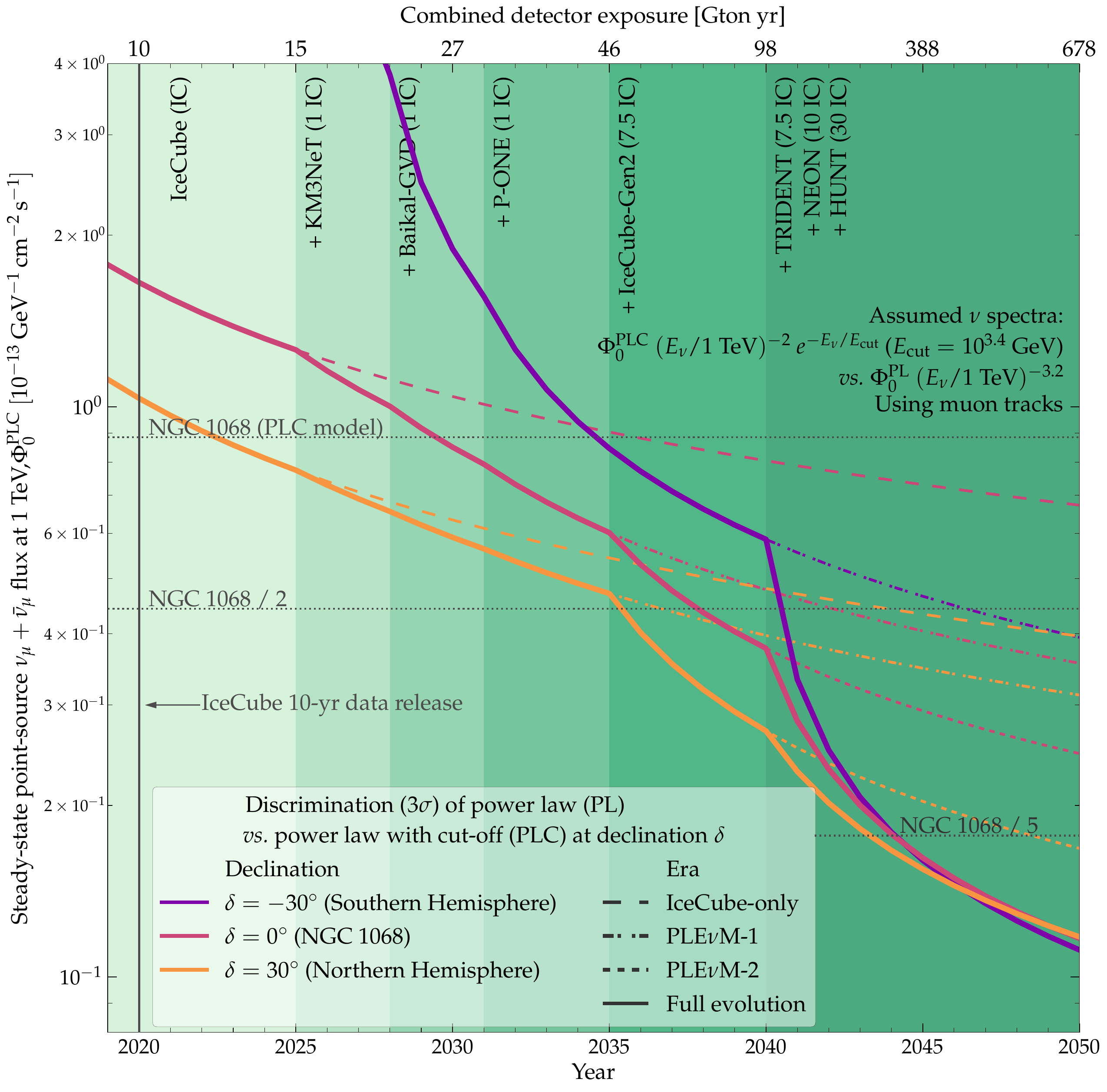}
 \caption{\textbf{\textit{Projected evolution of the discrimination potential ($3\sigma$) between a power-law neutrino spectrum \textit{vs.}~one with a high-energy cut-off.}}  Results are for a point source located at declination, $\delta$, emitting power-law-with-cut-off spectrum $\propto \Phi_0^{\rm PLC} \,E_\nu^{-2}\, e^{-E_\nu/E_{\rm cut}}$ [\equ{cut}].
 We find the value of $\Phi_0^{\rm PLC}$ that would yield discrimination between PL and PLC with a statistical significance of $3\sigma$ for a high-energy cut-off, employing muon tracks observed by one or more neutrino telescopes, using the methods in Sec.~\ref{sec:statistics}.  We show results for sources at three illustrative declinations: $\delta = -30^\circ$, 0, and $30^\circ$; \figu{source_discovery_vs_declination} shows results for other choices.  As benchmarks, we show the baseline flux level of NGC 1068 if it had a PLC spectrum~(Table~\ref{tab:params}), and 50\% and 20\% of it.   The start date of future detectors is staggered and follows the tentative timeline in Table~\ref{tab:detectors}. Each detector is a scaled version of IceCube (Table~\ref{tab:detectors}), translated to the detector location (\figu{skymap_detector_locations}); see Sec.~\ref{sec:method}. The top x-axis shows the accumulated exposure of \textbf{\textit{all}} available detectors up to the year of the bottom x-axis; it is only valid for the solid lines, \ie, ``Full evolution''. Table~\ref{tab:exposure} shows all summed exposures for the different eras. See Sec.~\ref{sec:results-ps_discovery} for details.  \textbf{\textit{By 2044, the cumulative exposure of a global network of neutrino telescopes could enable the identification of a high-energy neutrino cut-off even for dim neutrino sources located anywhere in the sky.}}}
 \label{fig:time_evolution_discrimination_pl_vs_plc}
\end{figure*}

\textbf{\textit{Evolution of the discrimination potential.---}}Figure~\ref{fig:time_evolution_discrimination_pl_vs_plc} shows, similarly to \figu{time_evolution_discovery_soft_spectrum}, the time evolution of the $3\sigma$ discrimination potential for NGC 1068, located at $\delta = 0$, and for a source with a similar soft spectrum, \ie, $\propto E_\nu^{-3.2}$, but located in the Southern Hemisphere, at $\delta = -30^\circ$, or in the Northern Hemisphere, at $\delta = 30^\circ$.  Like in \figu{source_discovery_vs_declination}, we fix the cut-off energy to $10^{3.4}$~GeV for illustration.  The evolution of the discrimination potential in \figu{time_evolution_discrimination_pl_vs_plc} follows a similar trend as the evolution of the discovery potential in \figu{time_evolution_discovery_soft_spectrum}.

To discriminate between the PL and PLC spectra for a source located at $\delta = 0$ and as bright as NGC 1068, we need about 27~Gton~yr of detector exposure, or about three times the exposure IceCube needed to discover \ngc~\cite{IceCube:2022der}.  According to our detector timeline, this could be tentatively achievable by 2030, combining the cumulative data from IceCube and two Northern Hemisphere detectors.  Otherwise, using IceCube alone, discrimination would require taking data until 2035. 

In parts of the Northern Hemisphere, discrimination between PL and PLC for sources as bright as \ngc may already be possible today without the need for detectors larger than IceCube.  
However, no source as bright as \ngc is known today in the Northern Hemisphere to which this analysis could be applied.
After 2035, discrimination becomes possible for sources half as bright as NGC 1068 by combining the cumulative data of IceCube, KM3NeT, Baikal-GVD, and P-ONE. 

In the Southern Hemisphere, discrimination between PL and PLC for sources as bright as NGC 1068 becomes possible by 2035 thanks to the combination of KM3NeT, Baikal-GVD, and P-ONE, with a combined exposure of 46~Gton~yr.
By 2041, \plenum-3 will be able to discriminate between spectra for sources all over the sky and half as bright as \ngc, thanks primarily to NEON, TRIDENT, and, especially, HUNT.  Achieving this will need a combined detector exposure of at least 157~Gton~yr.

\medskip

Figures~\ref{fig:discrimination_soft_spectrum_vs_declination} and \ref{fig:time_evolution_discrimination_pl_vs_plc} showcase, similarly to what Figs.~\ref{fig:source_discovery_vs_declination} and \ref{fig:time_evolution_discovery_soft_spectrum} do for discovery, another aspect of the \textbf{\textit{transformative gain garnered from a distributed network of neutrino telescopes: to enable the discrimination between alternative neutrino spectra, even for dim neutrino sources, anywhere in the sky}}.  According to our tentative detector timeline (Table~\ref{tab:detectors}), by the year 2044, with a combined detector exposure of 330~Gton~yr, it would be possible to discriminate between the PL and PLC neutrino spectra for sources only 20\% as bright as \ngc, regardless of their declination.

\subsection{Testing dedicated flux models of NGC 1068}
\label{sec:results_dedicated-models}

Earlier, in \figu{flux_models}, we showed the disk-corona (DC)~\cite{Kheirandish:2021wkm} and torus-wind (TW) flux models~\cite{Inoue:2022yak}, built to explain the neutrino emission from NGC 1068.  Below, we compute the potential to discover them over the atmospheric background and to distinguish them from the PL model.

\smallskip

\textbf{\textit{Discovering the DC and TW models.---}}Using our methods to compute event rates (Sec.~\ref{sec:event_rate-calculation}), the DC and TW flux models yield 40 and 31 muon tracks in IceCube, respectively, after a live time of 3186 days, the same used in the NGC 1068 analysis by the IceCube Collaboration~\cite{IceCube:2022der}.  These are fewer than the 98 tracks obtained earlier for the PL and PLC fluxes (\figu{signal_model}) because the DC and TW fluxes are lower at TeV energies.  

To assess the discovery potential, we use the same test statistic in \equ{ts_src} as before, but take the above mock event distributions generated using the DC and TW models as observed data and fit to them a PL flux with freely floating normalization, $\Phi_0$, and spectral index, $\gamma$.  This yields a best-fit value of $\Phi_0$ smaller than the baseline PL normalization of NGC 1068 (Table~\ref{tab:params}) by 30\% for the DC model and by 40\% for the TW model, and a best-fit value of $\gamma = 2.8$ in both cases, to be compared with the best-fit value of 3.2 of the baseline PL spectral index of NGC 1068 (Table~\ref{tab:params}).  The 68\%~C.L.~allowed regions of $\Phi_0$ and $\gamma$ (not shown) overlap with the allowed regions that we derived earlier using mock data generated with a PL flux (\figu{ngc_contour}), though only marginally.

We find that using 10 years of IceCube data alone, the DC  model could be discovered as an excess over the atmospheric background with a significance of $3.7\sigma$, and the TW flux model, with $2.9\sigma$.  In both cases, the significance is smaller than when fitting a PL flux to the experimental data (Sec.~\ref{sec:results-ps_discovery}), due again to the DC and TW models predicting a smaller flux at a few TeV.  With 14 years of IceCube plus 10 years of \plenum-1, both models would comfortably exceed the $5\sigma$ discovery threshold. 

\smallskip

\textbf{\textit{Distinguishing from a power law.---}}Next, we assess the significance with which, given an observed sample of neutrinos from NGC 1068, the DC and TW models can be distinguished from the PL model.

To do this, first, we mitigate the marginal incompatibility mentioned above between the allowed ranges of $\Phi_0$ and $\gamma$ obtained when assuming the DC and TW models to be true when generating mock event distributions and fitting them with a PL model, \textit{vs.}~the ranges obtained when assuming that the PL model is true when generating mock event distributions and fitting them also with it.
We do so by altering the flux normalization of the DC and TW models so that they match what we observe in experimental data (Sec.~\ref{sec:results-exp_data}).
We then fit a PL model to the DC and TW models, in turn, to find the values of $\Phi_0$ and $\gamma$ that best approximate the DC and TW models. 

We do this to accurately model the case where the DC or TW model represents the true flux of \ngc, in order to compute the significance of either of them against the PL model that we \textit{would have fit if the DC or TW model represented the truth.}
As a consequence, the distinction between the renormalized DC and TW models \textit{vs.}~the PL model that we compute below stems from a difference in the shape of their event energy distributions rather than a difference in their flux normalization.  

Then, we compute mock event distributions using the renormalized DC and TW models and take them as observed data.  We modify the test statistic in \equ{ts_discrimination_pl_vs_plc} by using the renormalized DC or TW flux model as the signal hypothesis (\ie, the numerator) and the PL model as the null hypothesis (\ie, the denominator).
Because the PL model parameters are not a subset of the DC or TW model parameters, Wilks' theorem does not hold in this case.
Therefore, we calculate the significance below based on mock experiments describing the null hypothesis made using the PL model.

We find that using 14 years of IceCube data alone, the distinction of the DC and TW models from the PL model is marginal: the DC model can be distinguished from it with a significance of about $1.5\sigma$, and the TW model, with about $1\sigma$.  Using \plenum-1 in addition, the distinction reaches $2.5\sigma$ and $2\sigma$, respectively; using \plenum-2 in addition, it reaches well beyond $3\sigma$ for both models.
With \plenum-3, we expect to detect around 1200 neutrinos for the TW model, 1600 for the DC model, or 4500 neutrinos for the PL model from \ngc.
At this level of detection rate, it may become feasible to fit piecewise power-law spectra within narrow energy intervals, to determine the energy spectrum directly rather than fitting to pre-established models.
Doing this is left for future work.

\medskip

Our results illustrate that \textbf{\textit{a global network of neutrino telescopes could test the predictions of dedicated models of the high-energy neutrino flux from steady-state point sources.}}


\subsection{Applying the \plenum tools to real IceCube data}
\label{sec:results-exp_data}

\begin{figure}[h!]
 \centering
 \includegraphics[width=\columnwidth]{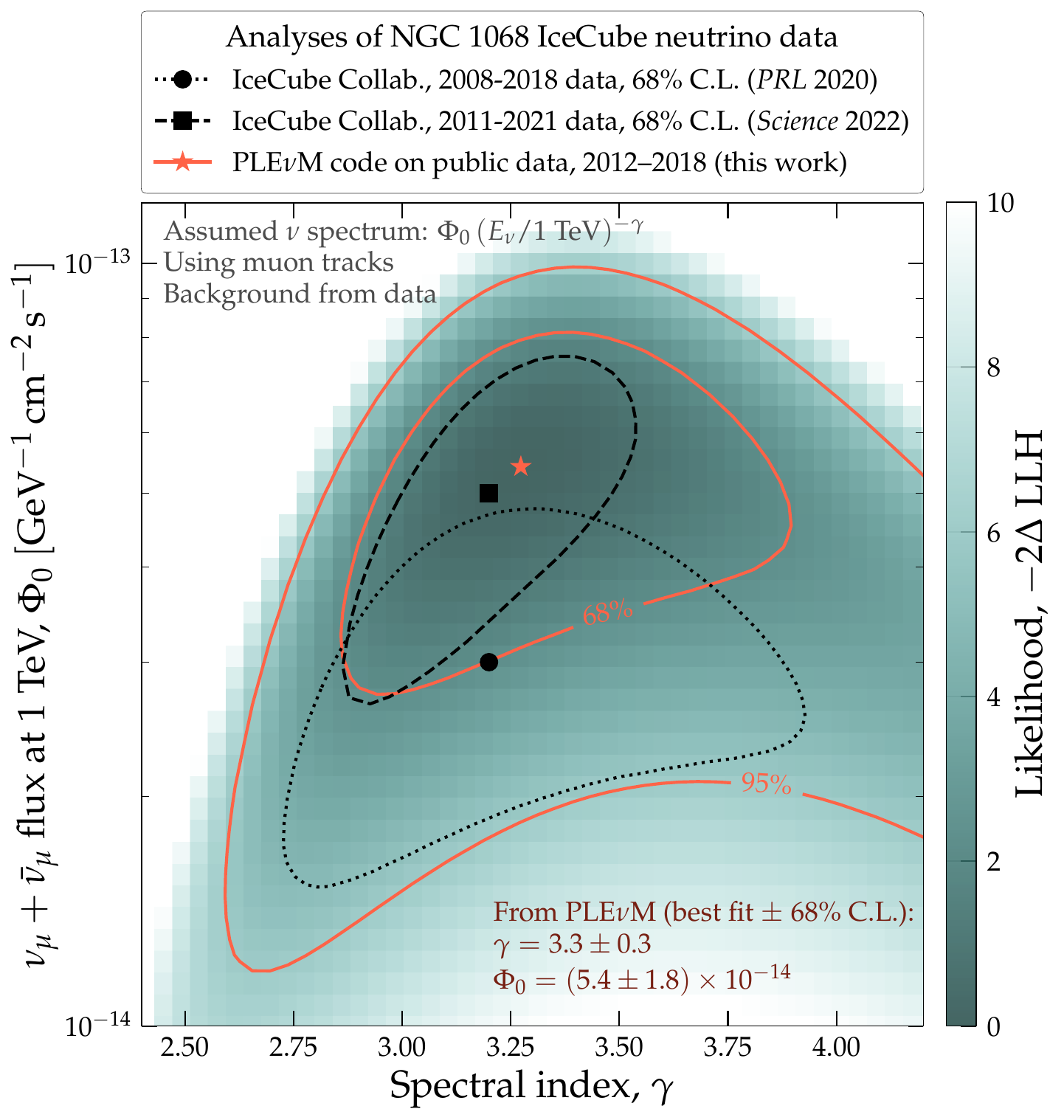}
 \caption{
 \textbf{\textit{Joint measurement of the NGC 1068 flux parameters using real, present-day IceCube data.}}  We use the methods introduced in Secs.~\ref{sec:method} and \ref{sec:statistics} to analyze public IceCube muon-track data collected in 2012--2018~\cite{IceCube:2019cia} (data-taking periods IC86-II to IC86-VII).   The source emits a power-law neutrino spectrum (Sec.~\ref{sec:flux_models}), \ie, $\Phi_0 \, (E_\nu/1~{\rm TeV})^{-\gamma}$.   For comparison, we show the results reported by the IceCube Collaboration in 2020~\cite{IceCube:2019cia} and 2022~\cite{IceCube:2022der}.  See Sec.~\ref{sec:results-exp_data} for details, including differences between our results and earlier ones.}
 \label{fig:exp_2Dscan}
\end{figure}

As a further test of our methods and their implementation in the \plenum~code, we use them to repeat our analyses above, but now using real data collected by IceCube.  We use the publicly available data sample of muon tracks collected in the period 2012--2018~\cite{IceCube:2019cia} using the latest IceCube event selection scheme.  This is the subset of the full sample during which the detector configuration and data-taking period are the ones for which our choices of IceCube effective area (Sec.~\ref{sec:method_effective_area}) and resolution functions (Sec.~\ref{sec:method_resolution}) apply.  The subset contains 761,162 events---about two-thirds of the full sample---and corresponds to a live time of \SI{2198.2}{days}, or just over 6 years---about 61\% of the full sample.

\smallskip

\textbf{\textit{Rediscovering NGC 1068.---}}We search for \ngc in the IceCube data using the same methods as before, with two prominent differences.  First, in our analysis of Asimov data, we estimated the atmospheric background using \textsc{Daemonflux} (\figu{atm_flux}).  Now, we estimate it from the data.  We do this by selecting all data within $3^\circ$ of the \textit{declination} of \ngc, then removing events within a radius of $3^\circ$ of \ngc, \ie, the events that are actually used in the analysis, and finally randomizing the remaining data in right ascension. This strategy is similar to what was done in the IceCube analysis in \Refe~\cite{IceCube:2019cia}.  
Second, as described above, we only use a subset of data with the latest, uniform event selection and the full detector configuration with 86 strings.
With this prescription, and using the same test statistic as before, \equ{ts_src}, \ie,  assuming a PL spectrum from \ngc, we rediscover it in the public IceCube data with a local $p$-value of \num{3.5E-4} (about $3.4\sigma$).
[The IceCube collaboration reports a local $p$-value of \num{1.8E-5} in~\cite{IceCube:2019cia}, which uses the full data release, and \num{1E-7} in~\cite{IceCube:2022der}, which uses the newer, updated data set].

Figure~\ref{fig:exp_2Dscan} shows the joint measurement of the PL parameters, $\Phi_0$ and $\gamma$, resulting from our fit, compared to the IceCube result reported in \Refe~\cite{IceCube:2022der}.  The allowed one-dimensional intervals are $\Phi_0 = (5.4 \pm 1.8) \cdot 10^{-14}$~GeV$^{-1}$~cm$^{-2}$~s$^{-1}$ and $\gamma = 3.3\pm 0.3$, where the errors are statistical only.  These intervals and also \figu{exp_2Dscan} show that our results are compatible with those reported by the IceCube Collaboration~\cite{IceCube:2019cia, IceCube:2022der}.  As a cross-check, we have found that using \textsc{Daemonflux} for the atmospheric background yields similar results.

The difference between our results and those reported by the IceCube Collaboration in \figu{exp_2Dscan} has multiple origins.  First, the different analyses shown in \figu{exp_2Dscan} use different event samples, so even if the analyses run on them were identical, we would expect the results to be somewhat different.  Second, we used a binned likelihood based on Poisson statistics instead of the unbinned likelihood methods applied by the IceCube Collaboration~\cite{IceCube:2019cia, IceCube:2022der};
we have not investigated the impact of using a binned \textit{vs.} unbinned likelihood.
Third, the public IceCube data release on which we base our work provides only binned instrument response functions---especially coarse in declination---instead of the full information used in internal IceCube analyses.
These are not fundamental limitations of the \plenum tool but merely a consequence of the simplifications made in the data release and its use in the present analysis.

\smallskip

\textbf{\textit{Looking for a high-energy cut-off.---}}We also fit the IceCube data using the PLC flux model, this time letting the value of $E_{\rm cut}$ float in addition to $\Phi_0$ and $\gamma$.  We do not find a value of $E_{\rm cut}$ that is below the maximum energy of the events close to \ngc, about $\SI{53.7}{TeV}$.  By fixing $\gamma = 2$, we find $E_{\rm cut} = \SI{3}{TeV}$, in agreement with what we had found using Asimov data ($E_{\rm cut} = 10^{3.4}~{\rm GeV} \approx 2.5~{\rm TeV}$; Appendix~\ref{app:plc_model_choice}) when requiring that the PLC model approximates the PL parameters reported in \Refe~\cite{IceCube:2022der}.  We find that there is no statistical power in the IceCube data that we use to reject the PL hypothesis, \ie, we find no indication of high-energy cut-off in the neutrino spectrum of \ngc. 

\smallskip

\textbf{\textit{Testing dedicated flux models.---}}Finally, we also fit the IceCube data with the DC and TW flux models, keeping the shape of their energy spectrum fixed (\figu{flux_models}) but allowing their normalization to float freely relative to their baseline values.  We find, for the DC model, a best-fit normalization 50\% higher than its baseline value; and, for the TW, a best-fit normalization 90\% higher than its baseline value.  This is consistent with our test on mock data (Sec.~\ref{sec:results_dedicated-models}), where we found a smaller flux normalization when fitting the baseline models with a PL spectrum.
Both flux models yield a significance of about $3.3\sigma$ with respect to the atmospheric background, nearly the same as for the PL model ($3.4\sigma$). Dedicated mock experiments confirm there is no statistical power to distinguish the DC and TW models from the PL model in the present experimental data.  


\subsection{Future developments}
\label{sec:results-future}

The \plenum framework is under constant development.
In addition to incorporating future improvements in energy and angular resolution, we envision two advances that would improve our calculations.

First, including event signatures other than the through-going muon tracks we have used---specifically, cascades and starting events---would enlarge the sample of detected events.  This would strengthen the source discovery potential~\cite{Sudoh:2023qrz}, especially in water-based neutrino telescopes, like the ongoing KM3NeT~\cite{KM3Net:2016zxf} and Baikal-GVD~\cite{Baikal-GVD:2019kwy}, where the angular resolution of cascades is expected to be as good as degree-scale.  (References~\cite{Fiorillo:2022rft, Liu:2023flr} contain projections based on combining events of different types.)

Second, including the instrument response functions specific to different detectors would more accurately represent their capabilities, which is important for future analyses based on real data.  In particular, detectors optimized for lower energies, \eg, KM3NeT-ORCA~\cite{KM3Net:2016zxf} or the IceCube Upgrade~\cite{Ishihara:2019aao}, might provide complementary information on soft-spectrum sources, which yield more lower-energy neutrinos.



\section{Summary and outlook}
\label{sec:summary}

The spectacular findings in the decade since the IceCube discovery of high-energy astrophysical neutrinos have provided us with new insight into high-energy astrophysics and particle physics.  Yet today, key questions remain unanswered; notably, what are the sources of the bulk of TeV--PeV neutrinos detected and whether they are the long-sought sources of ultra-high-energy cosmic rays.  Progress, while steady, is slowed by the fact that high-energy neutrino sources seem to be many and dim rather than few and bright and by the naturally low detection rate of neutrinos.  If a strategy is absent to surmount these hurdles, key questions might remain unanswered for a long time.

We have shown that the right strategy is at hand, made possible by new high-energy neutrino telescopes presently under construction, prototyping, and planning.  They are IceCube-sized or significantly larger---yielding substantially higher detection rates---and placed at various geographical locations---enabling them to search for astrophysical neutrino sources in different parts of the sky with prime sensitivity.
By combining their observations in a PLanEtary Neutrino Monitoring (\plenum) network, we have shown that immense improvements will be possible in the next 10--20 years: by 2040 (2045), a combined neutrino detector exposure up to 7 (28) times higher than what is available today (\figu{time_evolution_discrimination_pl_vs_plc}) and a greatly enlarged field of view of the neutrino sky (\figu{skymaps_event_rate}).

To explore this quantitatively, we have developed the \plenum code: a publicly available~\cite{github-plenum}, unified framework to analyze the combined data, mock and real, from multiple neutrino telescopes.  Presently, it considers, in addition to IceCube, TeV--PeV in-ice and in-water Cherenkov detectors Baikal-GVD and KM3NeT---currently under construction---IceCube-Gen2 and P-ONE---planned for the 2030s---and HUNT, NEON, and TRIDENT---planned for the 2040s (see Table~\ref{tab:detectors}).  

We have illustrated the upcoming power generated by combining neutrino telescopes by making projections for the discovery of point-like steady-state high-energy neutrino sources and for the characterization of their neutrino energy spectra.  To do so, we have used the projected detection of through-going muon tracks made by $\nu_\mu + \bar{\nu}_\mu$, modeled after the recent public IceCube 10-year data release. This allows us to capture the realistic experimental nuances involved in neutrino detection.  Further, we have adopted an 
energy resolution that approximates state-of-the-art methods in IceCube~\cite{IceCube:2022der}.  We have verified (Figs.~\ref{fig:ngc_contour} and \ref{fig:exp_2Dscan}) that our methods yield results that approximate those reported by IceCube~\cite{IceCube:2019cia} in the discovery of the first steady-state source of high-energy neutrinos, \ngc.

In our projections, we have considered three benchmark detector combinations representative of the imminent, near- and far-future (\figu{skymap_detector_locations}): \plenum-1, composed of IceCube, KM3NeT, Baikal-GVD, and P-ONE, achievable in the early 2030s; \plenum-2, composed of IceCube-Gen2, KM3NeT, Baikal-GVD, and P-ONE, achievable in the mid-2030s; and \plenum-3, composed of IceCube-Gen2, KM3NeT, Baikal-GVD, P-ONE, HUNT, NEON, and TRIDENT, achievable in the 2040s.  

Given the limited public availability of the detailed envisioned performance of future detectors, we have modeled each future detector as a scaled-up version of IceCube, relocated to the location of the detector on Earth.  These simplifying assumptions are temporary and can be revisited in the future.  Our main findings are:
\begin{description}[style=unboxed]
 \item[Higher detection rate (\figu{skymaps_event_rate})]
  Combining detectors will raise the global detection rate of high-energy neutrinos roughly by a factor of 2--4 in the near future with \plenum-1,  10 in the 2030s with \plenum-2, and tens in the 2040s with \plenum-3, according to our tentative detector timeline (Table~\ref{tab:detectors}).
 \item[Full-sky field of view (\figu{skymaps_event_rate_4fgl})]
  Combining detectors will expand the global field of view in high-energy neutrinos to the full sky, covering known neutrino sources and hundreds of known gamma-ray sources that might also be neutrino sources.
 \item[Finding dim sources (Figs.~\ref{fig:source_discovery_vs_declination},  \ref{fig:time_evolution_discovery_soft_spectrum}, \ref{fig:time_evolution_discovery_hard_spectrum})] 
  Adding 10 years of \plenum-1 data to present-day IceCube data would allow discovering, with $5\sigma$ statistical significance, soft-spectrum ($\propto E_\nu^{-3.2}$) steady-state neutrino sources half as bright as \ngc anywhere in the sky (\figu{source_discovery_vs_declination}).  With \plenum-2, sources as dim as 20\% of \ngc, or dimmer, could be discovered in the Northern Hemisphere; and, with \plenum-3, anywhere.  Using IceCube alone, achieving this would require taking data well past the year 2050 and would be even more difficult for Southern-Hemisphere sources (\figu{time_evolution_discovery_soft_spectrum}).  Discovering hard-spectrum sources ($\propto E_\nu^{-2}$) is more challenging (\figu{time_evolution_discovery_hard_spectrum}): \plenum-2 is needed to discover steady-state sources half as bright as the \txs in the Northern Hemisphere (\figu{source_discovery_vs_declination}) and \plenum-3 is necessary to discover them anywhere in the sky.
 \item[Precise neutrino spectra~(\figu{ngc_contour})]
  Using \plenum-3, the normalization, $\Phi_0$, and spectral index, $\gamma$, of a power-law neutrino spectrum from \ngc could be measured with a relative statistical error of 3.3\% and 0.72\%, respectively, an improvement of a factor of about 8 compared to the values obtained using 10 years of IceCube data (Table~\ref{tab:param_uncertainties}).
  \item[Identifying a high-energy cut-off~(Figs.~\ref{fig:discrimination_soft_spectrum_vs_declination}, \ref{fig:time_evolution_discrimination_pl_vs_plc})]
   Individually, neither present-day IceCube nor 10 years of future KM3NeT or Baikal-GVD can discriminate, with $3\sigma$ statistical significance, between a power-law neutrino spectrum ($\propto E_\nu^{-\gamma}$) and one augmented with a high-energy cut-off ($\propto E_\nu^{-\gamma} e^{-E_\nu/E_{\rm cut}}$, with $E_{\rm cut} \sim~{\rm TeV}$), as expected from cosmic-ray acceleration (\figu{discrimination_soft_spectrum_vs_declination}).  Adding 10 years of \plenum-1 data to present-day IceCube data would allow discrimination for sources as bright as \ngc, mainly in the Northern Hemisphere.  With \plenum-2, discrimination would be possible for Northern-Hemisphere sources half as bright as \ngc; and, with \plenum-3, for sources as dim as 20\% in parts of the Southern Hemisphere.
\end{description}

Our calculations highlight the discovery opportunities starting to become available to a growing field. These opportunities do not arise from the exploitation of individual experiments in isolation but from an open and collaborative community effort. Shared technological advancements and the development of common analysis techniques form the foundation for the future combination of results.  We present our forecasts and make our tools public in the hope of encouraging this process.

\medskip

\textbf{\textit{Code availability.---}} All code and calculations presented here are currently available with an open-source license at GitHub~\href{https://github.com/PLEnuM-group/Plenum}{\faGithubSquare}. This \href{https://github.com/PLEnuM-group/Plenum/blob/dev/notebooks/misc/plotting_standalone.ipynb}{notebook} 
reproduces all figures of this publication and links to all necessary data files. Please cite our code when you use it in your work.




\medskip

\section*{Acknowledgments}

We thank Ali Kheirandish for kindly providing the disk-corona model flux expectations for \ngc, Christian Haack for providing the toy model for generating effective areas, and Chad Finley, Erin O'Sullivan, and Ignacio Taboada for their suggestions.  MB is supported by the \textsc{Villum Fonden} under project no.~29388.  This work was also supported by the Science and Technology Facilities Council, part of the UK Research and Innovation (Grant Nos.~ST/W00058X/1 and ST/T004169/1), and by the UCL Cosmoparticle Initiative.


\appendix


\renewcommand{\theequation}{A\arabic{equation}}
\renewcommand{\thefigure}{A\arabic{figure}}
\renewcommand{\thetable}{A\arabic{table}}
\setcounter{figure}{0} 
\setcounter{table}{0}

\begin{figure*}[t!]
 \centering
 \includegraphics[width=\textwidth, trim={0.3cm 0 0 0.1cm 0.5cm}, clip]{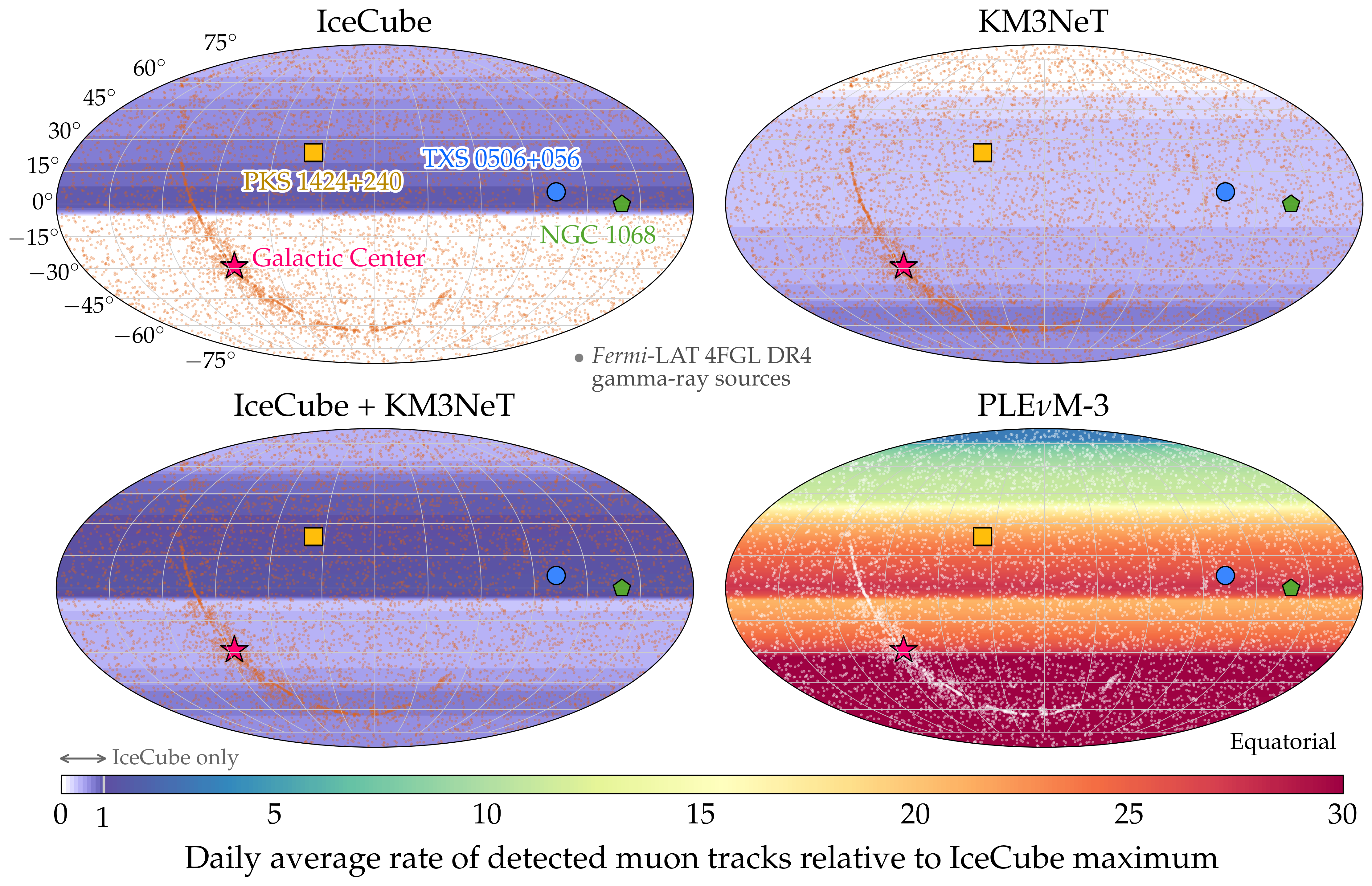}
 \caption{Expected daily \textit{\textbf{averaged}} rate of muon tracks detected in present and future neutrino telescopes, compared to the positions of known high-energy astrophysical sources.}
 \label{fig:integrated_skymaps_event_rate_4fgl}
\end{figure*}
\FloatBarrier
\section{A note on scaling effective areas}
\label{app:aeff_scaling}

As noted in~Sec.~\ref{sec:method_similar_detectors}, we scale up the effective areas of future, larger detectors with their respective volumes relative to IceCube's volume of about \SI{1}{km^3}.
While this is a good approximation for neutrino events that interact \textit{inside} the detector volume, the scaling becomes more complex for our data set of \textit{throughgoing muons}, which can have interaction vertices outside the detection volume.
Scaling instead with the surface area or the projected area orthogonal to the direction of muon direction was suggested as a possible improvement.

To evaluate alternative scaling choices, we employ a toy simulation for effective areas where we scale the height and radius of cylindrical detectors. 
The simulation is based on calculating transmission and interaction probabilities of neutrinos and the range of the produced muon. 
The simulation does not account for detection and selection efficiencies.
Even without these efficiencies, we find that it reproduces the features of the actual effective area of IceCube reasonably well in the Northern Hemisphere.
Still, it remains a stark simplification.

We analyze the number of expected events for a hard-spectrum and a soft-spectrum source at various declinations, for different detector locations on Earth, and for cylindrical detectors with varying heights and radii.
We focus on cylinders from \SIrange{0.75}{1.25}{km} in height and from \SIrange{0.5}{3}{km} in radius, thus approximating the detector geometries expected for the future telescopes.

Our results can be summarized as follows: no simple scaling of the detector effective area (proportional to volume, surface area, volume$^{2/3}$, projected area, radius, or radius squared) can describe all considered cases similarly well. 
For hard-spectrum sources, the number of events scales indeed roughly with the orthogonally projected detector area.  But for soft-spectrum sources it scales roughly as volume$^{0.9}$ (however, this scaling might be an overly specific result for our toy simulations).
Averaged over all the cases we explored, scaling with surface area seems the best choice, while the variation per case remains large.

Based on this investigation, we retain our choice of scaling the effective area with the detector volume to produce our results. This choice over-estimates the number of events by, at worst, around 50\% for soft-spectrum sources. As our analysis method performs up to 50\% worse than the IceCube state-of-the-art analysis, these two effects partially cancel each other out, such that our main conclusions remain robust.

As mentioned in Sec.~\ref{sec:method_similar_detectors}, the real effective areas of future detectors will not only depend on the overall detector geometry but also on other factors, such as the detection medium, spacing between detection modules, data selection, and event reconstruction.
Once more elaborate simulations become available for the future detectors, our forecasts can be revisited using more realistic estimates of their effective areas.


\section{Daily averaged event rates}
\label{app:averaged_rates}

\renewcommand{\theequation}{B\arabic{equation}}
\renewcommand{\thefigure}{B\arabic{figure}}
\renewcommand{\thetable}{B\arabic{table}}
\setcounter{figure}{0} 
\setcounter{table}{0} 

Same as \figu{skymaps_event_rate} and \figu{skymaps_event_rate_4fgl}, \figu{integrated_skymaps_event_rate_4fgl} shows the expected rate of muon track, but \textit{averaged} over Earth's daily rotation.

\section{PLC model choices}
\label{app:plc_model_choice}

\renewcommand{\theequation}{C\arabic{equation}}
\renewcommand{\thefigure}{C\arabic{figure}}
\renewcommand{\thetable}{C\arabic{table}}
\setcounter{figure}{0} 
\setcounter{table}{0} 

In Sec.~\ref{sec:flux_models} of the main text, we introduced the power-law flux model with a high-energy cut-off (PLC) to describe soft-spectrum sources akin to \ngc.  In Table~\ref{tab:params}, we presented the baseline values we adopted for the PLC model parameters, $\Phi_0$, $\gamma$, and $E_{\rm cut}$.  Below, we describe how we choose these baseline values.

\begin{figure}[h!]
 \centering
 \includegraphics[width=\columnwidth]{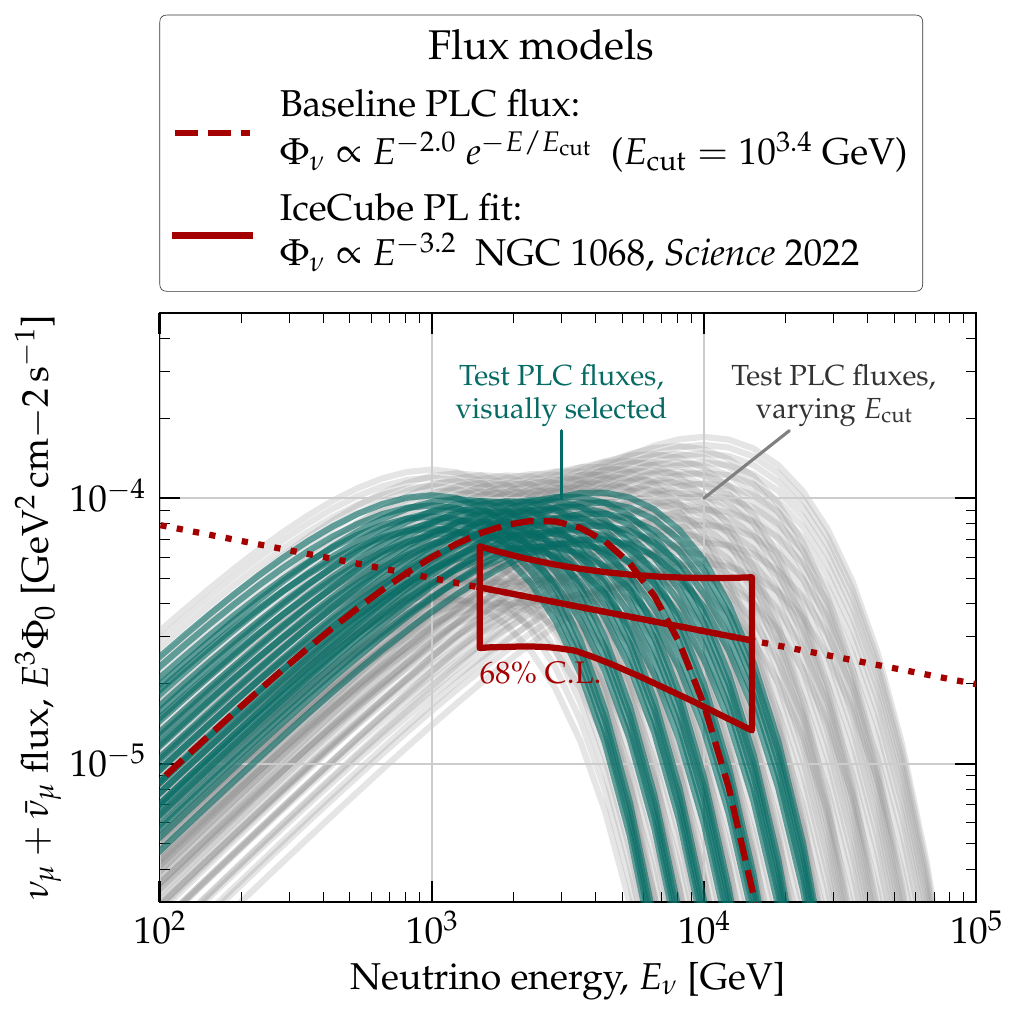}
 \caption{ 
 \textbf{\textit{Selecting our baseline PLC neutrino flux model.}}  Comparison of test PLC fluxes computed using \equ{cut} with different choices of the cut-off energy, $E_{\rm cut}$, against the IceCube PL fit to \ngc observations~\cite{IceCube:2022der}.  Of the test PLC fluxes, we choose as our baseline the one with $E_{\rm cut} = 10^{3.4}$~GeV.}
 \label{fig:model_matching}
\end{figure}

Figure \ref{fig:model_matching} outlines our procedure. First, we set $\gamma = 2$ and varied $E_{\rm cut}$ within 1--10~TeV.  (We describe our choice of value of $\Phi_0$ later.)  For each choice of $E_{\rm cut}$, we generated the resulting PLC flux.  
We fit the resulting PLC fluxes with a PL flux and compare it against the PL flux of \ngc reported by the IceCube Collaboration, which is $\propto E_\nu^{-3.2}$~\cite{IceCube:2022der}.  Then, we visually selected the PL fits within the IceCube \ngc 68\% C.L. allowed contour in \figu{model_matching}, and the underlying PLC fluxes they were fitted to.  This selection is qualitative, not meant to be rigorous.
Of these PLC fluxes, we chose the one with $E_{\rm cut} = 10^{3.4}~{\rm GeV} \approx 2.5~{\rm TeV}$, as, for this choice, the peak of the energy flux lies well within the energy range of 1.5--15~TeV in which IceCube observes \ngc.  Further, for this choice, the corresponding PL flux has $\gamma = 3.1$. close to the observed $\gamma = 3.2$.
We choose the value of $\Phi_0$ so that our PL fit $\propto E_\nu^{-3.1}$ approximates closely the IceCube \ngc PL fit.

The above procedure selects a baseline PLC model to test our capacity to distinguish it from the PL model (Sec.~\ref{sec:results-discrimination_pl_vs_plc}).  Our simple selection procedure cannot claim which PLC models are compatible with IceCube data in a statistically rigorous way; for this, see Sec.~\ref{sec:results-exp_data} in the main text and state-of-the-art analysis like the one in \Refe~\cite{Naab:2023xcz}. 

\FloatBarrier


\section{Discovery potential for hard-spectrum sources}
\label{app:discovery_hard}

\renewcommand{\theequation}{D\arabic{equation}}
\renewcommand{\thefigure}{D\arabic{figure}}
\renewcommand{\thetable}{D\arabic{table}}
\setcounter{figure}{0} 
\setcounter{table}{0} 

\begin{figure*}
 \centering\includegraphics[width=\textwidth]{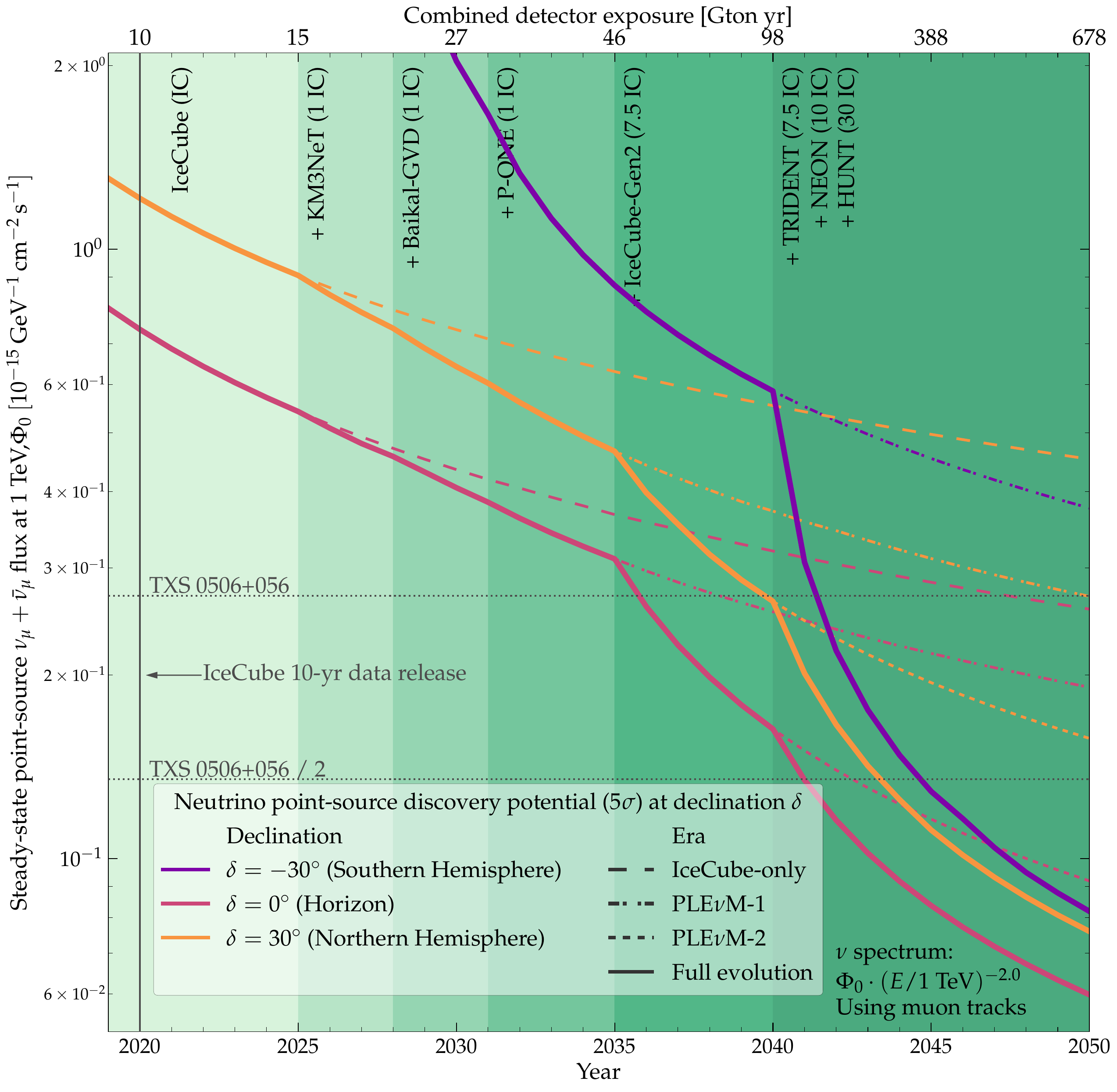}
 \caption{\textbf{\textit{Projected evolution of the discovery potential ($5\sigma$) of a steady-state point source of high-energy neutrinos with a hard energy spectrum.}}
 The source has a power-law spectrum, $\Phi_0 (E_\nu / 1~{\rm TeV})^{-\gamma}$, with $\gamma = 2.0$, as measured for \txs~\cite{IceCube:2022der}, and we find the value of $\Phi_0$ that would yield discovery with a statistical significance of $5\sigma$, employing muon tracks observed by one or more neutrino telescopes, using the methods in Sec.~\ref{sec:statistics}.  We show results for sources at three illustrative declinations: $\delta = -30^\circ$, 0, and $30^\circ$; \figu{source_discovery_vs_declination} shows results for other choices.  As benchmarks, we show the baseline flux level measured for \txs (Table~\ref{tab:params}), and 50\% of it. The start date of future detectors is staggered and follows the tentative timeline in Table~\ref{tab:detectors}.   Each detector is a scaled version of IceCube (Table~\ref{tab:detectors}), translated to the detector location (\figu{skymap_detector_locations}); see Sec.~\ref{sec:method}.  The top x-axis shows the accumulated exposure of \textbf{\textit{all}} available detectors up to the year of the bottom x-axis; it is only valid for the solid lines, \ie, ``Full evolution''. Table~\ref{tab:exposure} shows all summed exposures for the different eras. See Sec.~\ref{sec:results-ps_discovery} for details.
  }
 \label{fig:time_evolution_discovery_hard_spectrum}
 \end{figure*}

Figure \ref{fig:time_evolution_discovery_hard_spectrum} shows the time evolution of the discovery potential for hard-spectrum sources with energy spectrum $\propto E_\nu^{-2}$, akin to that of \txs.  This complements \figu{time_evolution_discovery_soft_spectrum} in the main text, which shows the discovery potential for a soft-spectrum source akin to \ngc.

\FloatBarrier

\renewcommand{\theequation}{E\arabic{equation}}
\renewcommand{\thefigure}{E\arabic{figure}}
\renewcommand{\thetable}{E\arabic{table}}
\setcounter{figure}{0} 
\setcounter{table}{0} 

\begin{table}[h!]
 \begin{ruledtabular}
  \caption{\textbf{Detector exposure per detector era and year.}  This table follows the tentative timeline and detector definitions in Table~\ref{tab:detectors} in the main text.}
  \label{tab:exposure}
  \renewcommand{\arraystretch}{1.1}
  \begin{tabular}{lrrrr}
   \multirow{2}{*}{Year} & \multicolumn{4}{c}{Detector exposure [\si{Gton~yr}]} \\
   \cline{2-5}
   & IceCube & \plenum-1 & \plenum-2 & \plenum-3 \\
   \hline
   2018 & 8.0 & $\cdots$ & $\cdots$ & $\cdots$ \\
   2019 & 9.0 & $\cdots$ & $\cdots$ & $\cdots$ \\
   2020 & 10.0 & $\cdots$ & $\cdots$ & $\cdots$ \\
   2021 & 11.0 & $\cdots$ & $\cdots$ & $\cdots$ \\
   2022 & 12.0 & $\cdots$ & $\cdots$ & $\cdots$ \\
   2023 & 13.0 & $\cdots$ & $\cdots$ & $\cdots$ \\
   2024 & 14.0 & $\cdots$ & $\cdots$ & $\cdots$ \\
   2025 & 15.0 & 15.0 & $\cdots$ & $\cdots$ \\
   2026 & 16.0 & 17.0 & $\cdots$ & $\cdots$ \\
   2027 & 17.0 & 19.0 & $\cdots$ & $\cdots$ \\
   2028 & 18.0 & 21.0 & $\cdots$ & $\cdots$ \\
   2029 & 19.0 & 24.0 & $\cdots$ & $\cdots$ \\
   2030 & 20.0 & 27.0 & $\cdots$ & $\cdots$ \\
   2031 & 21.0 & 30.0 & $\cdots$ & $\cdots$ \\
   2032 & 22.0 & 34.0 & $\cdots$ & $\cdots$ \\
   2033 & 23.0 & 38.0 & $\cdots$ & $\cdots$ \\
   2034 & 24.0 & 42.0 & $\cdots$ & $\cdots$ \\
   2035 & 25.0 & 46.0 & 46.0 & $\cdots$ \\
   2036 & 26.0 & 50.0 & 56.5 & $\cdots$ \\
   2037 & 27.0 & 54.0 & 67.0 & $\cdots$ \\
   2038 & 28.0 & 58.0 & 77.5 & $\cdots$ \\
   2039 & 29.0 & 62.0 & 88.0 & $\cdots$ \\
   2040 & 30.0 & 66.0 & 98.5 & 98.5 \\
   2041 & 31.0 & 70.0 & 109.0 & 156.5 \\
   2042 & 32.0 & 74.0 & 119.5 & 214.5 \\
   2043 & 33.0 & 78.0 & 130.0 & 272.5 \\
   2044 & 34.0 & 82.0 & 140.5 & 330.5 \\
   2045 & 35.0 & 86.0 & 151.0 & 388.5 \\
   2046 & 36.0 & 90.0 & 161.5 & 446.5 \\
   2047 & 37.0 & 94.0 & 172.0 & 504.5 \\
   2048 & 38.0 & 98.0 & 182.5 & 562.5 \\
   2049 & 39.0 & 102.0 & 193.0 & 620.5 \\
   2050 & 40.0 & 106.0 & 203.5 & 678.5 
  \end{tabular}
 \end{ruledtabular}
\end{table}

\FloatBarrier
\section{Exposure per detector configuration}
\label{app:exposure}

Table~\ref{tab:exposure} lists the detector exposure  per detector combination and year.
These values are used for the calculations summarized in Figs.~\ref{fig:time_evolution_discovery_soft_spectrum}, \ref{fig:time_evolution_discrimination_pl_vs_plc}, and \ref{fig:time_evolution_discovery_hard_spectrum}.
On any given year, the rightmost nonzero value in this table corresponds to the top x-axis in these plots, \ie, to the largest possible exposure per year when taking all available detectors together.
\balance

\clearpage
\bibliography{references.bib}  

\end{document}